\pdfoutput=1
\documentclass{jfm}

\usepackage{graphicx}
\usepackage{natbib}
\usepackage{journals}
\usepackage{amsmath}
\usepackage{upmath}
\usepackage{amssymb}
\usepackage{longtable}
\usepackage{rotating}
\usepackage{multirow}
\usepackage{amsbsy}
\usepackage[colorlinks=True,citecolor=blue,linkcolor=blue,%
			urlcolor=blue,pdftex]{hyperref}

\newcommand{\tbli}{\lambda_T^{i}}
\newcommand{\tblo}{\lambda_T^{o}}
\newcommand{\ubli}{\lambda_U^{i}}
\newcommand{\ublo}{\lambda_U^{o}}

\newcommand{\tempgrad}{\beta_T}
\newcommand{\lp}{\left(}
\newcommand{\rp}{\right)}

\newcommand{\modif}[1]{#1}

\usepackage{xcolor}

\def\vec#1{\ensuremath{\mathchoice{\mbox{\boldmath$\displaystyle#1$}}
{\mbox{\boldmath$\textstyle#1$}}
{\mbox{\boldmath$\scriptstyle#1$}}
{\mbox{\boldmath$\scriptscriptstyle#1$}}}}

\newcommand{\parallelsum}{\mathbin{\!/\mkern-4mu/\!}}

\newsavebox{\astrutbox}
\sbox{\astrutbox}{\rule[-5pt]{0pt}{20pt}}

\title[Scaling regimes in spherical shell rotating convection]{Scaling regimes 
in spherical shell rotating convection}

\author[T. Gastine, J. Wicht and J. Aubert]%
{Thomas Gastine$^{1,2}$%
  \thanks{Email address for correspondence: gastine@ipgp.fr},\ns
 Johannes Wicht$^2$, \ns Julien Aubert$^1$}

\affiliation{$^1$Institut de Physique du Globe de Paris, Sorbonne Paris Cit\'e, 
Universit\'e Paris-Diderot, UMR 7154 CNRS, 1 rue Jussieu, F-75005 Paris, 
France, \\ $^2$Max Planck Institut f\"ur Sonnensystemforschung, 
Justus-von-Liebig-Weg 3, 37077 G\"ottingen, Germany.}

\pubyear{2016}
\volume{650}
\pagerange{119--126}
\date{?; revised ?; accepted ?. - To be entered by editorial office}
\begin{document}

\maketitle

%
%
\begin{abstract}
Rayleigh-B\'enard convection in rotating spherical shells can be considered 
as a simplified analogue of many astrophysical and geophysical fluid flows. 
Here, we use three-dimensional direct numerical simulations to study this 
physical process. We construct a dataset of more than 200  numerical models
that cover a broad parameter range with Ekman numbers spanning $3\times 
10^{-7} \leq E \leq 10^{-1}$, Rayleigh numbers within the range $10^3 < Ra < 
2\times 10^{10}$ and a Prandtl number unity. \modif{The radius ratio $r_i/r_o$ 
is 0.6 in all cases and the gravity is assumed to be proportional to $1/r^2$.}
We investigate the scaling 
behaviours of both local (length scales, boundary layers) and global (Nusselt 
and Reynolds numbers) properties across various physical regimes from onset of 
rotating convection to weakly-rotating convection. Close to critical, the 
convective flow is dominated by a triple force balance between viscosity, 
Coriolis force and buoyancy. For larger supercriticalities, a small subset of 
our numerical data approaches the asymptotic diffusivity-free scaling of 
\modif{rotating convection $Nu\sim Ra^{3/2}E^{2}$} in a narrow fraction of 
the parameter space delimited by $6\,Ra_c \leq Ra \leq 0.4\,E^{-8/5}$. Using a 
decomposition of the viscous dissipation rate into bulk and boundary layer 
contributions, we establish a theoretical scaling of the flow velocity that 
accurately describes the numerical data. In rapidly-rotating turbulent 
convection, the fluid bulk is controlled by a triple force balance between 
Coriolis, inertia and buoyancy, while the remaining fraction of the dissipation 
can be attributed to the viscous friction in the Ekman layers. Beyond $Ra 
\simeq E^{-8/5}$, the rotational constraint on the convective flow is 
gradually lost and the flow properties continuously vary to match the regime 
changes between rotation-dominated and non-rotating convection. We show that 
the quantity $Ra E^{12/7}$ provides an accurate transition parameter to 
separate rotating and non-rotating convection.
\end{abstract}


\section{Introduction}

Convection-driven flows under the influence of rotation are an ubiquitous 
physical phenomenon in the fluid interiors of natural objects. The 
liquid iron core of terrestrial planets, the envelopes of gas giants, or the 
convective regions of rapidly-rotating cool stars harbour highly turbulent 
convective flows strongly constrained by the dominant role of the Coriolis 
force. Rayleigh-B\'enard convection (hereafter RBC) is a classical framework to 
examine the influence of rotation on turbulent convection. In its canonical 
form, rotating RBC consists of a planar fluid 
layer confined between two horizontal rigid plates separated from a distance 
$L$, rotating about the vertical axis with a constant rotation rate $\Omega$. 
In this setup configuration, convective motions are driven by a fixed imposed 
temperature contrast $\Delta T$ between the two plates. The dynamics is then 
controlled by three dimensionless numbers, namely the Rayleigh number $Ra = 
\alpha_T g L^3 \Delta T / \nu\kappa$, the Ekman number $E=\nu/\Omega L^2$, and 
the Prandtl number $Pr=\nu/\kappa$, where $\nu$ and $\kappa$ are the viscous 
and thermal diffusivities, $\alpha_T$ is the thermal expansivity and $g$ is the 
gravity. A combination of these parameters, named the convective Rossby number 
$Ro_c = Ra^{1/2}E/Pr^{1/2}$, is frequently employed as a reasonable proxy of 
the ratio between the global-scale buoyancy and Coriolis forces 
\citep{Gilman77}. The key issue in RBC is to explore the efficiency of the heat 
and momentum transports across the layer. Important quantities in this regard 
are the Reynolds number $Re$ and the dimensionless heat transport, defined by 
the Nusselt number $Nu = \mathcal{Q} L /\rho c_p \kappa \Delta T$, where 
$\mathcal{Q}$ is the total heat flux, $\rho$ is the density and $c_p$ the heat 
capacity. The Nusselt number is the most widely studied diagnostic since it can 
be easily measured experimentally and numerically and then compared to 
numerical and theoretical predictions. The understanding of the scaling 
dependence of $Nu$ upon the control parameters $Ra$, $E$ and $Pr$ is of 
paramount importance to identify the different regimes and to possibly 
extrapolate the scaling behaviours to natural objects.

Rotational constraints delay the onset of convection and 
the critical Rayleigh number increases with increasing rotation rates as 
$Ra_c\sim E^{-4/3}$ when $E\rightarrow 0$ \citep{Chandra61}. The 
convective pattern takes the form of elongated Taylor columns which have a 
typical horizontal size $\ell\sim E^{1/3}L$ and are aligned with the 
rotation axis. Beyond $Ra_c$, the heat transport rises much more rapidly than 
for non-rotating convection \citep[e.g.][]{Rossby69,Boubnov90}. Hence, if the 
Rayleigh number is continuously increased far beyond $Ra_c$ at a given Ekman 
number, the heat transfer properties will eventually transition to a state 
where rotational effects become secondary. In this weakly-rotating regime with 
$Ro_c \gg 1$, the scaling properties become essentially reminiscent to 
non-rotating RBC \citep[e.g.][]{Liu97,Zhong09}. The heat transport scaling is 
then expected to become independent of the Ekman number and to approach the 
scalings obtained in classical RBC, i.e. $Nu\sim Ra^{\nu_{\text{eff}}}$, 
with $0.27 \leq \nu_{\text{eff}} \leq 1/3$ for $10^5 \leq Ra \leq 10^{12}$ 
\citep[e.g.][]{Grossmann00, Funfschilling05, Ahlers09, Chilla12}.

A turbulent regime of rotating convection is expected when both $Ro_c \ll 1$ 
and $Re \gg 1$. This implies large supercriticalities in combination with low 
Ekman numbers to ensure that the rotational 
constraints are not lost. This parameter range  is therefore particularly 
difficult to explore with current-day laboratory experiments and numerical 
simulations \citep[see][]{Aurnou15}. In the following, we make the assumption 
that the heat transport scaling can be written as
\[
 Nu \sim Ra^{\alpha}E^{\beta}Pr^{\gamma}\,
\]
in this regime. In general though, the values of the exponents $\alpha$, $\beta$ 
and $\gamma$ might well continuously vary with $Ra$ similarly to classical RBC 
\citep{Grossmann00}. 
In contrast to non-rotating convection where the heat transport is controlled 
by diffusive processes in the thermal boundary layers, a dominant fraction of 
the temperature difference is accommodated in the fluid bulk when $Ro_c \ll 1$ 
\citep[e.g.][]{Boubnov90,Schmitz09,Kunnen10,Julien12,King13}. We can therefore 
hypothesise that the heat transport is controlled by 
the fluid bulk rather than by the thermal boundary layers that play an 
important role in the weakly-rotating limit. Because the viscous dissipation is 
rather weak in the fluid interior, \modif{we make the assumption that $Nu$ will 
thus become} independent of the diffusivities $\nu$ and $\kappa$ \modif{in the 
asymptotic limit of rapidly-rotating convection} \citep{Gillet06,Jones15}. 
From the definition of $Ra$, $E$, $Pr$ and $Nu$, this requirement yields the 
following combination of the scaling exponents
\[
 -\alpha+\beta+\gamma=0, \quad \alpha+\gamma=1\,.
 \label{eq:scalingExpos}
\]
\modif{In the limit of non-rotating convection, the dependence on $E$ vanishes
(i.e. $\beta=0$) and the previous relationship between the scaling exponents
relating $\alpha$ and $\gamma$ yields the so-called \emph{ultimate regime} of 
classical RBC $Nu\sim Ra^{1/2}Pr^{1/2}$ \citep{Kraichnan62}.}
Numerical models of rotating convection by \cite{King12} further indicate that 
the heat flux might only depend on the supercriticality $Ra/Ra_c$ when $Ro_c 
\ll 1$ \citep[see also][]{Julien12a,Stellmach14}. This second hypothesis yields 
$\beta = 4\alpha/3$ when $E \rightarrow 0$ and allows us to derive the 
following diffusivity-free scaling for rotating convection
\begin{equation}
 Nu \sim Ra^{3/2}E^2Pr^{-1/2}\,.
 \label{eq:asymptotic}
\end{equation}
\cite{Gillet06} derive the same equation under the hypothesis that a triple 
force balance between Coriolis force, inertia and buoyancy controls the 
asymptotic regime of rapidly-rotating turbulent convection \citep[see 
also][]{Stevenson79,Barker14}. The analysis carried out by \cite{Julien12a} 
leads to the same inviscid scaling in the framework of 
asymptotically-reduced equations expected to hold when $E \rightarrow 
0$. The evidence for such a low-$Ro_c$ scaling law is however strongly debated. 
Specifically, \cite{King12} suggest a much steeper heat transport scaling $Nu 
\sim Ra^3E^4$ for $E=\mathcal{O}(10^{-5})$ and $Ro_c=\mathcal{O}(0.1)$,
based on laboratory experiments \modif{in water} complemented by numerical 
simulations of rotating convection with rigid mechanical boundaries in 
cartesian coordinates \citep[see also][]{King13}. Similar numerical models by 
\cite{Schmitz09} that instead employ stress-free boundaries rather found $Nu 
\sim (Ra\,E^{4/3})^{1.22}$ for a comparable parameter range. This discrepancy 
implies that the viscous boundary layers might still have a direct influence on 
the heat transport even at low $E$. The recent comparative study by 
\cite{Stellmach14} for rigid and stress-free numerical models at lower 
Ekman numbers 
$E=\mathcal{O}(10^{-7})$ indeed reveals an active role of the Ekman boundary 
layers \citep[see also][]{Kunnen16,Plumley16}. While the stress-free models 
gradually approach the diffusivity-free scaling (\ref{eq:asymptotic}) 
\citep[see also][]{Barker14}, the Ekman pumping in the cases with rigid 
boundaries leads to increasing scaling exponents when $E$ is
decreased towards geophysical values \citep{Cheng15,Julien16}.
This prominent role of the boundary layers therefore questions the 
relevance of the inviscid scaling (\ref{eq:asymptotic}) for rigid boundaries.

Although the spherical geometry is more natural for studying rotating 
convection in astrophysical and geophysical objects, the majority of the 
rotating RBC laboratory experiments developed over the past three decades have 
been carried out in planar and cylindrical cells, in which the rotation axis 
and the gravity are aligned. The cartesian geometry also allows the computation 
of efficient local direct numerical simulations that operate at low Ekman 
numbers \citep[e.g.][]{Kunnen10,Ecke14,Horn15}. Hence, the fruitful 
interplay between numerical and laboratory experiments in planar or cylindrical 
geometry enabled a complementary coverage of the parameter space in the 
low-$Ro_c$ regime \citep[e.g.][]{Aurnou07,Stellmach14,Cheng15,Aurnou15}.
However, it remains unclear whether the planar RBC results can be directly 
applied to rotating convection in spherical geometry. Two specific features
of thermal convection in rotating spherical shells are indeed expected to yield 
significant dynamical differences with the planar or cylindrical RBC setups:
(\textit{i}) in most of the fluid volume of spherical shells, gravity is 
inclined with respect to the rotation axis; (\textit{ii}) due to both curvature 
and radial variations of the gravitational acceleration, spherical RBC
features a significant asymmetry between the hot and the cold bounding 
surfaces \citep[e.g.][]{Bercovici89,Jarvis93,Gastine15}.

Besides the micro-gravity experiments that operate at relatively large Ekman 
numbers due to their moderate sizes \citep[$E =\mathcal{O}(10^{-3})$, 
see][]{Hart86,Egbers03}, most of the laboratory investigations of spherical 
rotating RBC make use of the centrifugal force as a surrogate for the radial 
gravitational acceleration 
\citep{Busse74,Cardin94,SumitaOlson03,Shew05}.
The combined influence of the laboratory gravity and the centrifugal 
acceleration of the rotating vessel indeed allows to generate surfaces of 
gravity potential close to the spherical surfaces in the lower
hemisphere of the spherical shell \citep[see for a review][]{Cardin15}. This 
technique was first employed to explore the onset of convection 
in rotating spherical RBC \citep{Cordero92}. \cite{SumitaOlson03} studied 
the scaling behaviours of the heat transport for a relatively low Ekman number
($E\simeq 5\times 10^{-6}$) but with a strong convective forcing ($Ra\geq 
200\,Ra_c$). In this parameter range, they obtained $Nu \sim 
Ra^{0.41}$, a scaling exponent that is hardly steeper than non-rotating RBC and 
therefore suggests that Coriolis force only plays a minor role on the heat 
transfer.

Numerical simulations of three-dimensional convection in spherical shells have 
been developed since the late 1970s as a complementary approach to the 
laboratory experiments \citep[e.g.][]{Gilman77,Tilgner97,Christensen02}.
\cite{Christensen06}, later complemented by \cite{King10}, conducted a 
systematic parameter study of the scaling properties of convective dynamo 
models with rigid boundaries and reported the scaling $Nu \sim Ra^{6/5}E^{8/5}$ 
for $E=\mathcal{O}(10^{-5})$. These studies were however carried out in 
the presence of a self-sustained magnetic field that can possibly impinge on 
the heat transport. More recently, \cite{Yadav16} studied the 
influence of both the presence of a magnetic field and the nature of the 
mechanical boundary condition on the heat transfer in spherical RBC.
Magnetic and non-magnetic models were found to exhibit similar heat transfer 
scaling behaviours possibly because of the parameter space limitation to $E =  
\mathcal{O}(10^{-5})$ \citep[see also][]{Soderlund12,Garcia14}. Furthermore, 
these authors  did not observe the steep 
increase of the $Ra$-scaling exponent beyond $3/2$ reported in the cartesian 
calculations \citep[e.g.][]{King12} when rigid boundaries were employed.
Since the heat transport in spherical shells is dominated by the equatorial 
regions where the gravity is nearly perpendicular to the rotation axis 
\citep{Tilgner97,Yadav16}, it is not entirely surprising that the scaling 
properties differ from the cartesian models that would best represent the 
high-latitude dynamics of a spherical shell.

While $Nu$ is an ubiquitous diagnostic quantity studied in both laboratory 
experiments and numerical models; numerical calculations also enable the 
computation of additional diagnostics that can provide direct insights
on the dynamical regimes. In that regard, the scaling analysis of 
the convective flow speed $Re$ and the typical length scale 
$\ell$ can be used to disentangle the underlying force balance in rotating RBC 
\citep{Aubert01,KingBuffett13}. Two theoretical scalings for $Re$ and $\ell$ 
have been put forward. The first one, frequently called the 
inertial scaling of rotating convection, hypothesises a triple force balance 
between Coriolis, inertia and Archimedean forces \citep[CIA scaling, 
see][]{Stevenson79,Cardin94,Aubert01,Barker14}, while the second rather 
relies on a triple balance between viscosity, Archimedean and Coriolis forces 
\citep[VAC scaling, see][]{King13}. These two concurrent theories 
however lead to scaling exponents for 
$Re(Ra,E,Pr)$ close to each other. Using the numerical dataset by 
\cite{Christensen06}, \cite{KingBuffett13} even demonstrated that the
limited data can support both $Re$-scalings at
the same statistical confidence level. This study indicates a sizeable role 
played by viscosity in rotating RBC models with $E \geq 10^{-5}$.
To decrease the numerical cost of direct three-dimensional calculations, the 
quasi-geostrophic approximation of  spherical convection \citep[hereafter QG, 
see][]{Busse86,Cardin94} has been developed. In the limit of 
$E\rightarrow 0$ and $Ro_c \ll 1$, convection is strongly constrained by 
rotation and can be approximated by a quasi two-dimensional flow, which
allows to decrease the Ekman number to $E=\mathcal{O}(10^{-7})$ 
\citep{Aubert03,Gillet06,Guervilly10}. However, the scaling analyses carried 
out by \cite{Gillet06} and \cite{Guervilly10} did not show any clear evidence 
of convergence towards the diffusivity-free scalings for both $Nu$ and $Re$, 
even when $E \lesssim 10^{-6}$. Furthermore, the lack of a direct one-to-one 
comparison between the QG results and the fully three-dimensional computations 
makes the interpretation of these results difficult. Hence, the determination 
of an accurate scaling law for both $Re$ and $\ell$ in rapidly-rotating 
spherical RBC forms one of the main goals of this work.

The aims of this study are twofold: (\textit{i}) determine the boundaries of 
the different physical regimes of rapidly-rotating convection in spherical 
shells by means of three-dimensional numerical simulations; (\textit{ii}) 
establish the scaling behaviours of both the local (length scale, boundary 
layers) and the global ($Nu$ and $Re$) properties that hold within each of 
these different regimes. We conduct a systematic parameter study 
varying the Ekman number within the range $3\times 10^{-7} \leq E \leq 10^{-1}$ 
and the Rayleigh number within the range $10^3 \lesssim Ra \lesssim 2\times 
10^{10}$ for a unity Prandtl number spherical shell of radius ratio 
$r_i/r_o=0.6$. To do so, we construct a dataset of 227 rotating 
simulations that complements our previous non-rotating calculations 
\citep{Gastine15}. Recent improvements of pseudo-spectral numerical codes 
\citep{Schaeffer13} enabled us to decrease the Ekman number to values comparable 
to these used in present-day local cartesian calculations. Our dataset allows 
us to determine the regime boundaries for rotating convection and to check the 
validity of the diffusivity-free scaling (\ref{eq:asymptotic}) in the low-$E$ 
regime. We examine the scaling behaviours of seven different diagnostics: $Nu$ 
and $Re$, the viscous dissipation rate, the typical flow length scale, the 
interior temperature gradient and the viscous and the thermal boundary layer 
thicknesses.

In \S~\ref{sec:model}, we introduce the hydrodynamical model and the various 
diagnostic quantities of interest. The numerical results are presented in 
\S~\ref{sec:results}. We conclude with a summary of our findings 
in \S~\ref{sec:conclu}.

\section{Formulation of the hydrodynamical problem}
\label{sec:model}

\subsection{Governing equations}

We consider rotating convection of a Boussinesq fluid confined in a 
spherical shell that rotates at a constant frequency $\Omega$ about the $z$
axis. Convective motions are driven by a fixed temperature contrast $\Delta 
T=T_i-T_o$ between the inner radius $r_i$ and the outer radius $r_o$.
The boundaries are impermeable, no slip and held at constant temperatures. We 
adopt a dimensionless formulation of the Navier-Stokes equations using the shell 
thickness $L=r_o-r_i$ as the reference 
length scale and the viscous diffusion time $L^2 /\nu$ as the reference 
timescale. The temperature contrast $\Delta T$ defines the 
temperature scale and gravity is non-dimensionalised using its reference value 
at the outer boundary $g_o$. The dimensionless equations that govern convective 
motions for the velocity $\vec{u}$, the pressure $p$ and the temperature $T$ 
are then expressed by
\begin{equation}
  \vec{\nabla}\cdot\vec{\tilde{u}} = 0,
  \label{eq:divu}
\end{equation}
\begin{equation}
 \frac{\partial \vec{\tilde{u}}}{\partial t} + 
\vec{\tilde{u}}\cdot\vec{\nabla} \vec{\tilde{u}} 
+\dfrac{2}{E}\,\vec{e_z}\times\vec{\tilde{u}} = 
-\vec{\nabla} \tilde{p} + 
\frac{Ra}{Pr}\,\tilde{g}\,\tilde{T}\,\vec{e_r}+ 
\vec{\Delta}\vec{\tilde{u}},
 \label{eq:navier}
\end{equation} 
\begin{equation}
 \frac{\partial \tilde{T}}{\partial t} + 
\vec{\tilde{u}}\cdot\vec{\nabla} 
\tilde{T} = \frac{1}{Pr}\Delta \tilde{T},
\label{eq:temp}
\end{equation}
where the tildes designate the dimensionless variables. The units vectors in the 
radial and vertical directions are denoted by $\vec{e_r}$ and $\vec{e_z}$, 
respectively. We make the assumption of a centrally-condensed mass which yields 
a dimensionless gravity profile of the form $\tilde{g} = (r_o/r)^2$. This 
particular choice is motivated by the exact analytical relation between the 
buoyancy power and the so-called flux-based Rayleigh number when $\tilde{g} \sim 
1/r^2$ \citep[see below and][]{Gastine15}. This enables us to conduct an exact 
analysis of the viscous dissipation rate and to directly compare our numerical 
models with our previous calculations of non-rotating convection which employed 
the same gravity profile.

The dimensionless equations (\ref{eq:divu}-\ref{eq:temp}) are governed by the 
Ekman number $E$, the Rayleigh number $Ra$, the Prandtl number $Pr$ and the 
radius ratio of the spherical shell $\eta$ defined by
\begin{equation}
 E=\frac{\nu}{\Omega L^2},\quad Ra = \frac{\alpha_T g_o \Delta T 
L^3}{\nu\kappa},\quad 
Pr=\frac{\nu}{\kappa},\quad 
\eta=\frac{r_i}{r_o},
\end{equation}
where $\alpha_T$ is the thermal expansion coefficient, $\nu$ is the kinematic 
viscosity and $\kappa$ is the thermal diffusivity. In the following, we will 
also consider the convective Rossby number, $Ro_c$, expressed by
\begin{equation}
 Ro_c = \sqrt{\frac{\alpha_T g_o \Delta 
T}{\Omega^2 L}}=\frac{Ra^{1/2}E}{Pr^{1/2}},
\end{equation}
which provides a good proxy of the relative importance of buoyancy forcing and 
Coriolis force in rotating convection \citep[e.g.][]{Gilman77}.

\subsection{Numerical technique}

The numerical simulations carried out in this study have been computed using 
the open-source magnetohydrodynamics 
code MagIC\footnote{Available at \url{http://www.github.com/magic-sph/magic}} 
\citep{Wicht02,MagIC}. MagIC has been validated via several benchmark 
tests for convection and dynamo action in spherical shell geometry 
\citep{Christensen01,Jones11}. To solve the system of equations 
(\ref{eq:divu}-\ref{eq:temp}) in spherical coordinates
$(r,\theta,\phi)$, the velocity field is decomposed into a 
poloidal and a toroidal contribution
\[
 \vec{\tilde{u}} = \vec{\nabla}\times\left(\vec{\nabla}\times \tilde{W} 
\vec{e_r}\right) +
\vec{\nabla}\times \tilde{Z} \vec{e_r},
\]
where $\tilde{W}$ and $\tilde{Z}$ are the poloidal and toroidal potentials. The 
dimensionless unknowns $\tilde{W}$, $\tilde{Z}$, $\tilde{p}$ and $\tilde{T}$ 
are expanded in spherical harmonic functions up to degree
$l_{\text{max}}$ in the colatitude $\theta$ and longitude $\phi$ and in
Chebyshev polynomials up to degree $N_r$ in the radial direction.
\modif{The equations are time-advanced using an explicit second-order 
Adams-Bashforth scheme for the Coriolis acceleration and the non-linear terms 
and an implicit Crank-Nicolson algorithm for the remaining
linear terms. The explicit treatment of the Coriolis force implies that the 
dimensionless time-step size is limited to a fraction of the rotation period.}
For a comprehensive description of the numerical method and 
the associated spectral transforms, the reader is referred to 
\cite{Glatz1,Tilgner97} and \cite{Christensen15}.
For the most demanding calculations, MagIC uses the open-source library 
SHTns\footnote{Available at \url{https://bitbucket.org/nschaeff/shtns}} to 
speed-up the spherical harmonic transforms \citep{Schaeffer13}.

\subsection{Diagnostics}
\label{sec:diagnostics}

We introduce in the following several diagnostic properties
to quantify the impact of the different control parameters on 
the flow and temperature properties. We adopt several notations regarding 
averaging procedures. Overbars $\overline{\cdots}$ correspond to temporal 
averaging, angular brackets $\langle \cdots \rangle$ to spatial averaging over 
the entire spherical shell volume and $\langle \cdots \rangle_s$ to 
an average over a spherical surface
 \[
 \overline{f} = \dfrac{1}{\tau}\int_{t_0}^{t_0+\tau} f\,{\rm d} t, \quad
  \left\langle f \right\rangle = \frac{1}{V} \int_V 
f(r,\theta,\phi)\,{\rm d} V, \quad
\left\langle f \right\rangle_s = \frac{1}{4 \pi} \int_0^\pi \int_0^{2\pi} 
f(r,\theta,\phi)\sin\theta\,{\rm d}\theta\,{\rm d}\phi\,,
\]
where $V$ is the spherical shell volume and 
$\tau$ is the time averaging interval. 

For the sake of clarity, we introduce the notation $\vartheta$ to define the
time and horizontally-averaged radial dimensionless temperature profile
\[
 \vartheta(r) = \overline{\left\langle \tilde{T} \right\rangle_s}\, .
\]
The heat transport is characterised by the Nusselt number $Nu$, the ratio of 
the total heat flux to the heat carried by conduction
\begin{equation}
 Nu = \frac{\eta\,\mathcal{Q}\,L}{\rho c_p \kappa \Delta T} =
\frac{\displaystyle\left.\frac{\mathrm{d}\vartheta} { \mathrm { d } 
r}\right|_{r=r_o}}{\displaystyle\left.\frac{\mathrm{d} 
\tilde{T}_c}{\mathrm{d} r}\right|_{r=r_o}}= 
\frac{\displaystyle\left.\frac{\mathrm{d}\vartheta}{\mathrm{d} 
r}\right|_{r=r_i}}{\displaystyle\left.\frac{\mathrm{d} 
\tilde{T}_c}{\mathrm{d} r}\right|_{r=r_i}},
\end{equation}
where $\mathcal{Q}$ is the heat flux, $c_p$ is the heat capacity
and $\tilde{T}_c$ is the dimensionless conductive temperature profile, solution 
of 
\begin{equation}
  \frac{\rm d }{{\rm d} r}\left( r^2 \frac{{\rm d} \tilde{T}_c}{{\rm d} r} 
\right) 
= 0, \quad \tilde{T}_c(r_i)=1,  \quad \tilde{T}_c(r_o)=0\,.
\label{eq:dtcdr}
\end{equation}
The dimensionless kinetic energy $\tilde{E}_k$ is defined by
\[
 \tilde{E}_k = \dfrac{1}{2} \overline{\langle \tilde{u}^2 \rangle} = 
\sum_{l=1}^{l_{\text{max}}} \sum_{m=0}^{l} \mathcal{E}_l^{m},
\]
where $\mathcal{E}_l^{m}$ is the dimensionless kinetic energy density 
at a spherical harmonic degree $l$ and order $m$. In rotating  spherical 
shells, the axisymmetric zonal flow can represent a significant fraction of 
the total kinetic energy \citep[e.g.][]{Christensen02,Gastine12,Yadav16}. Since 
this flow does not directly contribute to the heat transfer, we decide to 
rather characterise the r.m.s. flow velocity by the time-averaged convective 
Reynolds number expressed 
by
\begin{equation}
 Re_c = \overline{\sqrt{2\sum_{l=1}^{l_{\text{max}}} \sum_{m=1}^{l} 
\mathcal{E}_l^{m}}}\,,
 \label{eq:rec}
\end{equation}
where the contribution from the axisymmetric flows ($m=0$) has been excluded.
The fluid bulk in rotating convection usually departs from the 
well-mixed convective interior obtained in classical RBC. Under the influence 
of rapid rotation, turbulent mixing is inhibited and strong interior 
temperature gradients can persist 
\citep[e.g.][]{Julien96,KingNature09,Stellmach14}.
We denote the bulk temperature gradient at mid-shell by
\begin{equation}
 \tempgrad = \left.\frac{\mathrm{d} \vartheta}{\mathrm{d} 
r}\right|_{r=r_m}, \quad r_m = \frac{1+\eta}{2(1-\eta)}\,.
 \label{eq:beta}
\end{equation}
Following \cite{Christensen06}, the typical flow length scale is determined 
from the time-averaged kinetic energy spectrum 
\begin{equation}
 \ell^{-1} = 
\overline{\lp\frac{L\,\displaystyle\sum_{l=1}^{l_{\text{max}}} \sum_{m=0}^{l}  
l\, \mathcal{E}_l^m(t)}{\pi\,\displaystyle\sum_{l=1}^{l_{\text{max}}} 
\sum_{m=0}^{l} \mathcal{E}_l^{m}(t)}\rp}\,.
\label{eq:lengthscalesdef}
\end{equation}

Several different approaches have been usually considered to define the 
thermal boundary layer thickness $\lambda_T$. \modif{They either rely on the 
mean radial temperature profile $\vartheta(r)$ 
\citep[e.g.][]{Verzicco99,Breuer04,Liu11} or on the root mean square of the 
temperature fluctuations \citep[e.g.][]{Julien12,King13,Kunnen16}.} For 
consistency with the definition adopted in our non-rotating calculations 
\citep{Gastine15}, $\tbli$ ($\tblo$) is defined here as the depth where the 
linear fit to the temperature profile near the inner (outer) boundary intersects 
the linear fit to the profile at mid-depth.
\modif{The comparison of this boundary layer definition with the estimate 
coming from the location of the peaks of the r.m.s. of the 
temperature fluctuations for a few selected cases yield similar boundary 
layer thicknesses.}
The viscous boundary layer thicknesses are estimated following a similar 
slope-intersection method: $\ubli$ ($\ublo$) is defined as the distance from the 
inner (outer) boundary where the linear fit to the horizontal velocity profile 
$u_h=\sqrt{u_\theta^2+u_\phi^2}$  near the inner (outer) boundary intersects the 
horizontal line passing through the maximum of $u_h$ \citep[see 
also][]{Breuer04}.

\subsection{Parameter choice and resolution checks}

We aim at studying the scaling behaviours of both the local (length scales, 
temperature gradient, boundary layers) and the global (Nusselt and Reynolds 
numbers) properties for each dynamical regimes of rotating convection in 
spherical shells. To achieve this goal, we build a dataset of 227 global 
spherical shell models that span the range $3\times 10^{-7} \leq E \leq 10^{-1}$ 
and $10^{3} < Ra < 2\times 10^{10}$. The radius ratio $\eta$ is kept 
fixed to $\eta=0.6$ and the Prandtl number to $Pr=1$ to limit the extension of 
the parameter space. This setup also allows a direct comparison with our 
previous non-rotating models \citep{Gastine15}. The numerical dataset 
constructed in this study thoroughly explores various 
dynamical regimes of rotating convection encompassing convection close to 
critical, turbulent quasi-geostrophic convection and weakly-rotating 
convection. The summary table~\ref{tab:runs} given in the appendix contains the 
calculations of the main control and diagnostic quantities computed for each 
numerical simulation.

Particular attention must be paid to the numerical resolution requirements for 
global models of convection \citep{Shishkina10,King12}, since under-resolution
can significantly degrade the accuracy of the obtained exponents when 
deriving asymptotic scaling laws \citep{Amati05}. The comparison of the viscous 
and thermal dissipation rates with the time-averaged Nusselt number provides a 
robust way to validate the numerical resolutions employed in our models 
\citep[e.g.][]{Stevens10,Lakkaraju12}. As demonstrated in our previous RBC 
calculations \citep{Gastine15}, the choice of a centrally-condensed mass (i.e.  
$\tilde{g}\sim 1/r^2$) ensures the following analytical relations between the 
dimensionless viscous dissipation rate $\tilde{\epsilon}_U$ and the flux-based 
Rayleigh number defined by $Ra_{\mathcal{Q}} = Ra(Nu-1)$:
\begin{equation}
 \tilde{\epsilon}_U = \overline{\left\langle (\vec{\nabla}\times 
\vec{\tilde{u}})^2 \right\rangle}= 
\frac{3}{1+\eta+\eta^2}\,\frac{Ra_{\mathcal{Q}}}{Pr^2},
\label{eq:epsV}
\end{equation}
and between the dimensionless thermal dissipation rate $\tilde{\epsilon}_T$ and 
the Nusselt number
\begin{equation}
 \tilde{\epsilon}_T = \overline{ \left\langle ( \vec{\nabla} \tilde{T} )^2 
\right\rangle} = 
\frac{3\eta}{1+\eta+\eta^2}\,Nu\,.
\label{eq:epsT}
\end{equation}
The related ratios 
\[
 \chi_{T} = \frac{(1+\eta+\eta^2)\,\tilde{\epsilon}_T}{3\eta\,Nu}, 
\quad
 \chi_{U} = 
\frac{(1+\eta+\eta^2)\,Pr^2\,\tilde{\epsilon}_U}{3\,Ra_{\mathcal{Q}}},
\]
can thus be used to check the adequacy of the spatial resolutions of the 
numerical models. As shown in table~\ref{tab:runs}, these ratios are very 
close to unity (within $3\%$) for all the simulations computed here, validating 
the grid resolutions. \modif{The temporal convergence of the numerical models 
has been ensured by running the simulations at least 50 convective overturn 
time.}

The numerical truncations employed here range from ($N_r=61,\ 
l_{\text{max}}=64$) for the case with the largest Ekman and lowest
Rayleigh numbers to ($N_r=641,\ l_{\text{max}}=1345$) for the numerical model 
with $E=3\times 10^{-7}$ and the highest $Ra$. To save computational 
resources, some of the most demanding cases (with $E \leq 3\times 10^{-6}$) 
have been computed on an azimuthally truncated spherical shell with a two-fold, 
four-fold or eight-fold symmetry (see table~\ref{tab:runs} for 
details). \modif{Since rapidly-rotating convection is dominated by small-scale 
structures, this assumption is not considered to have any significant impact on 
the results. In addition,} the comparison of test cases with or without 
symmetries showed no statistical differences. The total computational time spent 
to construct the present dataset of numerical models corresponds to roughly 13 
millions core hours of Intel Ivy Bridge CPUs.

\section{Numerical results}
\label{sec:results}

\begin{figure}
 \centering
 \includegraphics[width=\textwidth]{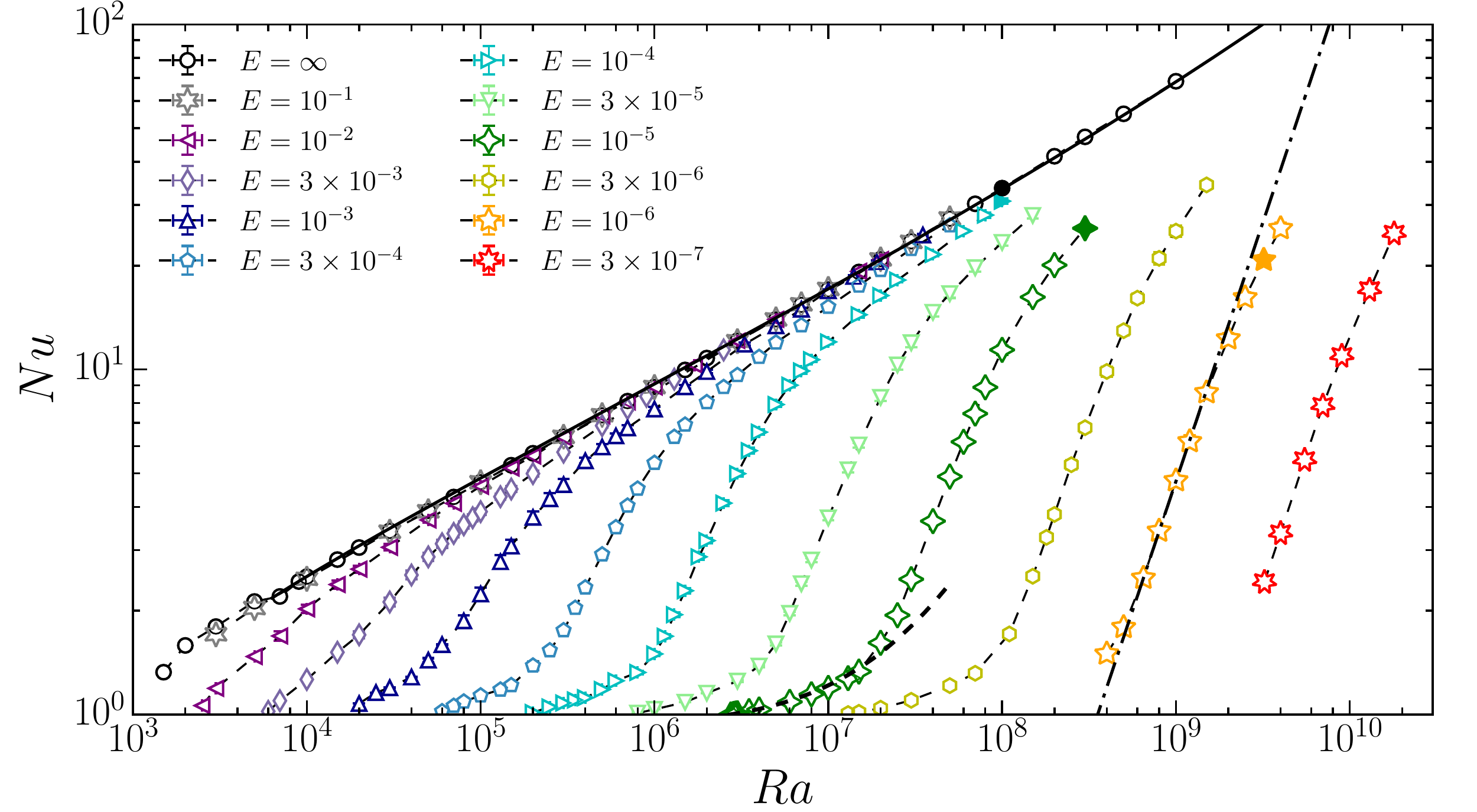}
 \caption{Nusselt number as a function of the Rayleigh number with Ekman 
numbers denoted by symbol shape (and colour online). The dashed black 
line corresponds to the weakly non-linear onset scaling 
$Nu-1=0.08\,(Ra/Ra_c-1)$  (Equation \ref{eq:NuWnl}) discussed in 
\S~\ref{sec:wnl}. The dot-dashed black line to the asymptotic scaling $Nu = 
0.15\,Ra^{3/2}E^{2}$ (Equation \ref{eq:numscalingnu}) in the rotation-dominated 
regime discussed in \S~\ref{sec:snl}. The solid black line corresponds to heat 
transfer scaling for non-rotating convection derived in \cite{Gastine15}. The 
four filled-in symbols correspond to the four cases highlighted in 
figure~\ref{fig:snaps}. The error bars correspond to one standard-deviation from 
the time-averaged Nusselt number (i.e. $Nu \pm \sigma$). For most of the 
numerical models, these error bars are smaller than the size of the symbols.}
 \label{fig:nura}
\end{figure}

Figure~\ref{fig:nura} shows $Nu$ as a function of $Ra$ for both rotating and 
non-rotating cases.  
In rotating convection, the onset is delayed and the critical Rayleigh number 
$Ra_c$ increases with decreasing Ekman number. Close to critical, $Nu$ 
initially increases slowly and roughly linearly with $Ra$ (dashed 
line). Beyond this weakly non-linear regime of rotating convection, i.e. when 
$Nu \gtrsim 2$, the heat transport increases more rapidly with $Ra$ than in the 
non-rotating RBC (dotted-dashed line). This behaviour is reminiscent to that 
observed in plane layer studies \citep[e.g.][]{Rossby69, 
KingNature09,Schmitz09,Cheng15}. Above a transitional Rayleigh number that will 
be determined below, the heat transfer data for each Ekman number then
tends towards the non-rotating behaviour. As shown in our previous study 
\citep{Gastine15}, the non-rotating data ($E =\infty$) can be accurately 
described by a dissipation analysis that follows the \cite{Grossmann00} theory 
(black circles and solid black line). The non-rotating RBC thus
defines an effective upper limit for the heat transport in rotating convection 
in spherical shells at large $Ra$. In contrast to plane layer calculations, the 
spherical shell data do not exhibit any overshoot of rotating heat transfer
beyond the non-rotating RBC scalings \citep{Liu97,Zhong09}.

\begin{figure}
 \centering
 \includegraphics[width=\textwidth]{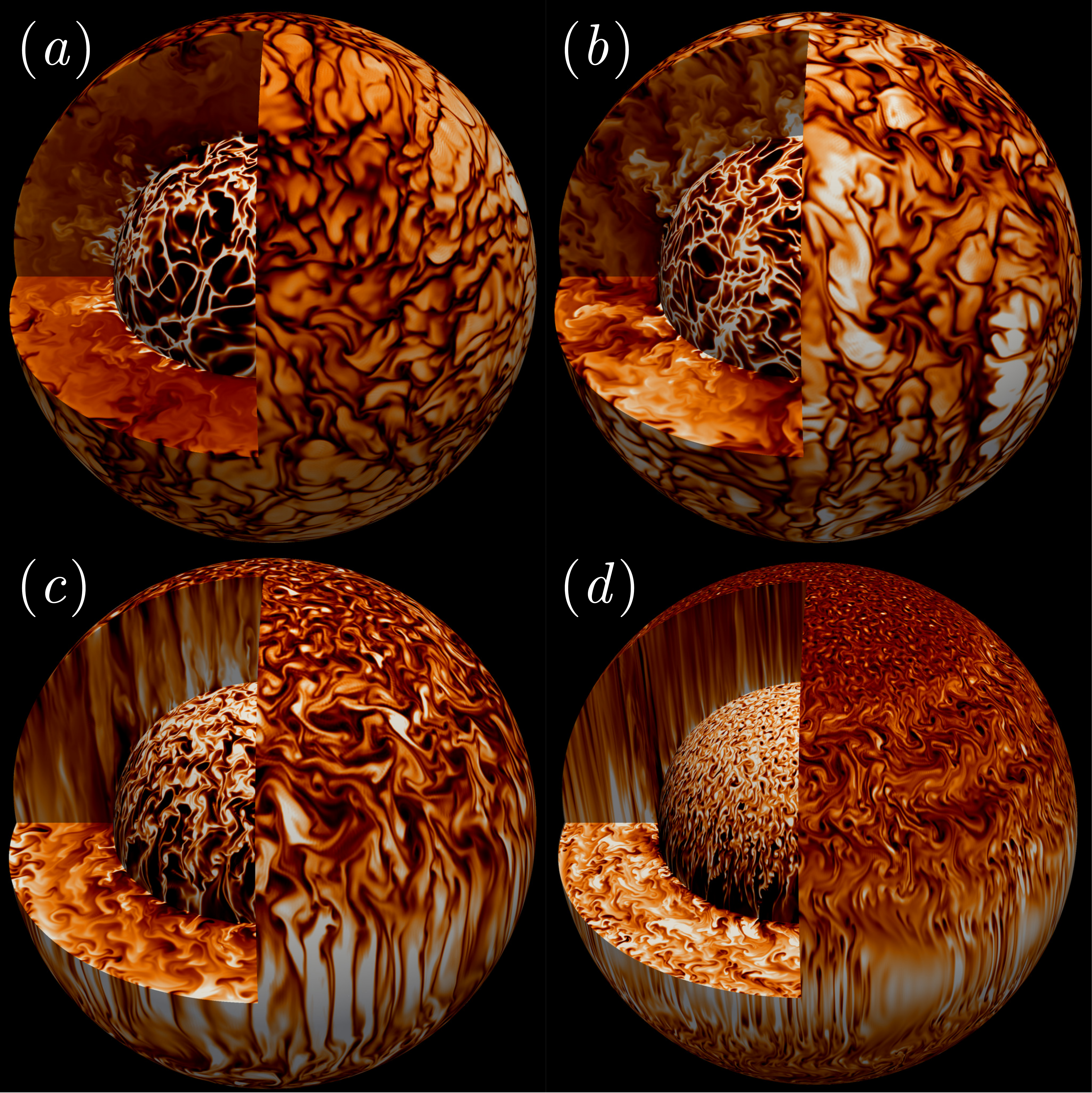}
 \caption{Meridional sections, equatorial cut and radial surfaces of the 
temperature fluctuations $\tilde{T}'=\tilde{T}-\vartheta$. The inner (outer) 
surface corresponds to the location of the inner (outer) thermal boundary layer 
$r=r_i+\tbli$ ($r=r_o-\tblo$). Color levels range from $-0.15$ (black) to $0.15$ 
(white). Panel (\textit{a}) corresponds to a non-rotating model with $Ra=10^8$ 
\citep{Gastine15}. Panel (\textit{b}) corresponds to a numerical model with 
$E=10^{-4}$ and $Ro_c=1$ (Case 146 in table~\ref{tab:runs}). Panel 
(\textit{c}) corresponds to a numerical model with $E=10^{-5}$ and $Ro_c=0.17$ 
(Case 191 in table~\ref{tab:runs}). Panel (\textit{d}) corresponds to a 
numerical model with $E=10^{-6}$ and $Ro_c=0.05$ (Case 219 in 
table~\ref{tab:runs}).}
 \label{fig:snaps}
\end{figure}

To illustrate the diversity of the dynamical regimes in rotating spherical 
shell RBC, figure~\ref{fig:snaps} shows equatorial, meridional 
and radial cuts of the dimensionless temperature fluctuation 
$\tilde{T}'=\tilde{T}-\vartheta$ for four selected cases. The filled-in symbols 
in figure~\ref{fig:nura} indicate their positions in the $Nu-Ra$ parameter 
space. The image in panel (\textit{a}) corresponds to a non-rotating model with 
$Ra=10^8$. Due to an efficient turbulent mixing, the fluid-bulk is nearly
isothermalised and most of the temperature fluctuations take the form of 
thermal plumes that depart from the thin boundary layers. As visible on both 
spherical surfaces, the plume network is organised in long and thin sheet-like 
structures. Panel (\textit{b}) corresponds to a numerical 
simulation with $E=10^{-4}$ and $Ra=10^8$, i.e. $Ro_c=1$. The turbulent flow 
remains essentially anisotropic and three-dimensional with a typical length 
scale comparable to that in the non-rotating model. At the connection points of 
the 
convective lanes network, we nevertheless observe the formation of vortical 
structures, which indicates the influence of the Coriolis force on the largest 
scale. Panel (\textit{c}) shows a numerical model with $E=10^{-5}$ and 
$Ra=3\times 10^8$. In the equatorial region, columnar structures aligned with 
the rotation axis are observed. At high-latitudes, the integrity of the 
convective features is disrupted. This gradual loss of geostrophy is expected 
when the buoyancy forcing increases in strength with respect to the rotational 
constraint (here $Ro_c=0.17$).  The image in panel (\textit{d}) corresponds to 
a numerical model with $E=10^{-6}$ and $Ro_c=0.05$. As compared to the 
previous model, the stronger rotational influence on the flow results in a more 
pronounced alignment of the convective features with the rotation axis.  
Decreasing the Ekman number from $10^{-5}$ (panel \textit{c}) to $10^{-6}$ 
(panel \textit{d}) is accompanied by an obvious decrease of the typical 
convective flow length scale. In the rapidly-rotating regime when $Ro_c \ll 1$, 
Coriolis forces inhibit the turbulent mixing and the fluid bulk therefore 
departs from an isothermal state.

\subsection{Weakly non-linear regime of rotating convection}
\label{sec:wnl}

\begin{figure}
 \centering
 \includegraphics[width=\textwidth]{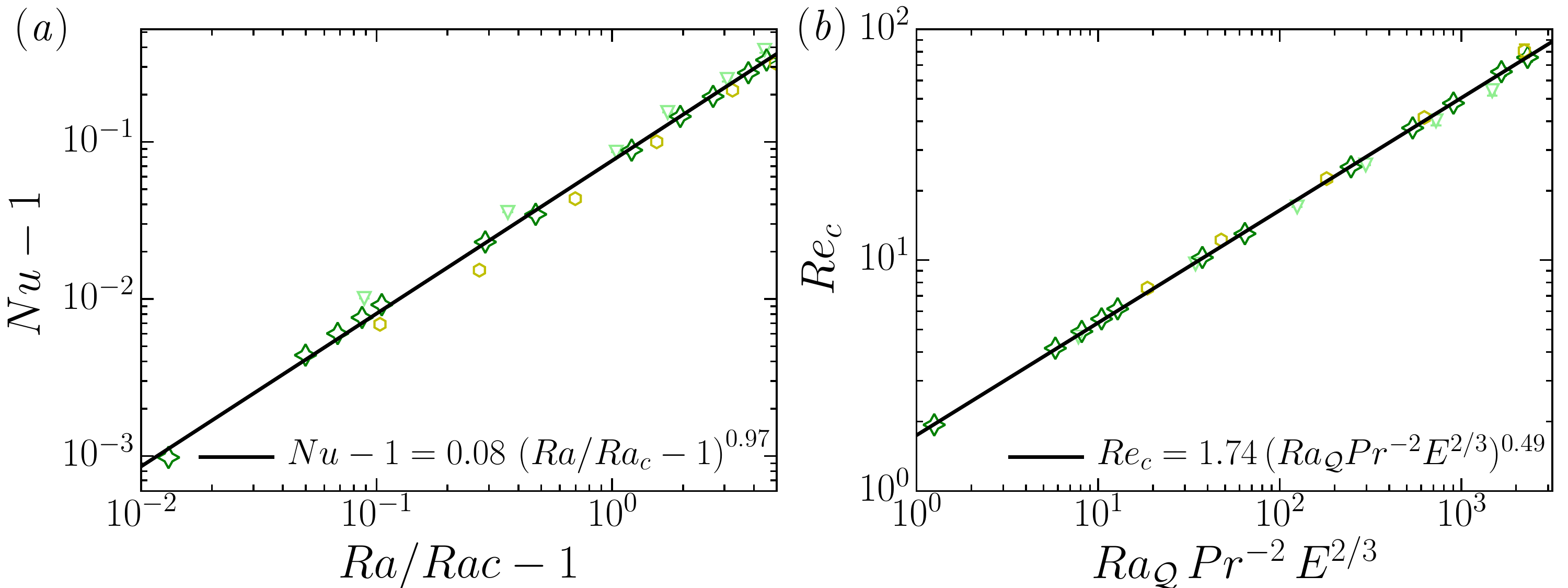}
 \caption{(\textit{a}) Nusselt number $Nu-1$ as a function of 
$Ra/Ra_c-1$. (\textit{b}) Reynolds number $Re_c$ as a function of 
$Ra_\mathcal{Q}Pr^{-2}E^{2/3}$. Only the cases with $E \leq 3\times 10^{-5}$ and 
$Ra \leq 6\,Ra_c$ are displayed in this figure. In both panels, the solid black 
lines correspond to the least-square fits to the data. The symbols have the 
same meaning as in figure~\ref{fig:nura}.}
 \label{fig:wNL}
\end{figure}

In rotating spherical shells, the convective flow at onset takes the form of 
prograde drifting thermal Rossby waves that first develop in the vicinity of 
the tangent cylinder \citep[e.g.][]{Busse70,Dormy04}. The critical Rayleigh 
number $Ra_c$ and azimuthal wavenumber $m_c$ follow
\begin{equation}
  \quad Ra_c \sim E^{-4/3}, \quad m_c \sim E^{-1/3}\,.
 \label{eq:critical}
\end{equation}
The exact values of $Ra_c$ and $m_c$ for the different Ekman numbers considered 
in this study are given in table~\ref{tab:rac}. For marginally supercritical 
Rayleigh numbers, the weakly non-linear perturbation analysis carried out by 
\cite{Busse86} and 
\cite{Gillet06} predicted that the heat transport increases linearly with the 
supercriticality, i.e.:
\begin{equation}
 Nu-1 \sim \dfrac{Ra}{Ra_c}-1\,.
 \label{eq:NuWnl}
\end{equation}
Figure~\ref{fig:wNL}(\textit{a}) shows $Nu-1$ as a 
function of $Ra/Ra_c-1$ for the numerical simulations with $E\leq 
3\times 10^{-5}$ and $Ra\leq 6\,Ra_c$. The best-fit to the data yields $Nu-1 = 
0.076(\pm0.003)(Ra/Ra_c-1)^{0.973(\pm0.019)}$. In spite of a remaining weak 
dependence on the Ekman number, the scaling (\ref{eq:NuWnl}) holds reasonably 
well for the numerical models with $Ra\leq 6\,Ra_c$.

A useful tool for establishing a scaling for the typical convective flow 
speed ($Re_c$) is the linear relation between the viscous dissipation rate 
$\tilde{\epsilon}_U$ and the flux-based Rayleigh number $Ra_{\mathcal{Q}}$ 
given in (\ref{eq:epsV}). For a laminar flow close to the onset of 
convection, $\epsilon_U$ can be approximated by
\begin{equation}
 \epsilon_U \sim \nu \frac{U_c^2}{\ell_\perp^2},
 \label{eq:epsvac}
\end{equation}
where $\ell_\perp$ is the typical flow length scale and $U_c$ the typical 
velocity. Close to the onset of convection, $\ell_\perp$ 
remains close to the critical azimuthal wavenumber, such that
\begin{equation}
 \ell_\perp / L \sim 1/m_c \sim E^{1/3}\,.
 \label{eq:lperpvac}
\end{equation}
Combining (\ref{eq:epsV}), (\ref{eq:epsvac}) and (\ref{eq:lperpvac})
yields a scaling law for the convective flow speed $Re_c\sim U_c L/\nu$
\begin{equation}
 Re_c \sim Ra_{\mathcal{Q}}^{1/2}Pr^{-1}E^{1/3}\,.
 \label{eq:revac}
\end{equation}
This scaling relation is sometimes called the \emph{Visco-Archimedean-Coriolis}
(hereafter VAC) scaling and can also be directly derived from the balance 
between these three forces \citep{Aubert01,KingBuffett13,King13}. 
Figure~\ref{fig:wNL}(\textit{b}) shows $Re_c$ versus 
$Ra_\mathcal{Q}Pr^{-2}E^{2/3}$ 
for the numerical simulations with $E\leq 3\times 
10^{-5}$ and $Ra \leq 6\,Ra_c$. The 
least-square fit to the data yields $Re_c = 1.744(\pm 0.041)
(Ra_{\mathcal{Q}}\,E^{2/3})^{0.487(\pm0.004)}$, in excellent agreement with the 
theoretical scaling.


\begin{figure}
 \centering
 \includegraphics[width=\textwidth]{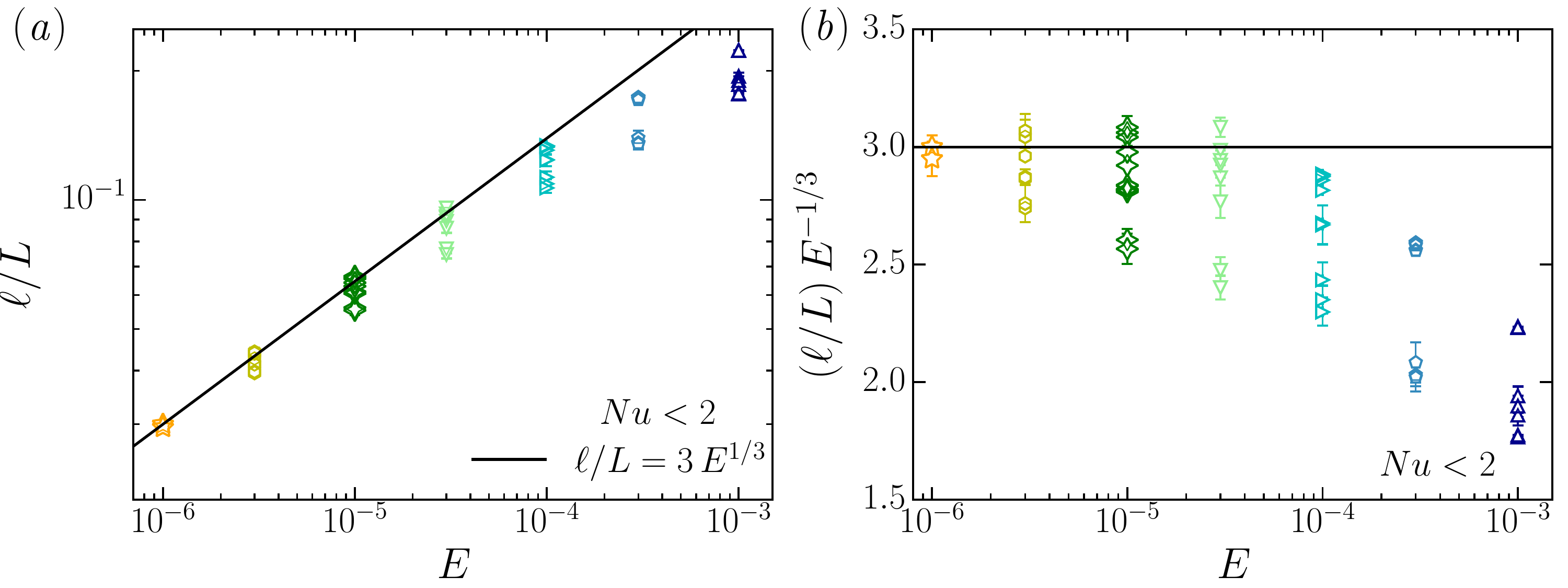}
 \caption{(\textit{a}) Average flow length scale 
$\ell/L$ calculated using (\ref{eq:lengthscalesdef}) as a function of the 
Ekman number for the numerical models close to the onset of convection with 
$Nu \leq 2$ and $E \leq 10^{-3}$. (\textit{b}) Corresponding compensated 
$\ell/L$ scaling. In both panels, the solid black lines correspond to 
$\ell/L = 3\,E^{1/3}$ (Equation~\ref{eq:lperpvac}). The error bars illustrate 
the r.m.s. fluctuations and correspond to one standard-deviation from the 
time-averages. The symbols have the same meaning as in figure~\ref{fig:nura}.}
 \label{fig:lengthscalesOnset}
\end{figure}

Figure~\ref{fig:lengthscalesOnset}(\textit{a}) shows the average flow 
length scale $\ell/L$ plotted as a function of $E$ for the numerical 
models close to the onset of convection. The typical scale of convection
gradually approaches the theoretical scaling $\ell/L \sim E^{1/3}$ when $E\leq 
10^{-5}$ and the additional $Ra$ dependence vanishes. The compensated scaling 
displayed in panel (\textit{b}) suggests the asymptotic scaling $\ell \simeq 
3\,E^{1/3}\,L$. The cases with the largest Ekman number (i.e. $E \geq 10^{-4}$) 
significantly depart from this asymptotic law. This is not surprising 
since the scaling $m_c \sim E^{-1/3}$ is expected to hold for 
asymptotically-small Ekman numbers (see table~\ref{tab:rac}).

\subsection{Non-linear regime of rotating convection}
\label{sec:snl}

\subsubsection{Nusselt number scaling}

\noindent As can be seen on figure~\ref{fig:nura}, the weakly non-linear 
scaling (\ref{eq:NuWnl}) provides an accurate description of the heat transport 
behaviour up to $Nu \simeq 1.5$, which roughly corresponds to $Ra\simeq 
6\,Ra_c$.
When $Nu > 2$, the heat transport increases much more rapidly and enters the 
non-linear regime of rotating convection.

\begin{figure}
 \centering
 \includegraphics[width=\textwidth]{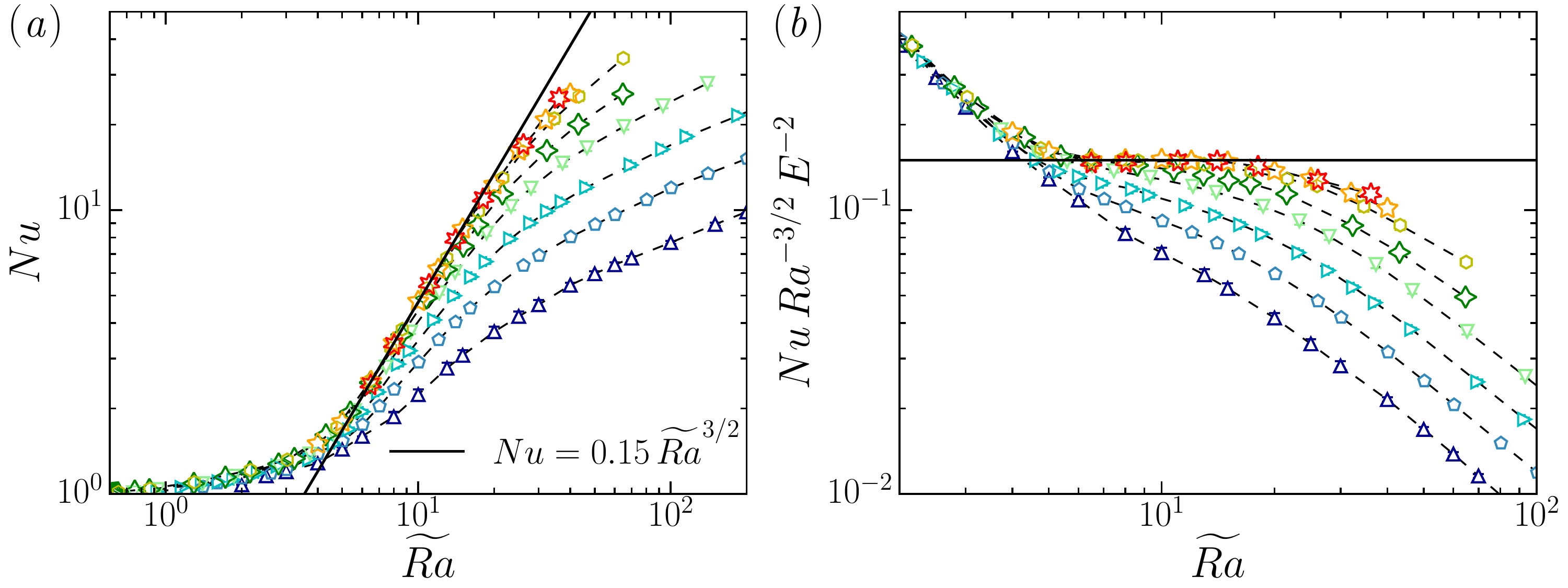}
 \caption{(\textit{a}) Nusselt number as a function of
$\widetilde{Ra}=RaE^{4/3}$. (\textit{b}) Corresponding compensated $Nu$ 
scaling. In both panels, the solid black 
line corresponds to the scaling $Nu = 0.15\, Ra^{3/2}E^{2}$ 
(Equation~\ref{eq:numscalingnu}). For clarity, 
only the cases with $E\leq 10^{-3}$ and $\widetilde{Ra} < 200$ are displayed 
in this figure. The symbols have the same meaning as in figure~\ref{fig:nura}.}
 \label{fig:nuraek}
\end{figure}

\begin{figure}
 \centering
 \includegraphics[width=9cm]{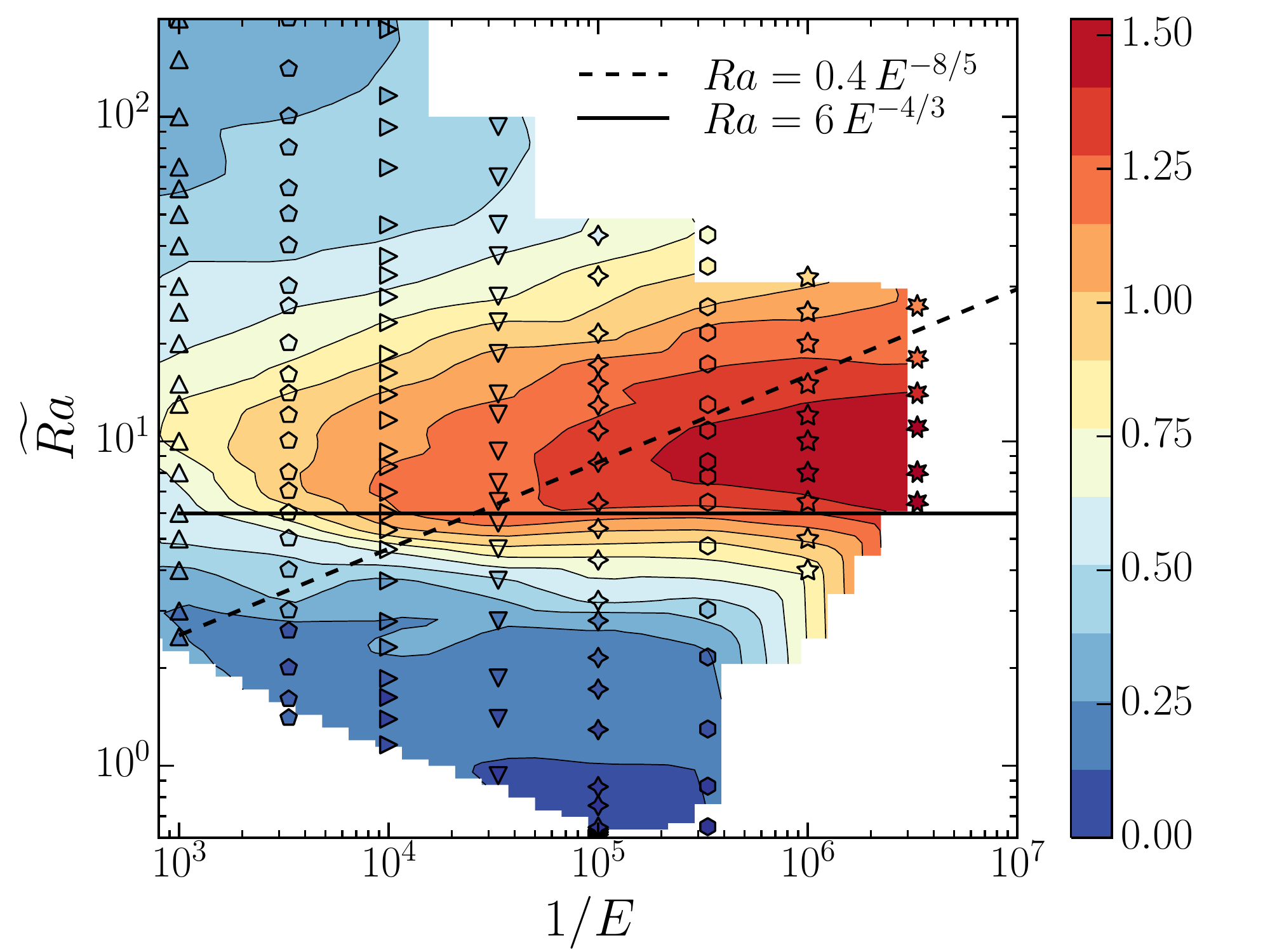}
 \caption{Isocontours of the local effective exponent $\alpha_\text{eff}$ 
of the $Nu =Ra^{\alpha_\text{eff}}E^{\beta_\text{eff}}$ scaling in the 
$(1/E,\widetilde{Ra})$-plane. The symbol shape corresponds to the Ekman number 
as in the previous figures, while the symbol color scales here with the value 
of local slope $\alpha_{\text{eff}}$. The solid black line corresponds to the 
upper limit of the weakly non-linear regime $Ra=6\,Ra^{4/3}$, while the dashed 
black line corresponds  to the dynamical boundary $Ra=0.4\,E^{-8/5}$ derived in 
figure~\ref{fig:rotationDominated}.}
 \label{fig:NuExpLoc}
\end{figure}

The plane layer numerical simulations of rotating convection by \cite{King12} 
and \cite{Stellmach14} as well as the asymptotically-reduced  theoretical models 
by \cite{Julien12a} indicate that the convective heat flux only 
depends on the supercriticality $Ra/Ra_c$ when $Ro_c \ll 1$. To check whether 
our numerical calculations actually support such a behaviour at low Ekman 
numbers, 
figure~\ref{fig:nuraek}(\textit{a}) shows $Nu$ as a function of 
$\widetilde{Ra}=RaE^{4/3}$, since $Ra_c\sim E^{-4/3}$. For $\widetilde{Ra} < 
6$, $Nu$ slowly increases with the Rayleigh number and follows the previously 
described weakly non-linear regime. Beyond this point, a fast 
steepening of the slope is observed and the function $Nu=f(\widetilde{Ra})$ 
starts to show an additional dependence on $E$: while the large Ekman number 
cases quickly depart from a pure function of $\widetilde{Ra}$, the low Ekman 
number simulations continue to follow a law of the form $Nu = f(\widetilde{Ra})$ 
up to higher $\widetilde{Ra}$ values. The exact location of this departure 
point gradually shifts to increasing $\widetilde{Ra}$ with decreasing $E$.
Figure~\ref{fig:nuraek}(\textit{b}) shows the compensated scalings of 
$Nu$ and reveals that the steepest part of the heat transfer 
law approaches the diffusivity-free scaling $Nu = 0.15\widetilde{Ra}^{3/2}$, at 
least in the range $6\leq \widetilde{Ra} \leq 20$. To further assess the 
validity domain of the diffusivity-free scaling (\ref{eq:asymptotic}) in our 
numerical dataset, we introduce the local effective exponents 
$\alpha_{\text{eff}}$ and $\beta_{\text{eff}}$ of the $Nu(Ra,E)$ law:
\begin{equation}
 \alpha_{\text{eff}} = \lp\frac{\partial \ln Nu}{\partial \ln Ra}\rp_{E}; \quad
\beta_{\text{eff}} = \lp\frac{\partial \ln Nu}{\partial \ln E}\rp_{Ra}\, .
\label{eq:alphaeff}
\end{equation}
The local exponent $\beta_{\text{eff}}$ is unfortunately difficult to constrain 
since our dataset only samples sparse variations of the Ekman number 
(half-decade sampling). However, the dense coverage in $Ra$ for each Ekman 
number subset allows to estimate the variations of $\alpha_{\text{eff}}$.  
Figure~\ref{fig:NuExpLoc} shows isocontours of $\alpha_{\text{eff}}$ in a 
$(1/E,\widetilde{Ra})$-plane. Beyond the weakly non-linear regime delimited by 
the horizontal solid black line, the effective slope gradually increases with 
decreasing $E$. It reaches a maximum value $\alpha_{\text{eff}}\simeq 3/2$ for 
a small region around $E \lesssim 10^{-6}$ and $\widetilde{Ra} \simeq 10$. 

The numerical simulations carried out by \cite{Cheng15} in cartesian 
coordinates revealed a continuous increase of the local 
exponent $\alpha_{\text{eff}}$ with decreasing $E$ and exhibit much steeper 
scaling beyond $\alpha_{\text{eff}}\simeq 3$ \citep[see 
also][]{Stellmach14,Kunnen16}. This phenomenon has 
been attributed to the sizeable role of Ekman pumping that enhances the heat 
transfer even in the low Ekman number regime \citep{Julien16}. This steep 
scaling is not observed in absence of Ekman boundary layers, for 
instance when stress-free mechanical boundary conditions are 
adopted \citep{Stellmach14,Barker14,Kunnen16,Plumley16}. In that case, local 
exponents close to $\alpha_{\text{eff}}\simeq 3/2$ are recovered at low Ekman 
numbers.

The scaling exponents obtained in our simulations in spherical geometry for 
comparable Ekman numbers $E =\mathcal{O}(10^{-7})$ remain bounded by the 
diffusivity-free scaling  $\alpha_{\text{eff}} = 3/2$ \citep[see 
also][]{King10}.  Although a further increase of $\alpha_{\text{eff}}$ for
$E<3\times 10^{-7}$ cannot be \textit{a priori} ruled out, the compensated 
scaling shown 
in figure~\ref{fig:nuraek}(\textit{b}) suggests a gradual approach to 
$\alpha=3/2$, which seems to leave little room for further increase.
This suggests that our simulations approach the asymptotic diffusivity-free 
behaviour for the parameter range $E\leq 10^{-6}$ and $6 \leq \widetilde{Ra} 
\leq 20$.


Clearly, though, the validity domain of the diffusivity-free scaling 
(\ref{eq:asymptotic}) remains relatively narrow in our set of numerical models.
Beyond the steep scaling with $\alpha_{\text{eff}} \simeq 3/2$, our numerical 
calculations show a transition to a shallower heat transfer scaling at 
larger supercriticality $\widetilde{Ra}$. The characterisation of the upper 
bound of the rotation-dominated regime of rotating convection has been a 
long-standing question. The ratio of the global-scale Coriolis and buoyancy 
forces, approximated by the value of $Ro_c$, has long been postulated to 
control the transition from rotation-dominated convection to non-rotating 
convection \citep[e.g.][]{Gilman77,Zhong10,Stevens13a}. Using a combination of 
laboratory experiments and local numerical simulations in cartesian 
coordinates, \cite{KingNature09} however demonstrated that a simple $Ro_c$ 
criterion does not correctly capture the regime transition, which happens while 
the bulk Rossby number is still much smaller than unity \citep[see 
also][]{Cheng15}.  They instead suggested that this transition is controlled by 
the competing thickness of the thermal and Ekman boundary layers. The resulting 
crossover was found to follow $RaE^{3/2} \sim 1$ \citep{King12}. However, 
numerical experiments that employed stress-free boundary conditions, in which 
viscous boundary layers are not present, yield a similar regime transition, 
which questions the role of the Ekman layer on the regime change 
\citep[e.g.][]{Schmitz09}.

\begin{figure}
 \centering
 \includegraphics[width=8cm]{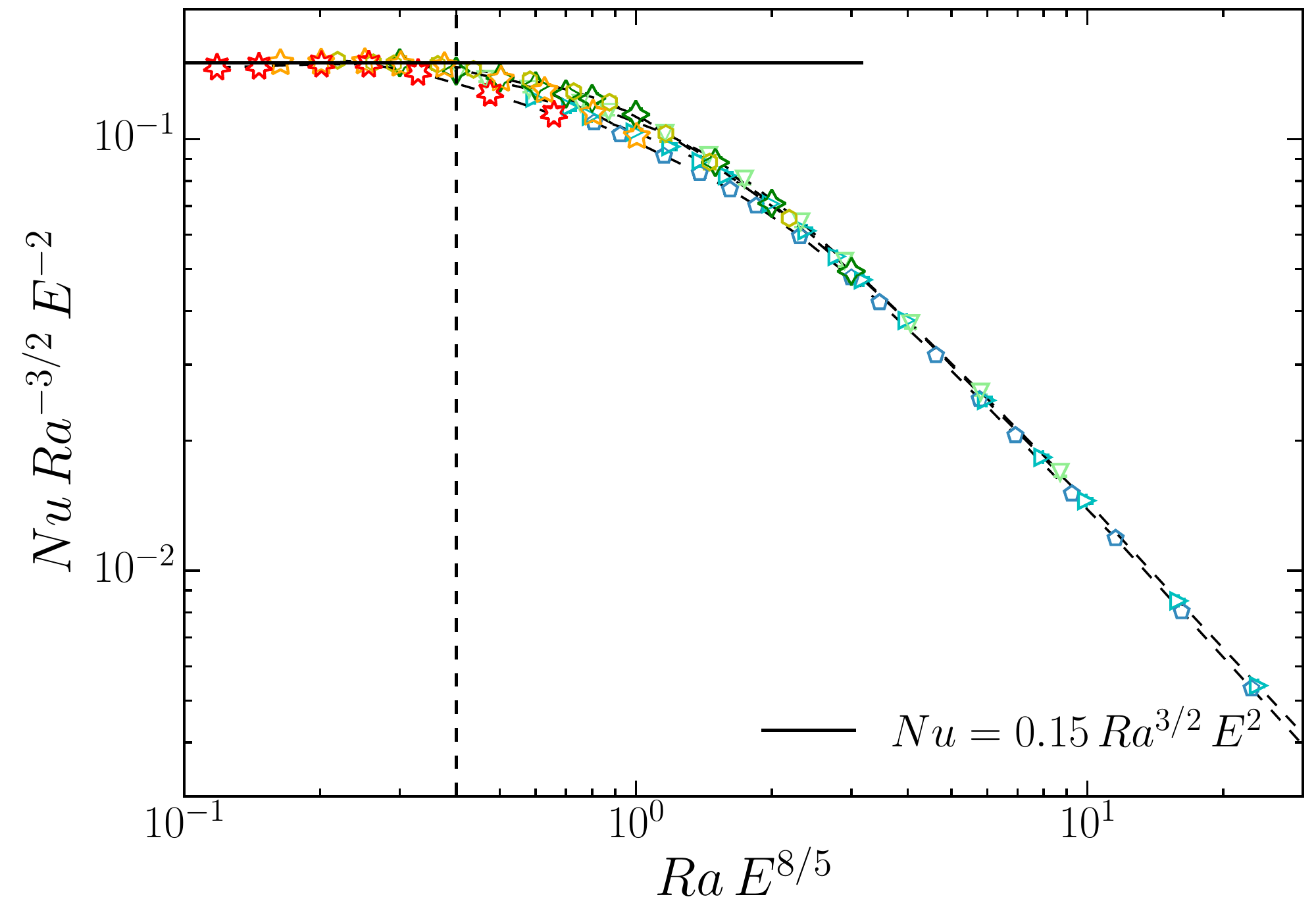}
 \caption{Nusselt number compensated by $Ra^{3/2}E^2$ as a function of 
$RaE^{8/5}$.  Only the cases with $E \leq 3\times 10^{-4}$ and $Nu \geq 2$ 
are displayed in this figure. The solid black line corresponds to the 
diffusivity-free scaling $Nu = 0.15\,Ra^{3/2}E^2$ 
(Equation~\ref{eq:numscalingnu}) already displayed in 
figure~(\ref{fig:nuraek}), while the vertical line corresponds to the 
dynamical boundary scaling $RaE^{8/5} = 0.4$. The symbols have the same 
meaning as in figure~\ref{fig:nura}.} 
 \label{fig:rotationDominated}
\end{figure}

\cite{Julien12} thus argued that the transition from 
\emph{rotationally-constrained} to what they call \emph{rotationally-influenced} 
convection is rather controlled by the dynamics 
of the thermal boundary layers. In particular, they proposed that the 
diffusion-free scaling (\ref{eq:asymptotic}) breaks down when the thermal 
boundary layer is no longer in geostrophic balance. In other words, 
$Nu\sim \widetilde{Ra}^{3/2}$ would hold as long as the local convective Rossby 
number in the thermal boundary layer $Ro_\lambda$ remains smaller than unity.  
The definition of $Ro_\lambda$ leads to
\[
 {Ro}_\lambda = \frac{Ra_\lambda^{1/2}\,E_\lambda}{Pr^{1/2}}=\frac{(\alpha_T g 
\Delta T_\lambda)^{1/2}}{\Omega \lambda^{1/2}},
\]
where $\lambda$ is the thermal boundary layer thickness. Since
\[
 Nu \sim \frac{\Delta T_\lambda}{\lambda}\frac{L}{\Delta T},
\]
using (\ref{eq:asymptotic}), $Ro_\lambda \sim 1$ thus yields
\begin{equation}	
 Ra\,E^{8/5}Pr^{-3/5} \sim 1\,.
 \label{eq:raek85}
\end{equation}
To test the applicability of this transition parameter, 
figure~\ref{fig:rotationDominated} shows the heat transfer data normalised by 
the diffusivity-free scaling (\ref{eq:asymptotic}), $NuRa^{-3/2}E^{-2}$, 
versus $RaE^{8/5}$ for the numerical models with $E\leq 3\times 10^{-4}$ and 
$Nu \geq 2$. The numerical simulations that fulfill $RaE^{8/5} < 0.4$ lie very 
close to the geostrophic scaling (\ref{eq:asymptotic}). A best fit to the 16 
cases that satisfy this criterion indeed yields 
$NuRa^{-3/2}E^{-2}=0.149(\pm 0.002)$. The combination of the two criteria 
$RaE^{4/3} > 6$ and $RaE^{8/5} < 0.4$ defines a wedge in the parameter 
space highlighted by the two black lines visible in figure~\ref{fig:NuExpLoc} 
where the scaling
\begin{equation}
 Nu = 0.149(\pm 0.002)Ra^{3/2}E^2
 \label{eq:numscalingnu}
\end{equation}
holds. The transition parameter (\ref{eq:raek85}) introduced by \cite{Julien12} 
thus enables us to successfully demarcate the rotation-dominated regime and to 
efficiently collapse the heat transfer data across the regime transition.

\subsubsection{Reynolds number scaling}

\noindent To establish  a scaling relation for the typical convective flow 
speed $Re_c$, 
we examine first the variations of the viscous dissipation rate 
$\epsilon_U$. In the limit of vanishing Ekman number, the viscous dissipation 
is expected to primarily occur in the bulk of the fluid \citep{Busse70}. 
However, in the parameter range currently accessible  to numerical models (i.e. 
$E \geq \mathcal{O}(10^{-7}$)), dissipation by friction through the Ekman 
boundary layers might still contribute substantially to the total viscous 
dissipation rate \citep[e.g.][]{Aubert01}. We thus follow the same strategy that 
has been fruitfully introduced by \cite{Grossmann00} for non-rotating RBC and
decompose the viscous dissipation rate into a fluid bulk contribution 
$\epsilon_U^{bu}$ and a boundary layer contribution $\epsilon_U^{bl}$:
\begin{equation}
 \epsilon_U = \epsilon_U^{bu}+ \epsilon_U^{bl} = 
\frac{\nu}{V}\lp\int_{\mathcal{T}} \lp 
\vec{\nabla}\times\vec{u}\rp^2 \mathrm{d}V + \int_{V\setminus\mathcal{T}} \lp 
\vec{\nabla}\times\vec{u}\rp^2 \mathrm{d}V\rp,
\end{equation}
where the integration domain is defined by $\mathcal{T} =\lbrace r_i+\ubli 
\leq r \leq r_o-\ublo; 0\leq \theta \leq \pi; 0\leq\phi\leq 2\pi\rbrace$, where 
$\ubli$ ($\ublo$) are the thicknesses of the inner (outer) viscous boundary 
layer. We then establish scaling relations for the two contributions 
$\epsilon_U^{bu}$ and $\epsilon_U^{bl}$ as functions of $Re_c$ and $E$, starting
with the kinetic dissipation rate in the bulk of the fluid which can be 
estimated by
\[
 \epsilon_U^{bu} \sim \nu \frac{U_c^2}{\ell_{\text{diss}}^2},
\]
where $\ell_{\text{diss}}$ is the viscous dissipation length scale defined such 
that the local Reynolds number associated with this length scale is unity, i.e. 
$u_\text{diss}\,\ell_{\text{diss}}/\nu \sim 1$. 
Following \cite{Davidson13}, we also introduce a small-scale vorticity 
$\omega_{\text{diss}}$ defined by $\omega_{\text{diss}} \sim 
u_\text{diss}/\ell_\text{diss}$, which yields
\begin{equation}
 \epsilon_U^{bu} \sim U_c^2\,\omega_{\text{diss}}\,.
 \label{eq:epsBulk}
\end{equation}
At this stage, the study of the viscous dissipation rate needs to be 
complemented by a force balance analysis in order to estimate the scaling of 
$\omega_{\text{diss}}$. A  viscosity-free scaling of rotating convection was 
first proposed by \cite{Stevenson79} and \cite{Ingersoll82} and later further 
developed by \cite{Aubert01}. This scaling, known as the inertial scaling of 
rapidly-rotating convection, hypothesises a triple force balance between 
Coriolis, inertia and Archimedean force (hereafter CIA). Taking the curl of the 
Navier-Stokes equation (\ref{eq:navier}) in its dimensional form, the CIA 
triple balance yields
\[
 \Omega \frac{\partial\vec{u}}{\partial z}\sim \vec{u} 
\cdot\vec{\nabla}\vec{\omega} \sim \vec{\nabla}\times \lp \alpha_T T 
g\,\vec{e_r} \rp \,,
\]
where the factor two in the vortex stretching term has been omitted.
We now introduce the two integral length scales $\ell_\perp$ and 
$\ell_{\parallelsum}$, respectively perpendicular and parallel to the 
rotation axis. Since rapidly-rotating convective patterns take the form of 
columnar structures aligned with the rotation axis, $\ell_\perp \ll 
\ell_{\parallelsum}$. Denoting $U_c$ the typical velocity, $\Theta$ the typical 
thermal perturbation and $\omega_c$ the typical vorticity, one gets
\begin{equation}
 \Omega\,\frac{U_c}{\ell_{\parallelsum}}
 \sim U_c\,\frac{\omega_c}{\ell_\perp} 
\sim \alpha_T g\,\frac{\Theta}{\ell_\perp}\,.	
\label{eq:cia}
\end{equation}
We now assume that the axial dimension of the convection rolls 
$\ell_{\parallelsum}$ can be approximated by the container length scale $L$. 
This assumption combined with the balance between Coriolis force and inertia in 
(\ref{eq:cia}) yields
\begin{equation}
 \ell_\perp/L \sim \omega_c/\Omega\,. 
 \label{eq:lperp}
\end{equation}
We further assume that the convective heat flux per unit area $\mathcal{Q}$ 
scales as
\[
 \mathcal{Q} \sim \rho c_p U_c \Theta \sim (Nu-1)\kappa \rho c_p \Delta T/L\,,
\]
which implies that the correlation between thermal perturbation and velocity 
are independent of $E$, $Pr$ and $Ra$ \citep{Gillet06}. The balance between 
inertia and buoyancy in (\ref{eq:cia}) then leads to
\[
 Ra_{\mathcal{Q}}Pr^{-2} \sim (L^4/\nu^3)\, U_c^2\,\omega_c\,.
\]
Since on time average $(\nu^3/L^4)\,\epsilon_U \sim Ra_{\mathcal{Q}}Pr^{-2}$, 
it follows 
from (\ref{eq:epsBulk}) $\omega_{\text{diss}} \sim \omega_c$. The 
proportionality relation between the typical time scale at the integral length 
$\ell_\perp$ and at the dissipation scale $\ell_{\text{diss}}$ is a distinctive 
feature of two-dimensional turbulence \citep[e.g.][]{DavidsonBook}. Finally 
assuming that $\omega_c \sim U_c/\ell_\perp$, one can derive the following 
expression for the viscous dissipation rate in the fluid bulk
\modif{
\begin{equation}
\epsilon_U^{bu} = \frac{U_c^3}{\ell_\perp} = 
\frac{\nu^3}{L^4}\,\frac{Re_c^{5/2}}{E^{1/2}}\,.
\label{eq:epsbulk}
\end{equation}}

\begin{figure}
 \centering
 \includegraphics[width=\textwidth]{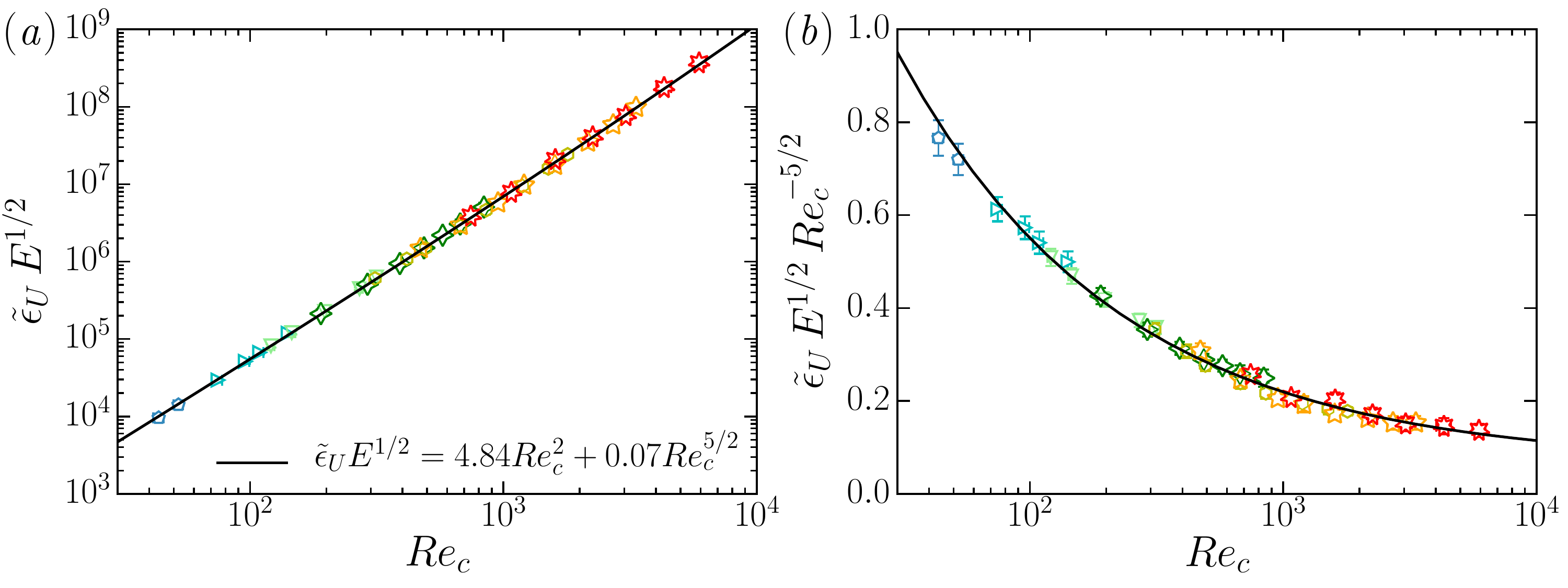}
 \caption{(\textit{a}) Dimensionless viscous dissipation rate 
$\tilde{\epsilon}_U$ as a function of the Reynolds number $Re_c$. (\textit{b}) 
Corresponding compensated $\tilde{\epsilon}_U$ scaling. The solid black lines 
correspond to the theoretical scaling derived in (\ref{eq:dissip_theory}) with 
$a=0.066$ and $b=4.843$. Only 
the cases that fulfill the conditions $Nu > 2$ and $Ra E^{8/5} <1$ have been 
selected on this figure. The error bars illustrate the r.m.s. fluctuations and 
correspond to one standard-deviation from the time-averages. The symbols have 
the same meaning as in figure~\ref{fig:nura}.}
 \label{fig:dissip}
\end{figure}

The same procedure can be applied to the viscous dissipation in the Ekman 
boundary layers with the difference that the typical length scale is now the 
boundary layer thickness $\lambda_U$. Furthermore, the dissipation 
occurs only in the fraction of the fluid volume occupied by the viscous 
boundary layers. In the limit of $\lambda_U \ll L$, this 
volume fraction can be approximated by the ratio $\lambda_U/L$, thus neglecting 
any curvature effect. This yields
\[
 \epsilon_U^{bl} \sim \nu \frac{U_c^2}{\lambda_U^2}\,\frac{\lambda_U}{L}\,.
\]
Since $\lambda_U/L \sim E^{1/2}$ (see below in \S~\ref{sec:ekLayers}), the 
viscous dissipation by friction through the boundary layers can be expressed by
\begin{equation}
 \epsilon_U^{bl} \sim \frac{\nu^3}{L^4} \,\frac{Re_c^2}{E^{1/2}}\,.
\label{eq:epsbl}
\end{equation}
Combining the equations (\ref{eq:epsbulk}) and (\ref{eq:epsbl}) allows us to 
derive the following scaling behaviour for the dimensionless viscous 
dissipation rate
\begin{equation}
 \tilde{\epsilon}_{U} = 
a\,\frac{Re_c^{5/2}}{E^{1/2}}+b\,\frac{Re_c^{2}}{E^{1/2}}\,.
 \label{eq:dissip_theory}
\end{equation}
To test the validity of this scaling, we calculate $\tilde{\epsilon}_U$ and 
$Re_c$ in our numerical simulations and directly fit the expression 
(\ref{eq:dissip_theory}). This leaves only the two prefactors $a$ and 
$b$ as free fitting parameters. Figure~\ref{fig:dissip}(\textit{a}) shows 
$\tilde{\epsilon}_U E^{1/2}$ as a function of $Re_c$ for the 41 cases that 
fulfill $Nu >2$ and $RaE^{8/5} <1$. The best-fit to the numerical data yields
$a=0.066(\pm0.003)$ and $b=4.843(\pm0.096)$. The expression 
(\ref{eq:dissip_theory}) provides an excellent agreement with the data, 
further confirmed by the bulk compensated scaling 
$\tilde{\epsilon}_U E^{1/2} Re_c^{-5/2}$  shown in panel (\textit{b}). 
Explicitly including the viscous friction that occurs in the Ekman boundary 
layers thus allows to very accurately describe the numerical data over a broad 
range of parameter $3\times 10^{-4}\leq E  \leq 3\times 10^{-7}$, $10^{2} \leq 
Re_c \leq 10^{4}$. We also note that a stricter restriction to the 16 turbulent 
cases that satisfy the criterion $RaE^{8/5} <0.4$ and thus lie within the wedge 
displayed in figure (\ref{fig:NuExpLoc}) yields very similar values for $a$ and 
$b$. The expression (\ref{eq:dissip_theory}) also allows to estimate the 
transition value of $Re_c$ beyond which the viscous dissipation that occurs in 
the interior of the fluid will dominate the dissipation in boundary layers.  
Figure~\ref{fig:dissip}(\textit{b}) demonstrates that our dataset nearly 
reaches this point where the bulk scaling alone provides a good description.
Using the best fitting parameters $a$ and $b$ yields $Re_c = (b/a)^2 \simeq 
5500$. Beyond this value, the 
viscous dissipation in the Ekman layers becomes secondary and the dynamical 
behaviour will then gradually tend towards the inertial scaling of rotating 
convection.

\begin{figure}
 \centering
 \includegraphics[width=\textwidth]{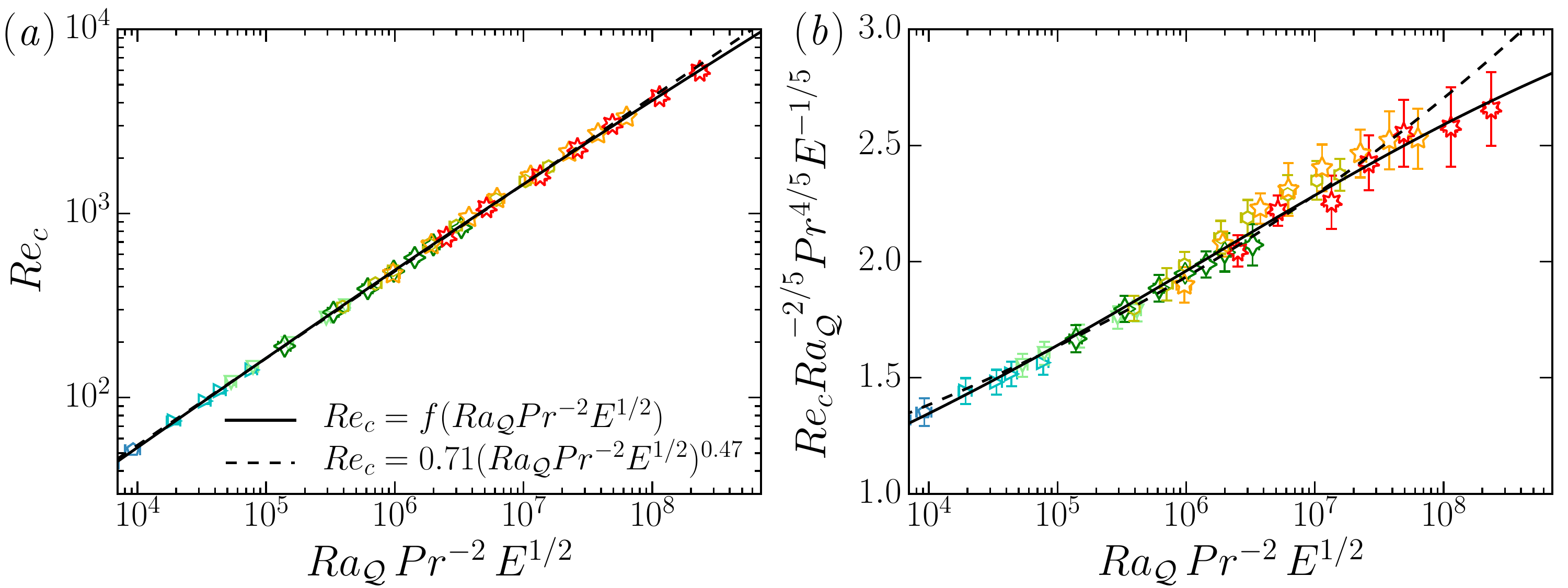}
 \caption{(\textit{a}) Reynolds number $Re_c$ as a function of 
$Ra_\mathcal{Q}Pr^{-2}E^{1/2}$. (\textit{b}) Corresponding compensated 
$Re_c$ scaling. The solid black lines correspond to the theoretical 
scaling derived in (\ref{eq:recfunc}), while the dashed black lines 
correspond to the least-square fit to the data. Only the cases that fulfill the 
conditions $Nu > 2$ and $RaE^{8/5} <1$ have been selected on this figure. The 
error bars illustrate the r.m.s. fluctuations and correspond to one 
standard-deviation from the time-averages. The symbols have the same 
meaning as in figure~\ref{fig:nura}.}
 \label{fig:rey_scaling}
\end{figure}

Assuming that the viscous dissipation either entirely occurs in the fluid bulk 
(i.e. $\epsilon_U \simeq \epsilon_U^{bu}$) or in the Ekman layers 
(i.e. $\epsilon_U\simeq\epsilon_U^{bl}$) allows to define the two 
end-member scaling relations:
\begin{equation}
Re_c^{bu} \leq Re_c \leq Re_c^{bl},
\label{eq:reybounds}
\end{equation}
where
\begin{equation}
Re_c^{bl} \sim Ra_{\mathcal{Q}}^{1/2}Pr^{-1}E^{1/4}
\label{eq:rey_bl}
\end{equation}
and
\begin{equation}
Re_c^{bu} \sim Ra_{\mathcal{Q}}^{2/5}Pr^{-4/5}E^{1/5}\,.
\label{eq:rey_cia}
\end{equation}
We note that (\ref{eq:rey_cia}) corresponds to the convective velocity 
scaling classically obtained in the CIA triple balance 
\citep[e.g.][]{Aubert01,Gillet06,Barker14}, while the scaling (\ref{eq:rey_bl}) 
has the same dependence on $Ra_\mathcal{Q}$ as the VAC triple balance 
(\ref{eq:revac}) but carries a different $E$-scaling exponent.
Once $a$ and $b$ have been determined, the equation 
(\ref{eq:dissip_theory}) can be numerically solved to derive the combined 
scaling law for $Re_c$:
\begin{equation}
 Re_c = f(Ra_{\mathcal{Q}}Pr^{-2}E^{1/2})\,.
 \label{eq:recfunc}
\end{equation}
For comparison purposes, we also compute a simple best fit to the data which 
yields $Re_c=0.708(\pm0.017)(Ra_{\mathcal{Q}}Pr^{-2}E^{1/2})^{0.473\pm0.002}$. 
Figure~\ref{fig:rey_scaling}(\textit{a}) shows $Re_c$ versus 
$Ra_{\mathcal{Q}}Pr^{-2}E^{1/2}$ for the 41 cases already shown in figure 
\ref{fig:dissip}, while figure~\ref{fig:rey_scaling}(\textit{b}) shows the 
corresponding bulk compensated scaling. 

Previous parameter studies of rotating convection in spherical shells either 
carried out under the quasi-geostrophic assumption \citep{Gillet06,Guervilly10} 
or in fully three-dimensional models \citep{Christensen02,KingBuffett13} 
obtained steeper exponents than the $2/5$ scaling expected in the classical CIA 
balance (Equation~\ref{eq:rey_cia}). This discrepancy has usually been 
attributed to the significant role played by viscosity in numerical models and 
thus prompted several authors to rather describe their dataset using the VAC 
scaling hypothesis (\ref{eq:revac}). Since the scaling exponents are relatively 
close to each other, $2/5$ versus $1/2$ regarding the dependence on 
$Ra_{\mathcal{Q}}$\footnote{Note however that the Ekman number dependence is not 
the same for these two scalings.}, discriminating one scaling from another 
remains however a difficult task \citep{KingBuffett13}. Here we show that the 
separation of $\epsilon_U$ into boundary layers and bulk contributions
allows to very accurately describe the numerical data. Moreover, in the range 
of $Re_c$ covered by our numerical dataset (i.e. $ 10^2 \leq Re_c < 10^4$), the 
simple power law 
$Re_c=0.708(\pm0.017)(Ra_{\mathcal{Q}}Pr^{-2}E^{1/2})^{0.473\pm0.002}$ 
provides a statistically nearly indiscernible fit quality 
from the scaling theory (\ref{eq:recfunc}). Deviations only become more 
significant at the high-end of the parameter range, where the power law 
fails to accurately capture the complex dependence of $Re_c$ upon 
$Ra_{\mathcal{Q}}Pr^{-2}E^{1/2}$. This explains why previous 
analyses that reduced the scaling behaviours of $Re_c$ to such simple power 
laws obtained scaling exponents robustly larger than $2/5$ 
\citep[e.g.][]{Gillet06,KingBuffett13}.

\subsubsection{Flow length scales and interior temperature gradients}

\begin{figure}
 \centering
 \includegraphics[width=\textwidth]{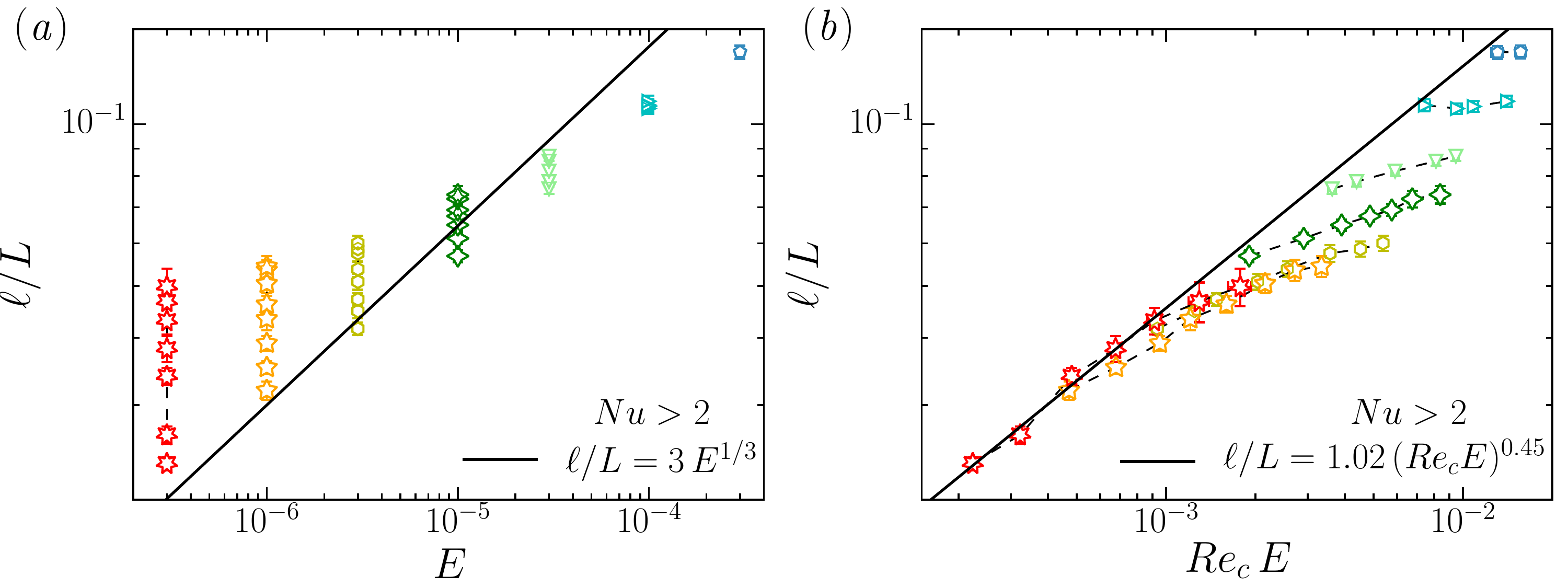}
 \caption{(\textit{a}) Average flow length scale $\ell/L$ calculated using 
(\ref{eq:lengthscalesdef}) as a function of the 
Ekman number for the numerical models with $E\leq 3\times 10^{-4}$ and $Nu > 
2$. For comparison, the weakly non-linear scaling obtained in 
figure~\ref{fig:lengthscalesOnset} has been plotted as a solid black line.
(\textit{b}) Corresponding compensated $\ell/L$
scaling. Only the 41 cases that satisfy $Nu > 2$ and $RaE^{8/5} < 1$
have been selected on this figure. The solid black line corresponds to the 
best-fit to the cases with $E=3\times 10^{-7}$. The error bars illustrate the 
r.m.s. fluctuations and correspond to one standard-deviation from the 
time-averages. The symbols have the same meaning as in figure~\ref{fig:nura}.}
 \label{fig:lengthscalesNL}
\end{figure}

\noindent Using the equations (\ref{eq:lperp}) and (\ref{eq:rey_cia}), the 
inertial theory of rotating convection predicts the following scaling for the 
integral flow length scale $\ell_\perp$
\begin{equation}
 \ell_\perp/L \sim \lp U_c / \Omega L \rp^{1/2} \sim \lp Re_c E \rp^{1/2},
 \label{eq:lperpcia}
\end{equation}
which is generally referred to as the Rhines scaling \citep{Rhines75}. Using 
the numerical models computed by \cite{Christensen06}, several analyses of the 
typical flow length scale suggest that the VAC scaling (\ref{eq:lperpvac}) might 
possibly hold beyond the weakly non-linear regime of rotating convection 
\citep[e.g.][]{KingBuffett13,Oruba14}. 
Figure~\ref{fig:lengthscalesNL}(\textit{a}) shows the calculations of the 
average flow length scale $\ell/L$ plotted as a function of $E$ for the 
numerical models with $E\leq 3\times 10^{-4}$ and $Nu >2$. The $\ell \sim 
E^{1/3}\,L$ scaling that accurately describes 
the numerical data close to onset of convection (see 
figure~\ref{fig:lengthscalesNL}) fails to properly capture the length scale 
variations in the non-linear regime. Especially, at low 
Ekman numbers ($E \leq 3\times 10^{-6}$), $\ell$ exhibits a strong additional 
dependence on the convective forcing and increases with the supercriticality 
(see 
Table~\ref{tab:runs} for details). Figure~\ref{fig:lengthscalesNL}(\textit{b}) 
shows the calculations of $\ell$ versus $Re_cE$ for the 41 
numerical models already shown in figures \ref{fig:dissip} and 
\ref{fig:rey_scaling}. A best-fit to the cases with 
$E=3\times 10^{-7}$ that fulfill $RaE^{8/5}<0.4$ 
yields $\ell/L = 1.016(\pm0.212)(Re_c E)^{0.45(\pm 0.027)}$, reasonably close 
to the theoretical scaling (\ref{eq:lperpcia}). Since the dissipation that 
occurs in the boundary layers still plays a substantial role in the velocity 
scaling (\ref{eq:dissip_theory}), it is not surprising that (\ref{eq:lperpcia}) 
is only approached for the lowest Ekman number considered in 
this study.

\begin{figure}
 \centering
 \includegraphics[width=\textwidth]{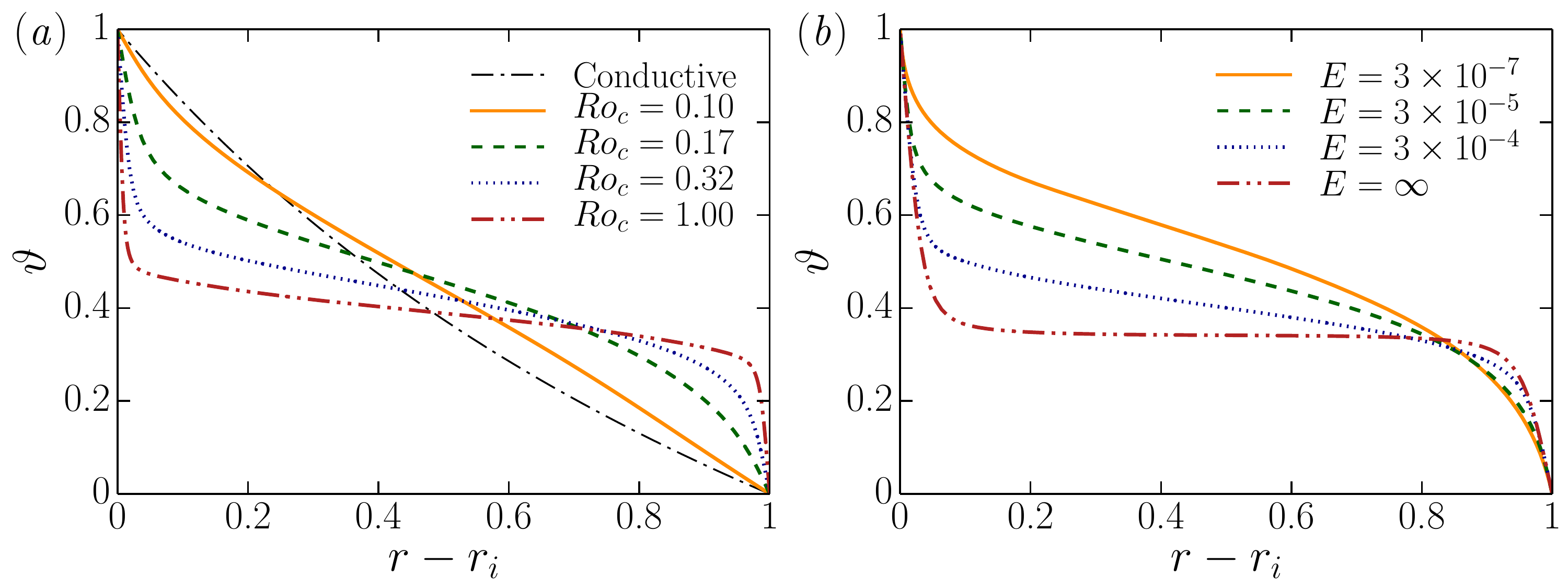}
 \caption{(\textit{a}) Radial profile of the time and horizontally-averaged 
temperature $\vartheta(r)$ for different convective Rossby number 
$Ro_c$ with a fixed Ekman number of $E=10^{-4}$. \modif{The dash-dotted black 
line corresponds to the conductive temperature profile solution of 
(\ref{eq:dtcdr}).}
(\textit{b}) Radial profile of the time and horizontally-averaged 
temperature $\vartheta(r)$ for different Ekman numbers with approximately the 
same Nusselt number $10.3<Nu<11$.}
 \label{fig:temp_ek} 
\end{figure}

While non-rotating convection undergoes an efficient turbulent mixing which 
results in an isothermal fluid interior \citep[e.g.][]{Verzicco99}, the 
prominent role played by the Coriolis force in rotating 
convection impedes the mixing. Both experiments 
\citep[e.g.][]{Boubnov90,Kunnen10} 
and direct numerical simulations \citep[e.g.][]{Julien96,King12,Stellmach14}
have indeed revealed that a large-scale interior temperature 
gradient can be maintained in the rapidly-rotating regime \citep[see 
also][]{Gillet06,King10}. Figure~\ref{fig:temp_ek} shows  
the time and horizontally averaged temperature  profile $\vartheta(r)$ for 
several rotating and non-rotating models. At a fixed Ekman number, an increase 
of the supercriticality (panel \textit{a}) changes the temperature distribution 
from a nearly conductive state (solid orange line) to a nearly isothermal fluid 
bulk (dot-dashed line). At larger $Ro_c$ (broken lines), the rotational 
constraint decreases and thin thermal boundary layers start to emerge and tend 
to accommodate an increasing fraction of the temperature drop. As visible on 
panel (\textit{b}), the transition from rotating to non-rotating convection is 
thus associated with a gradual lowering of the average temperature gradient and 
a transition to a physical regime where boundary layers entirely control the 
heat transfer.

\begin{figure}
 \centering
 \includegraphics[width=\textwidth]{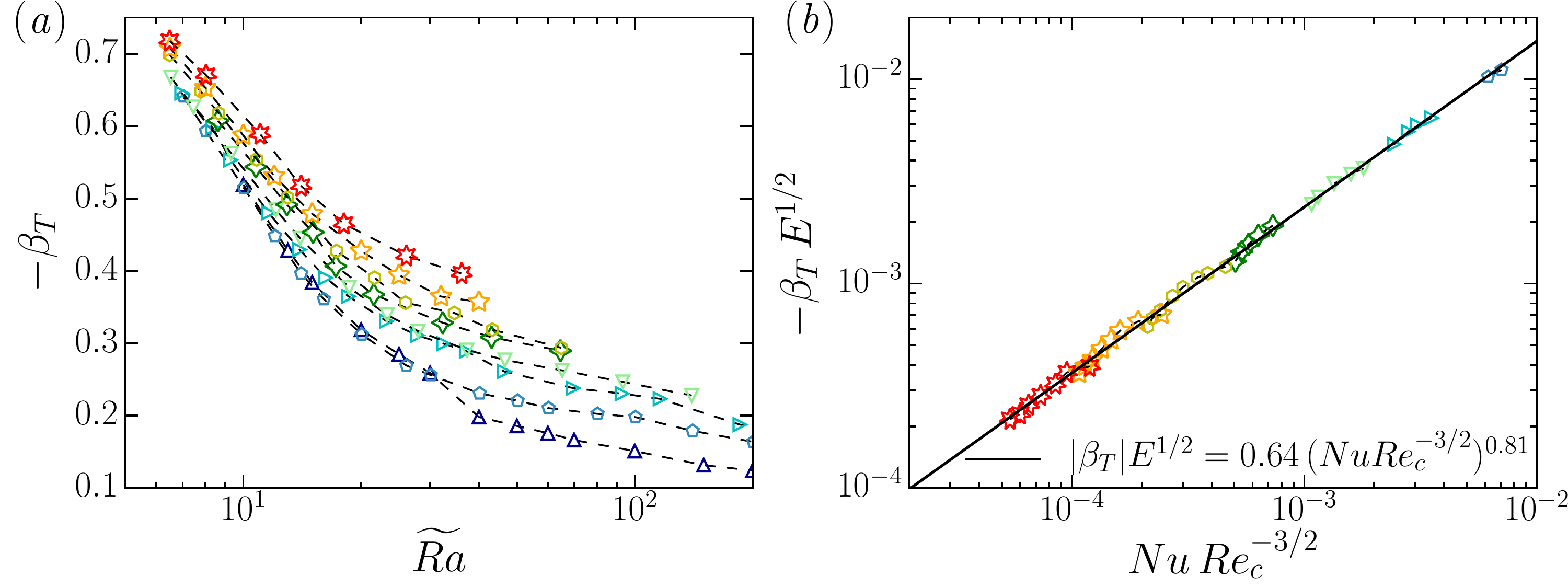}
 \caption{(\textit{a}) Temperature gradient at mid-shell radius 
$\tempgrad$ versus $\widetilde{Ra}$ for the numerical models with $Nu > 2$ and 
$E \leq 10^{-3}$. (\textit{b}) Absolute temperature gradient at mid-shell 
renormalised by $E^{-1/2}$ versus $NuRe_c^{-3/2}$  for the 
numerical models with $RaE^{8/5} < 1$ and $Nu > 2$. The solid black line 
corresponds to the best-fit to the data. The symbols have the same 
meaning as in figure~\ref{fig:nura}.}
 \label{fig:beta}
\end{figure}

Measuring the interior temperature gradient $\tempgrad$ (\ref{eq:beta}) 
thus helps to characterise the heat transfer regime. 
Figure~\ref{fig:beta}(\textit{a}) shows $\tempgrad$ versus $\widetilde{Ra}$ for 
the set of numerical simulations with $E\leq 10^{-3}$. An increase of the 
supercriticality $\widetilde{Ra}$ leads to a gradual decrease of $|\tempgrad|$.
The thermal profiles in the fluid bulk tend to a thermally well-mixed interior 
(i.e. vanishing $\tempgrad$) when the rotational influence on the flow 
decreases. In contrast to the asymptotic calculations by \cite{Julien12}, our 
dataset does not show any evidence for a direct scaling relation between 
$\tempgrad$ and $\widetilde{Ra}$. However, following \cite{Stevenson79} and 
\cite{Barker14}, the interior temperature gradient can be estimated by
\[
 \tempgrad \sim \frac{\Theta}{\ell_\perp}\frac{L}{\Delta T}\,,
\]
in the rapidly-rotating limit.
Using (\ref{eq:lperpcia}) for the length scale scaling and approximating 
$\Theta/\Delta T$ by $Nu/Re_c$ then yields
\begin{equation}
 \tempgrad \sim Nu\,Re_c^{-3/2}E^{-1/2}\,.
\label{eq:betacia}
\end{equation}
Figure~\ref{fig:beta}(\textit{b}) shows that this scaling provides a reasonable
description of the 41 numerical cases already discussed in figures 
\ref{fig:dissip} and \ref{fig:rey_scaling}. The best-fit to the data however 
gives a shallower exponent than expected from theory: $\tempgrad E^{1/2} = 
0.643(\pm0.027)(Nu Re_c^{-3/2})^{0.812(\pm0.005)}$. This discrepancy 
is not surprising since the underlying $\ell_\perp$ scaling that enters 
(\ref{eq:betacia}) is only partially realised in our set of simulations (see 
figure \ref{fig:lengthscalesNL}). Lowering the Ekman number further might help 
to ascertain this scaling. Indeed, the local numerical calculations carried 
out in cartesian coordinates by \cite{Barker14}  demonstrates the validity of 
(\ref{eq:betacia}) once the influence of the \modif{viscous} boundary layers 
has been minimised. Replacing the expected asymptotic scalings for $Nu$ 
(Equation~\ref{eq:asymptotic}) and $Re_c$ (Equation~\ref{eq:rey_cia}) in 
(\ref{eq:betacia}) 
yields $\tempgrad \sim  Pr$. This implies that $\tempgrad$ should saturate on a 
constant value in the turbulent quasi-geostrophic regime of 
rotating convection. Local numerical calculations carried out in cartesian 
coordinates by \cite{Stellmach14} indeed obtain a saturation of $\tempgrad$ for 
$E=\mathcal{O}(10^{-7})$ and $\widetilde{Ra} \sim 100$.

\subsubsection{Thermal boundary layers}

\begin{figure}
 \centering
 \includegraphics[width=\textwidth]{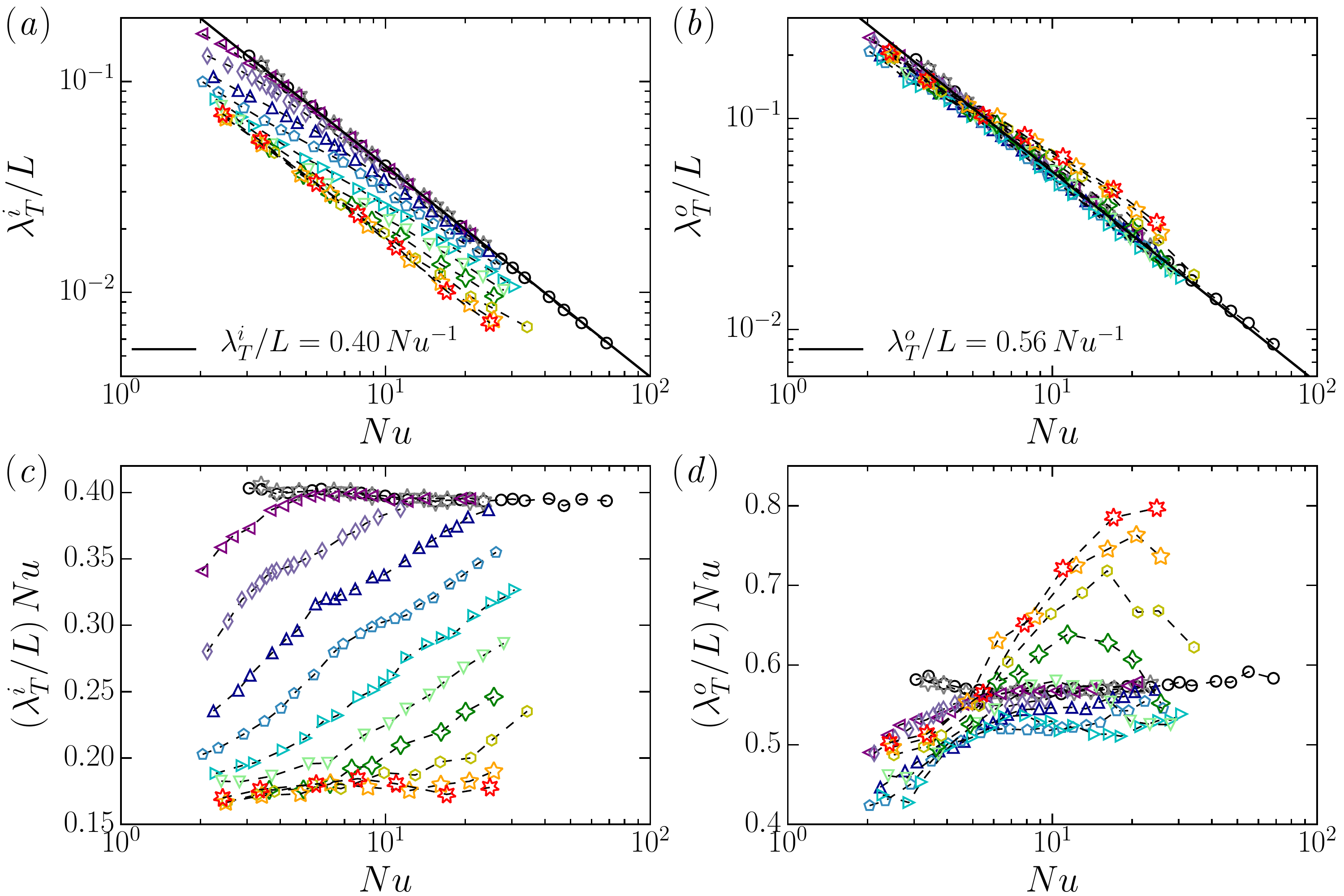}
 \caption{(\textit{a,b}) Thermal boundary layer thickness at the inner (outer)
boundary $\tbli$ ($\tblo$) as a function of the Nusselt number. The solid 
black lines correspond to the theoretical scalings for non-rotating convection 
derived in \citep{Gastine15}. (\textit{c,d}) Compensated $\tbli/L$ ($\tblo/L$) 
scaling. The symbols have the same meaning as in figure~\ref{fig:nura}.}
 \label{fig:tempLayers}
\end{figure}

\noindent Since the heat transport in the thermal boundary layers is controlled 
by diffusion, the Nusselt number provides an estimate for the ratio between the 
thermal boundary layer thickness and the related temperature drop:
\[
 Nu =  \eta \frac{\Delta T_i}{\tbli}\frac{L}{\Delta T} = 
\frac{1}{\eta}\frac{\Delta T_o}{\tblo}\frac{L}{\Delta T},
\]
where $\tbli$ ($\tblo$) is the thickness of the thermal boundary layer at the 
inner (outer) boundary and $\Delta T_i$ ($\Delta T_o$) is the associated 
temperature contrast. This yields the following scalings for the thermal 
boundary layer thicknesses at both boundaries:
\begin{equation}
 \frac{\tbli}{L} = \eta \frac{\Delta T_i}{\Delta T}\frac{1}{Nu}; \quad 
\frac{\tblo}{L} =
\frac{1}{\eta} \frac{\Delta T_o}{\Delta T}\frac{1}{Nu}\,.
\end{equation}
In non-rotating convection in spherical geometry, the thermal boundary layer 
are asymmetric and thus $\Delta T^{i} / \Delta T^{o} \neq 1$. In a previous 
study, we demonstrated that the ratio $\Delta T^i/\Delta T^o$ only depends on 
the radius ratio $\eta$ and on the gravity profile $\tilde{g}(r)$
\citep{Gastine15}. Thus, the scaling for the boundary 
layer thickness in non-rotating RBC is
\begin{equation}
\tbli/L \sim \tblo/L \sim Nu^{-1}\,.
\label{eq:tblrot}
\end{equation}
where only the scaling prefactors depend on the geometry and on the gravity 
distribution of the spherical shell. In rotating convection however, the 
temperature drops at each thermal boundary layer are likely to also depend on 
$Ra$ and $E$ and hence a non trivial additional dependence on $Nu$ might be 
expected (see 
figure~\ref{fig:temp_ek}). Figure~\ref{fig:tempLayers} shows 
\modif{the calculations of 
$\tbli$ and $\tblo$ using the slope-intersection method described in 
\S~\ref{sec:diagnostics}} versus $Nu$ for both rotating and non-rotating 
simulations.
\modif{As pointed out by \cite{Julien12}, other methods to estimate the 
thermal boundary layers in rotating convection might yield different  
thicknesses. A comparison between the values obtained from the slope 
intersection method with the location of the peaks of the r.m.s. of the 
temperature fluctuations for a few selected cases however yield similar 
boundary layer thicknesses \citep[for a comparison, see][]{Gastine15}.}
Large Ekman number models ($E \geq 10^{-2}$) essentially behave similarly to 
the non-rotating cases. At the inner boundary of the spherical shell, the cases 
with intermediate Ekman numbers $3\times 10^{-5} \leq E \leq 3\times 
10^{-3}$ exhibit a lower prefactor and a slightly shallower slope than 
$Nu^{-1}$ particularly visible in the compensated plot displayed in panel 
(\textit{c}). The cases with $E \leq 3\times 10^{-6}$, however seem to follow 
again a scaling $\tbli/L\sim Nu^{-1}$, though with a much smaller prefactor 
($\simeq 0.17$ instead of $\simeq 0.4$) than the non-rotating cases. The thermal 
boundary layer at the outer boundary behaves quite differently. On average, 
$\tblo$ remains much closer to the values found in the non-rotating models and 
can thus be roughly approximated by $\tblo/L \sim Nu^{-1}$. 
However, a secondary, intricate dependence on the Ekman number remains, as 
visible on the compensated scaling of $\tblo$ displayed in panel (\textit{d}).
For both boundary layers, the $Nu^{-1}$ scaling thus provides an acceptable 
first-order approximation of the thermal boundary layer thickness 
\modif{measured by the slope intersection method}, though subtle dependences on 
the Ekman number can impinge on both the prefactor and the scaling exponent.

\subsubsection{Viscous boundary layers}
\label{sec:ekLayers}

\begin{figure}
 \centering
 \includegraphics[width=\textwidth]{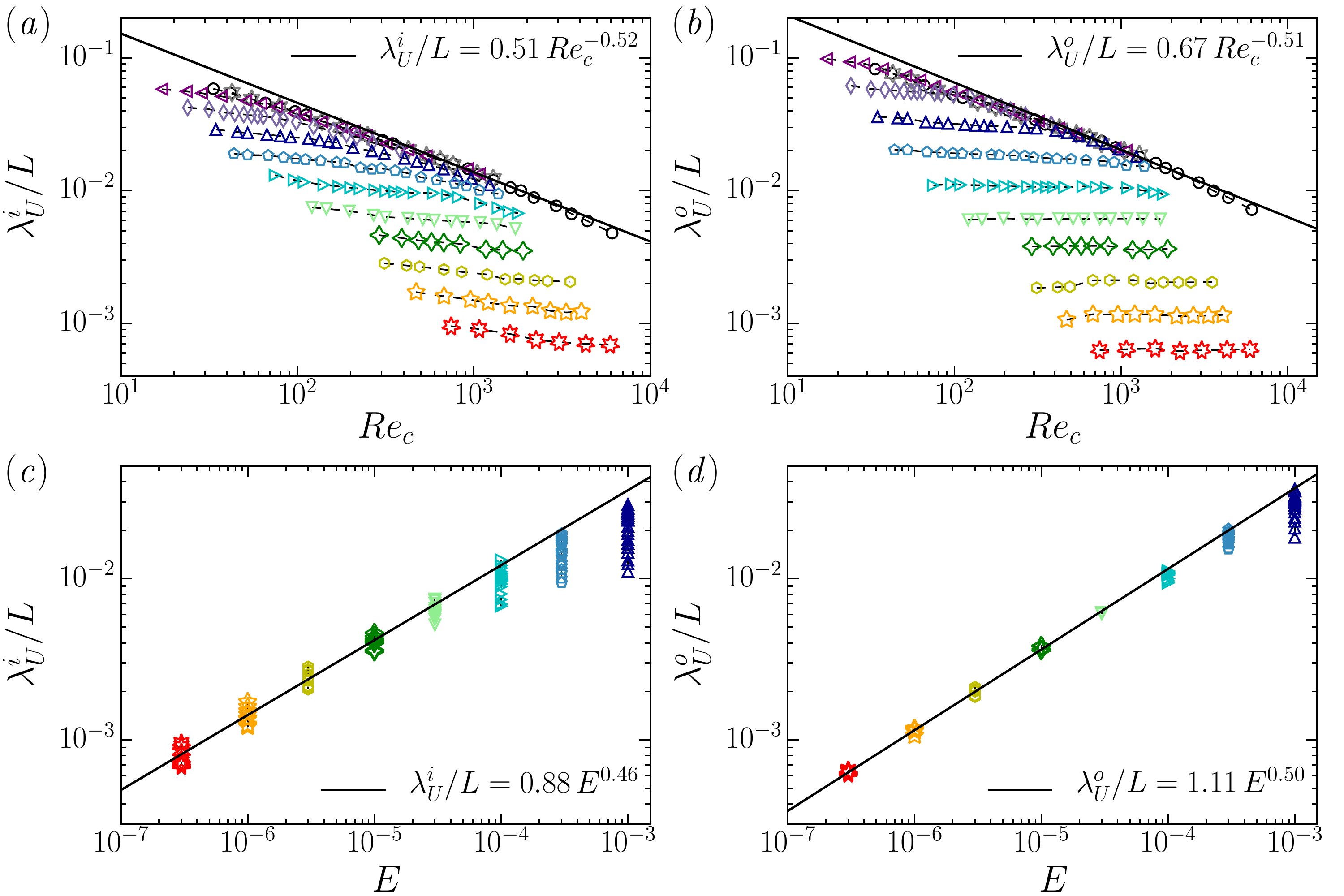}
 \caption{(\textit{a,b}) Viscous boundary layer thickness at the inner (outer)
boundary $\ubli$ ($\ublo$) as a function of $Re_c$. The solid lines 
correspond to the scalings obtained for non-rotating convection derived in 
\citep{Gastine15}. (\textit{c,d}) Viscous boundary layer thickness at the inner 
(outer) boundary as a function of the Ekman number.  The solid black lines 
correspond to the best-fit to the 41 cases that fulfill $RaE^{8/5} < 1$ and $Nu 
>2$. The symbols have the same meaning as in figure~\ref{fig:nura}.}
 \label{fig:ekLayers}
\end{figure}

\noindent In the range of Rayleigh numbers explored here ($10^{3}\leq Ra \leq 
10^{11}$), 
the viscous boundary layers in non-rotating convection are usually assumed to be 
laminar and follow the Prandtl-Blasius boundary layer theory 
\citep{Prandtl04,Blasius08}. Balancing the inertia of the fluid bulk with the
viscous forces in the boundary layers yields
\begin{equation}
 \lambda_U/L \sim Re^{-1/2}\,.
 \label{eq:RBCbl}
\end{equation}
In contrast, owing to the primary role of Coriolis force, the viscous 
boundary layers in rotating convection, called the Ekman layers, differ greatly 
from non-rotating convection. A scaling relation for the Ekman layers can be 
derived by considering a force balance between Coriolis force and viscosity in 
the limit of $E\rightarrow 0$ 
\citep[e.g.][]{Greenspan68}:
\begin{equation}
 \vec{\Omega} \times \vec{u} \sim \nu \vec{\Delta} \vec{u} \quad \rightarrow 
\quad
 \lambda_U/L \sim E^{1/2}\,.
 \label{eq:ekmanLayers}
\end{equation}
Figure~\ref{fig:ekLayers} shows the calculations of the viscous boundary layer 
thicknesses for both the inner and the outer spherical shell boundaries as a 
function of $Re_c$ and $E$. Here we only examine the mean viscous boundary 
layer, though in spherical geometry the Ekman layer thickness might also 
depend on the latitude \citep{Greenspan68}. When the Ekman number is large 
($E\geq 3\times 10^{-3}$), the influence of rotation on the viscous boundary 
layer remains secondary. $\ubli$ and $\ublo$ are then essentially the same as
the non-rotating cases (panels \textit{a} and \textit{b}).
As already shown in our previous study \citep{Gastine15}, when $Re_c \geq 250$, 
the scaling exponents are in close agreement with the Prandtl-Blasius scaling 
(\ref{eq:RBCbl}). At lower Ekman numbers ($E\leq 3\times 10^{-5}$), the viscous
boundary layer thicknesses become almost independent of $Re_c$ and approach the 
theoretical scaling for Ekman layers (panels \textit{c} and \textit{d}). 
Indeed, the best-fit to the numerical cases that fulfill $RaE^{8/5}< 1$ and 
$Nu >2$ yields $\ubli/L=0.875(\pm0.084)\,E^{0.465(\pm0.008)}$ and 
$\ublo/L=1.109(\pm0.042)\,E^{0.497(\pm0.003)}$, in good agreement with 
(\ref{eq:ekmanLayers}). 

\subsection{Transitional regime: from rotating to non-rotating convection}

\begin{figure}
 \centering
 \includegraphics[width=\textwidth]{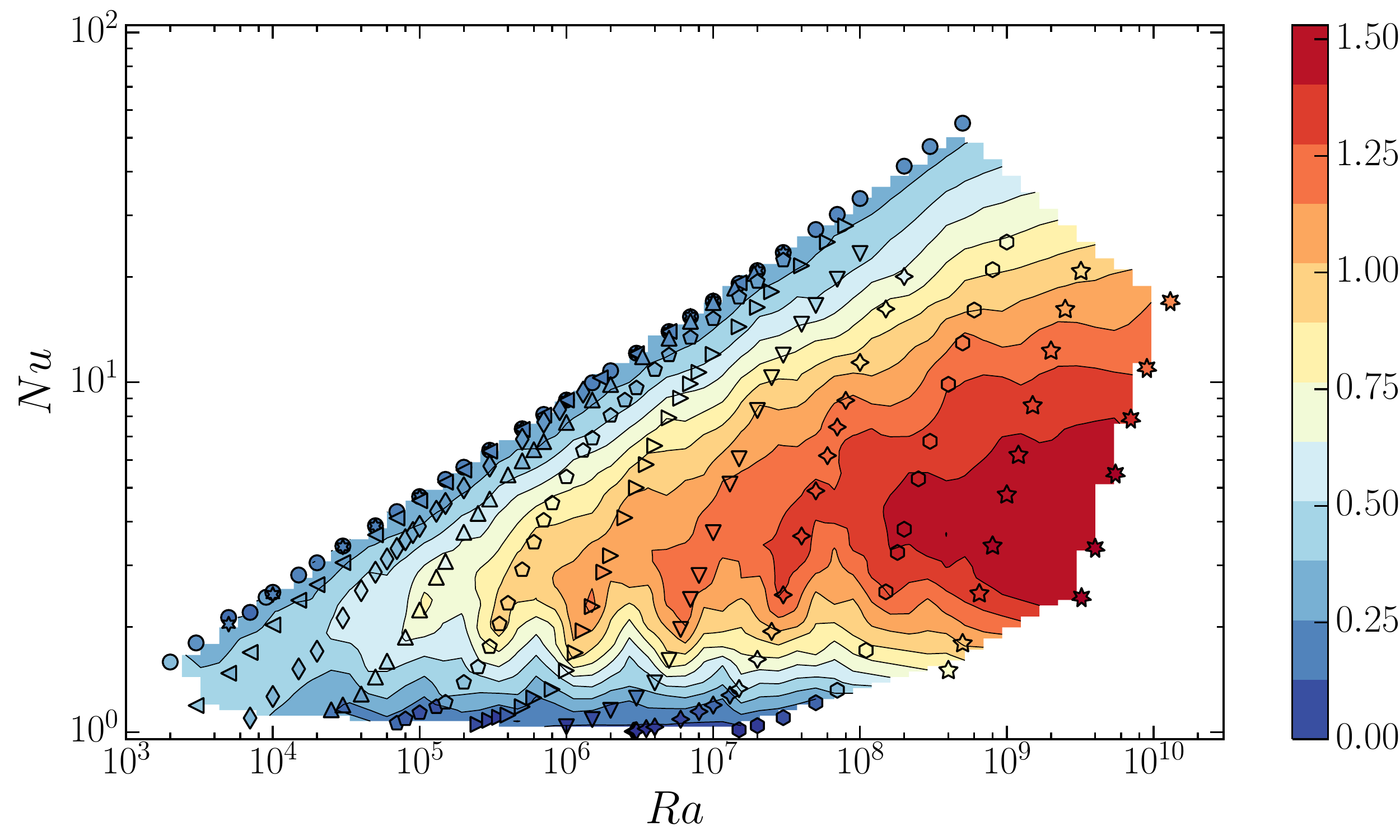}
 \caption{Isocontours of the local effective exponent $\alpha_\text{eff}$ 
of the $Nu =Ra^{\alpha_\text{eff}}E^{\beta_\text{eff}}$ scaling in the 
$(Ra,Nu)$-plane. The symbol shape corresponds to the Ekman number 
as in the previous figures, while the symbol color scales here with the value 
of local slope $\alpha_{\text{eff}}$.}
 \label{fig:alphaContour}
\end{figure}

In the previous sections, we mainly explored the range of parameters where the 
convective flow is strongly constrained by rotation. We showed that
the criterion $RaE^{8/5} \leq \mathcal{O}(1)$ posited by \cite{Julien12a} 
provides an effective way to characterise the upper bound of this physical 
regime. Beyond this point, i.e. when $Ra > E^{-8/5}$, convection is
still influenced by rotation but does no longer operate in the turbulent 
quasi-geostrophic regime. This defines the lower bound of the 
\emph{transitional regime} of rotating convection.
To illustrate the continuous variations of the physical properties within this 
regime, figure~\ref{fig:alphaContour} shows isocontours of the local 
slope $\alpha_{\text{eff}}$ of the Nusselt scaling in a $(Ra, Nu)$-plane.
Beyond the rapidly-rotating regime of
convection in which $\alpha_{\text{eff}} \simeq 3/2$ (dark red area), a rapid 
decrease of the local exponent is observed. The heat 
transport is not characterised by one single scaling exponent 
$\alpha_{\text{eff}}$ but rather exhibits continuous changes, similarly to what 
is observed in classical RBC \citep[e.g.][]{Grossmann00,Funfschilling05}. In 
addition, $Nu$ does not follow a pure function of the supercriticality 
$\widetilde{Ra}$ any longer and the local variations of $\beta_{\text{eff}}$ 
thus decorrelate from those of $\alpha_{\text{eff}}$ (see 
figure~\ref{fig:nuraek}). The numerical models with higher $E$ and $Ra$ in 
which the influence of Coriolis force becomes negligible then transition to a 
slow growth of $Nu$ with $Ra$ close to the values obtained in non-rotating 
convection (i.e. $\alpha_{\text{eff}}<0.4$, $\beta_{\text{eff}}\simeq 0$). This 
defines the upper bound of the transitional regime. 

To determine a transition criterion that only depends on the control 
parameters, we hypothesise that the transition between rapidly-rotating and 
weakly-rotating convection can be defined as the intersection between the steep 
asymptotic diffusivity-free scaling for rotating convection 
(\ref{eq:asymptotic}) and the shallow scaling for non-rotating convection 
$Nu_{\text{NR}}\sim Ra^{\nu_{\text{eff}}}$. Following \cite{Cheng15} we thus 
designate 
$Ra_T$ for the upper bound of the transitional regime by
\[
 Ra_T \sim E^{-4/(3-2\nu_{\text{eff}})}\,,
\]
where the unconstrained $Pr$ dependence has been dropped.
Since the local slope for non-rotating heat transfer increases from roughly 
$\nu_{\text{eff}}=0.27$ to $\nu_{\text{eff}}=0.32$ between $Ra=10^5$ and 
$Ra = 10^9$ in non-rotating spherical shells \citep{Gastine15}, $Ra_T$ can range 
from $E^{-1.62}$ to $E^{-1.69}$. In addition, the \cite{Grossmann00} theory 
predicts a further increase of $\nu_{\text{eff}}$  beyond 
$Ra=10^9$ to reach $\nu_\text{eff}=1/3$ \citep{Chilla12}. We thus rather 
decide to adopt 
$Nu_{\text{NR}}\sim Ra^{1/3}$ as the physically-motivated asymptotic scaling 
for 
non-rotating convection. This yields
\begin{equation}
 Ra_T \sim E^{-12/7}\,,
 \label{eq:ratnonrot}
\end{equation}
a value that lies within the range predicted by 
\cite{Ecke14}. The transitional regime of rotating convection thus covers the
broad parameter range  $E^{-8/5} \leq Ra \leq E^{-12/7}$ in which 
the flow properties will continuously vary to match the regime changes between 
rotation-dominated and non-rotating convection. Due to the intricate nature of 
the force balance with possible crossovers of leading forces, the derivation of 
asymptotic scalings inherent in this transitional regime is
challenging. In the following, we thus rather check whether $Ra_T$ 
(\ref{eq:ratnonrot}) represents an 
accurate transition parameter that separates the regime of rotating convection
from the gradual transition to non-rotating behaviour.

\subsubsection{Nusselt and Reynolds numbers}

\begin{figure}
 \centering
 \includegraphics[width=8.5cm]{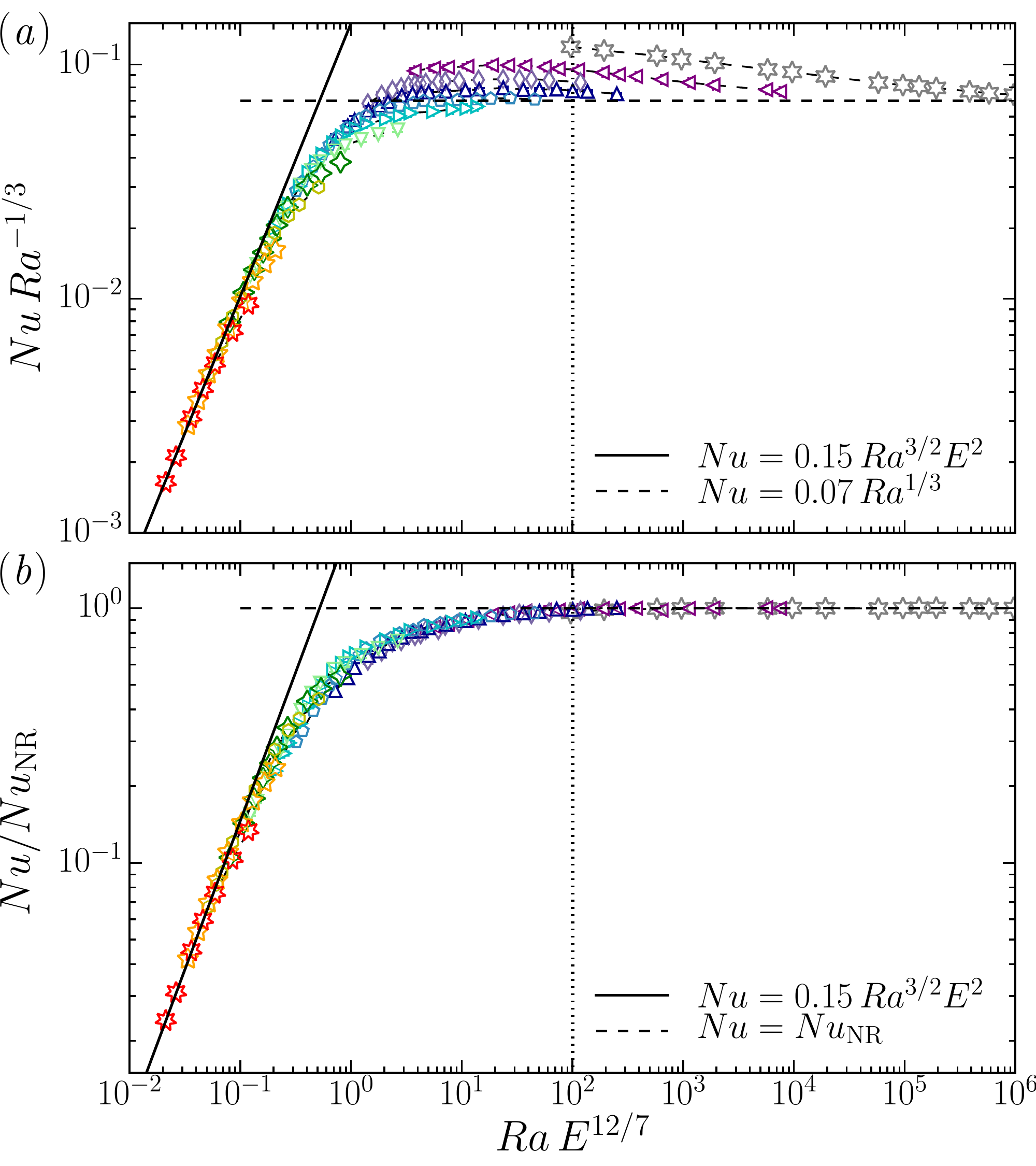}
 \caption{(\textit{a}) Nusselt number $Nu$ renormalised by $Ra^{1/3}$ as a 
function of $RaE^{12/7}$ (Equation~\ref{eq:ratnonrot}). The solid black line 
corresponds to the asymptotic 
scaling  for rotating convection (\ref{eq:numscalingnu}), while the dashed line 
corresponds to the tentative asymptotic scaling for non-rotating convection. 
(\textit{b}) Nusselt number $Nu$ renormalised by $Nu_{\text{NR}}$ as a function 
of 
$RaE^{12/7}$. $Nu_{\text{NR}}$ corresponds to the Nusselt number of the 
non-rotating 
cases at the same Rayleigh number. On both panels, the weakly non-linear cases 
with $Nu < 2$ have been excluded. The dotted vertical lines correspond to the
Rayleigh number $Ra=100\,E^{-12/7}$ beyond which $Nu$ adopts a scaling 
behaviour close to the non-rotating RBC. The symbols have the same meaning as 
in figure~\ref{fig:nura}.}
 \label{fig:nutransition}
\end{figure}

\noindent Figure~\ref{fig:nutransition} shows the calculations of the Nusselt 
number 
compensated by the heat transfer scaling for non-rotating convection as a 
function of the transition parameter (\ref{eq:ratnonrot}). In panel 
(\textit{a}), $Nu$ is compensated on the $y$-axis by the asymptotic scaling 
$Nu_{\text{NR}}\sim Ra^{1/3}$. Well into the rapidly-rotating regime (i.e. 
$RaE^{12/7} 
\ll 1$), the heat transfer follows the $Nu\sim Ra^{3/2}E^2$ scaling (solid 
line). The growth of $RaE^{12/7}$ is then accompanied by a lowering of the 
heat transfer slope which levels off around the non-rotating asymptotic scaling 
when  $RaE^{12/7} \gg 1$ (dotted vertical line). Similarly to \cite{King12}, 
the data are more 
scattered in the weakly-rotating regime. This is expected since the asymptotic 
scaling $Nu_{\text{NR}}\sim Ra^{1/3}$ used to renormalise the data is not yet 
realised in 
the range of Rayleigh numbers covered by our numerical dataset.
A much better collapse of the data can nevertheless be achieved by 
normalising $Nu$ with the actual calculations of $Nu_{\text{NR}}$, the Nusselt 
numbers of 
the corresponding non-rotating simulations at the same Rayleigh numbers (panel 
\textit{b}). Since a direct computation of the non-rotating models with 
$Ra>10^9$ becomes numerically intractable, the scaling laws 
derived in \cite{Gastine15} have been used to estimate $Nu_{\text{NR}}$ for the 
cases 
with $Ra > 10^9$. This renormalisation allows to better describe the range 
$10\leq RaE^{12/7} \leq 10^4$ and to remarkably collapse all the rotating 
simulations with $Nu >2$ on one single curve. 

\begin{figure}
 \centering
 \includegraphics[width=\textwidth]{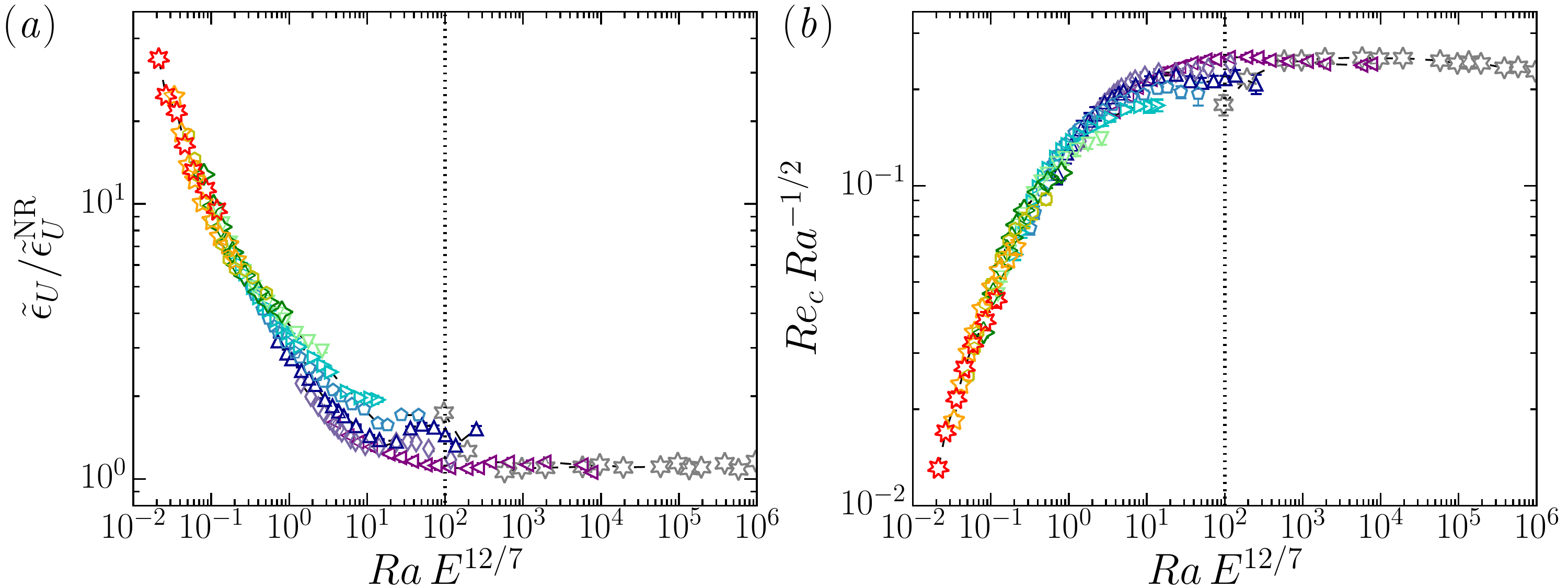}
 \caption{(\textit{a}) Dimensionless viscous dissipation rate 
$\tilde{\epsilon}_U$ normalised by the scaling for the viscous dissipation rate 
obtained in non-rotating convection $\tilde{\epsilon}_U^{\text{NR}} = 
7.08\,Re_c^{5/2}+0.25\,Re_c^3$ \citep[see][]{Gastine15} as a function of 
$RaE^{12/7}$ (Equation~\ref{eq:ratnonrot}). (\textit{b}) Reynolds number $Re_c$ 
renormalised by $Ra^{1/2}$ as a function of $RaE^{12/7}$. On both panels, the 
weakly non-linear cases with $Nu < 2$ have been excluded.  The dotted vertical 
lines correspond to the Rayleigh number $Ra=100\,E^{-12/7}$ beyond which 
$\tilde{\epsilon}_U$ and $Re_c$ adopt scaling behaviours close to the 
non-rotating RBC. The symbols have the same meaning as in 
figure~\ref{fig:nura}.}
 \label{fig:reytransition}
\end{figure}

We can check whether the regime transition criterion (\ref{eq:ratnonrot}) 
also applies to the scalings of the dimensionless viscous dissipation rate 
$\tilde{\epsilon}_U$ and of
the flow speed $Re_c$. In non-rotating convection, $\tilde{\epsilon}_U$ can 
also be decomposed into fluid bulk and boundary layers contributions. 
\cite{Grossmann00} 
predicted a scaling of the form $\tilde{\epsilon}_U^{\text{NR}} \sim 
a\,Re_c^{3}+b\,Re^{5/2}$, which has been found to also accurately describe
the numerical models of non-rotating convection in spherical shells 
\citep{Gastine15}. Figure~\ref{fig:reytransition}(\textit{a}) thus shows 
$\tilde{\epsilon}_U$ renormalised by $\tilde{\epsilon}_U^{\text{NR}}$ found by 
\cite{Gastine15} as a function of 
the transition parameter. In non-rotating convection, the dependence of the 
Reynolds number 
upon $Ra$ cannot be simply reduced to a simple power law but rather exhibits 
continuous changes in the local slope exponent, analogous to the Nusselt number 
scaling. We thus follow the same procedure as for the Nusselt number and 
renormalise $Re_c$ by the expected asymptotic scaling from the 
\cite{Grossmann00} theory in the limit of large $Ra$, i.e. 
$Re_c^{\text{NR}}\sim 
Ra^{1/2}$. Figure~\ref{fig:reytransition}(\textit{b}) shows the calculations of 
$Re_cRa^{-1/2}$ versus $RaE^{12/7}$. Except for the intermediate parameter 
range $1 \leq RaE^{12/7} \leq 10^2$, the numerical data for both 
$\tilde{\epsilon}_U$ and $Re_c$ are well collapsed on a single curve. We note 
that the scatter of the data in panel (\textit{b}) could be possibly further 
reduced by rather considering the actual values of $Re_c^{\text{NR}}$ calculated 
in the non-rotating models. Beyond $RaE^{12/7}=\mathcal{O}(10^2)$ (dotted 
vertical lines), both $\tilde{\epsilon}_U$ and $Re_c$ 
adopt scaling behaviours close to the classical non-rotating RBC.

\subsubsection{Flow length scale and interior temperature gradient}

\begin{figure}
 \centering
 \includegraphics[width=\textwidth]{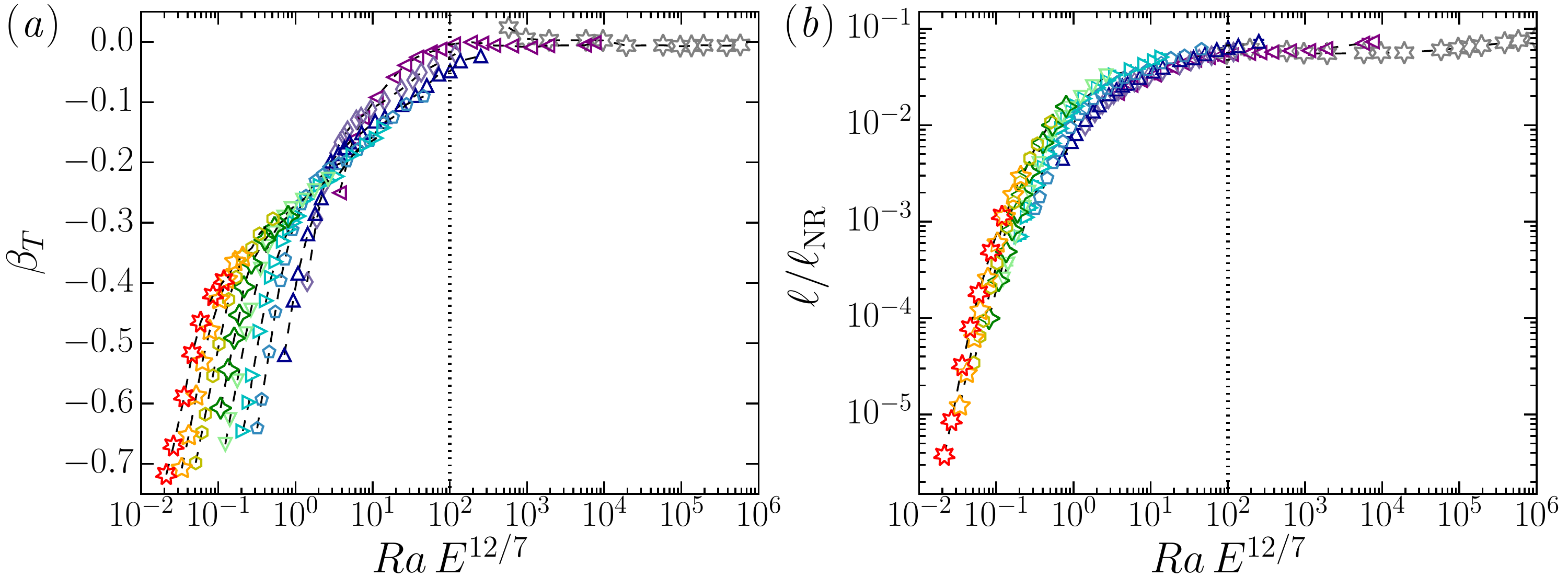}
 \caption{(\textit{a}) Temperature gradient at mid-shell $\tempgrad$  as a 
function of $RaE^{12/7}$ (Equation~\ref{eq:ratnonrot}). (\textit{b}) Convective 
flow lengthscale $\ell$ 
normalised by the scaling for the plume spacing obtained in non-rotating 
convection $\ell_{\text{NR}}$ as a function of $RaE^{12/7}$ \citep{Gastine15}. 
On both 
panels, the weakly non-linear cases with $Nu < 2$ have been excluded. The dotted 
vertical lines correspond to the Rayleigh number $Ra=100\,E^{-12/7}$ beyond 
which $\tempgrad$ and $\ell$ adopt scaling behaviours close to the non-rotating 
RBC. The symbols have the same meaning as in figure~\ref{fig:nura}.}
 \label{fig:lbetatransition}
\end{figure}

\noindent We now turn to discussing the variations of the bulk temperature 
gradient $\tempgrad$ and average flow lengthscale $\ell$ across the regime 
transition from rotating to 
non-rotating convection. Figure~\ref{fig:lbetatransition}(\textit{a}) shows 
$\tempgrad$ as a function of $RaE^{12/7}$. In contrast to the previous 
quantities, we observe a significant scatter of the data in the 
rapidly-rotating regime. As discussed above and similarly to the findings by 
\cite{King13}, $\tempgrad$ depends on $Ra$ and $E$ in a complex fashion that is 
not captured by the 
transition parameter (\ref{eq:ratnonrot}). In the weakly-rotating limit 
(i.e. $RaE^{12/7} > 10^2$), the bulk of the fluid becomes thermally 
well-mixed leading to $\tempgrad \simeq 0$. 

The study of the flow length scale necessitates the 
determination of an asymptotic scaling for $\ell$ in the non-rotating regime 
of convection. In classical RBC experiments with rigid sidewalls, the bulk flow 
is dominated by a large scale circulation (LSC) pattern. The average flow 
length scale can therefore be approximated by the vertical size of the 
container $L$. In spherical shells however, the absence of sidewalls precludes 
the formation of  such a global and single cell LSC structure. The typical flow 
length scale can instead be estimated by the average inter-plume distance 
\citep{King13}. In our previous study \citep{Gastine15}, we established 
the following scaling for the plume spacing
\begin{equation}
\ell_{\text{NR}}/L \sim Ra^{1/2}Nu^{-5/2}\,.
\label{eq:lengthscalesRBC}
\end{equation}
Figure~\ref{fig:lbetatransition}(\textit{b}) shows the calculations of $\ell$ 
renormalised by $\ell_{\text{NR}}$ versus the transition parameter $RaE^{12/7}$. 
Though 
the collapse of the data, especially when $RaE^{12/7} <1$, is not as good 
as those obtained for $Nu$, $\tilde{\epsilon}_U$ and $Re_c$, the transition 
parameter (\ref{eq:ratnonrot}) provides an effective way to capture the 
transition of the mean flow length scale from rotating 
(equation~\ref{eq:lperpcia}) to non-rotating (equation~\ref{eq:lengthscalesRBC}) 
convection. For $RaE^{12/7}>100$ (dotted vertical line), $\ell$ approaches the 
non-rotating scaling $\ell_{\text{NR}}$ (\ref{eq:lengthscalesRBC}).

\subsubsection{Thermal and viscous boundary layers}

\begin{figure}
 \centering
 \includegraphics[width=\textwidth]{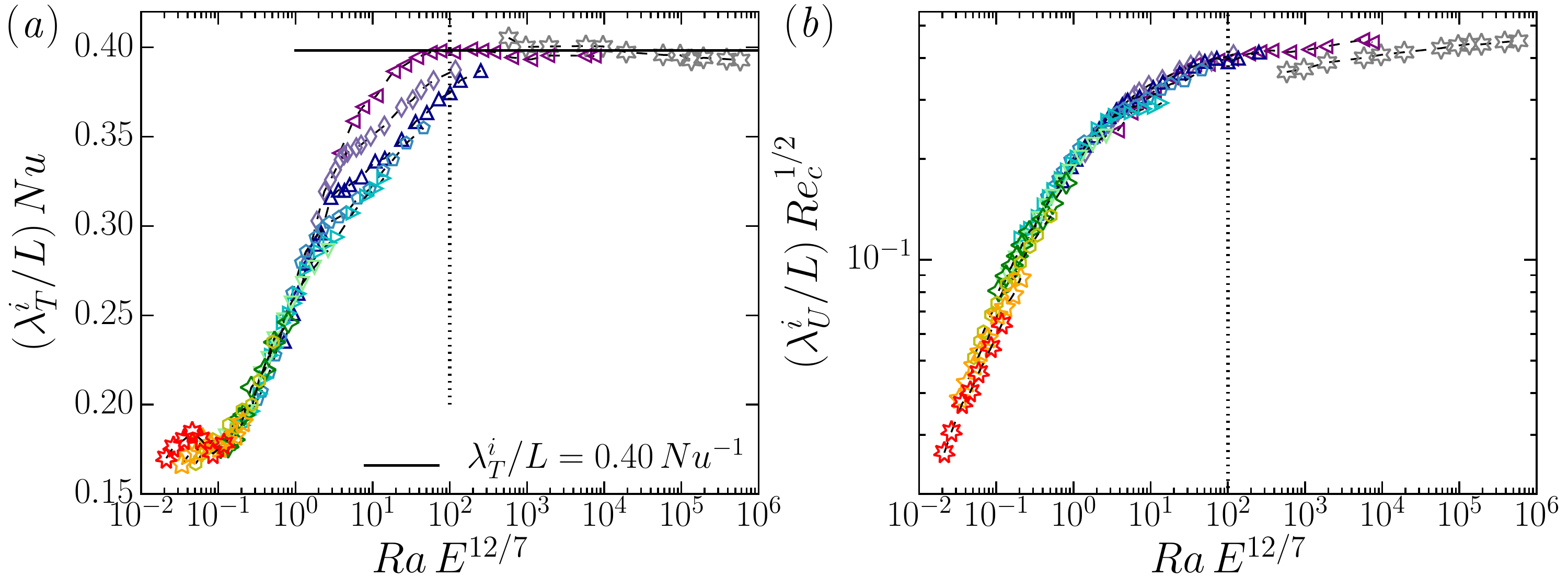}
 \caption{(\textit{a})Thermal boundary layer thickness at the inner
boundary ($\tbli$) normalised by $Nu^{-1}$ as a function of $RaE^{12/7}$ 
(Equation~\ref{eq:ratnonrot}). The solid black line 
corresponds to the theoretical scaling obtained in \cite{Gastine15} for 
non-rotating convection (\textit{b}) Viscous boundary layer thickness at the 
inner boundary ($\ubli$) normalised by $Re_c^{-1/2}$ as a function of 
$RaE^{12/7}$. The dotted vertical lines correspond to the Rayleigh number 
$Ra=100\,E^{-12/7}$ beyond which $\tbli$ and $\ubli$ adopt scaling behaviours 
close to the non-rotating RBC. The symbols have the same meaning as in 
figure~\ref{fig:nura}.}
 \label{fig:bLayersTransition}
\end{figure}

\noindent In figure~\ref{fig:bLayersTransition} we examine calculations of the 
thermal 
and viscous boundary layer thicknesses at the inner boundary renormalised by 
their asymptotic scalings in non-rotating convection plotted versus 
$RaE^{12/7}$. Once again, the renormalisation of the axes allows us to 
collapse the boundary layer thicknesses for all $E$ on one single curve.
The transition parameter $RaE^{12/7}$ therefore provides an accurate way of 
distinguishing the variations of most of the physical quantities considered in 
this study across the regime change from rotating to non-rotating convection. 

Beyond $RaE^{12/7} =\mathcal{O}(10^2)$, all the diagnostic 
quantities for the rotating  models becomes statistically indiscernible from the 
non-rotating behaviour. This value can therefore be adopted to define the upper 
bound of the transitional regime.

\section{Conclusion and outlooks}
\label{sec:conclu}

\begin{figure}
 \centering
 \includegraphics[width=8.5cm]{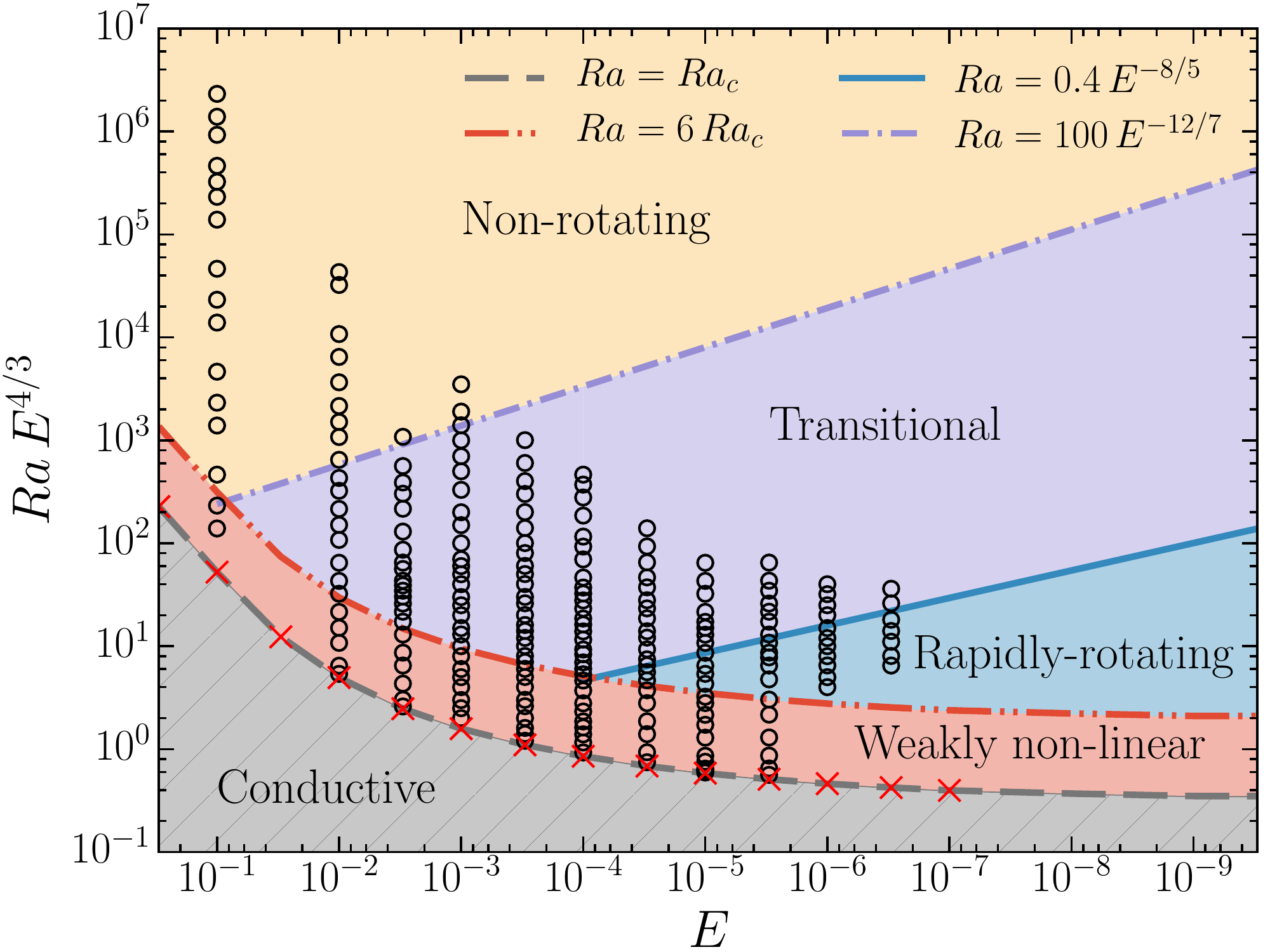}
 \caption{Regime diagram that summarises the regime transitions. The black 
circles correspond to the numerical simulations carried out in this study, 
while 
the red crosses mark the critical Rayleigh numbers $Ra_c$ given in 
table~\ref{tab:rac}.}
 \label{fig:regime}
\end{figure}

We have studied rotating convection in spherical shells by means of
three-dimensional direct numerical simulations. We have constructed a dataset of 
more than 200 models that cover a broad parameter range with Ekman numbers 
spanning $3\times 10^{-7} \leq E \leq 10^{-1}$, Rayleigh numbers within the 
range $10^3 < Ra < 2\times 10^{10}$. The Prandtl number $Pr$ is one and the 
radius ratio $r_i/r_o$ is 0.6 in all cases.
Figure~\ref{fig:regime} shows the parameter space covered by our 
numerical dataset as well as the regime boundaries of rotating convection 
derived in this work. We have studied seven different diagnostic quantities and 
investigated their scaling properties across the regime changes from onset of 
rotating convection to weakly-rotating convection. These quantities encompass 
the Nusselt number $Nu$, the Reynolds number $Re_c$, the dimensionless viscous 
dissipation rate $\tilde{\epsilon}_U$, the interior temperature gradient 
$\tempgrad$, the average flow length scale $\ell$ and the thermal and viscous 
boundary layer thicknesses $\lambda_T$ and $\lambda_U$. The scaling behaviours 
of these seven quantities of interest are summarised in 
table~\ref{tab:results}.

\begin{sidewaystable}
\centering
\vspace{14cm}
 \begin{tabular}{ccccccc}
 \hline \\
 \multirow{3}{*}{Regime}          & \multicolumn{2}{c}{Weakly non-linear}  & 
\multicolumn{2}{c}{Rapidly-rotating}
& \multicolumn{2}{c}{Non-rotating}  \\
	& \multicolumn{2}{c}{$ Ra_c <Ra < 6\,Ra_c$}       & 
\multicolumn{2}{c}{$6\,Ra_c <Ra < 0.4\,E^{-8/5}$} &  
\multicolumn{2}{c}{$Ra>100\,E^{-12/7}$} \\
& \multicolumn{2}{c}{VAC} &\multicolumn{2}{c}{bulk CIA+Ekman friction} 
& \multicolumn{2}{c}{IA} \\	
  \hline \\
	& \multirow{2}{*}{Scaling} & \multirow{2}{*}{Reference} & 
\multirow{2}{*}{Scaling} & \multirow{2}{*}{Reference} & 
\multirow{2}{*}{Scaling} & Reference\\ 
   & & & & & & \citep{Gastine15} \\ \\
  \multirow{2}{*}{ $Nu$ } & \multirow{2}{*}{$a\lp\dfrac{Ra}{Ra_c}-1\rp+1$ 
} & Equation~(\ref{eq:NuWnl})  &  
\multirow{2}{*}{$\dfrac{Ra^{3/2}E^2}{Pr^{1/2}}$} & 
Equation~(\ref{eq:asymptotic}) &  \multirow{2}{*}{$Ra^{0.27}$ to $Ra^{1/3}$} & 
\multirow{2}{*}{Figures~\ref{fig:nura} and \ref{fig:nutransition}} \\ 
  & &   Figure~\ref{fig:wNL}(\textit{a})  & & 
Figures~\ref{fig:nuraek}-\ref{fig:NuExpLoc} & &  \\ \\
   \multirow{2}{*}{$\tilde{\epsilon}_U$} & 
\multirow{2}{*}{$\dfrac{Re_c^2}{E^{2/3}}$} & 
\multirow{2}{*}{Equation~(\ref{eq:epsvac})} &
\multirow{2}{*}{$a\dfrac{Re_c^{2}}{E^{1/2}}+b\dfrac{Re_c^{5/2}}{E^
{1/2}}$} & Equation~(\ref{eq:dissip_theory}) &
\multirow{2}{*}{$a\,Re_c^{5/2}+b\,Re_c^3$} & 
\multirow{2}{*}{Figure~\ref{fig:reytransition}(\textit{a})} \\
  & & & &  Figure~\ref{fig:dissip} & & 
\\ \\
   \multirow{2}{*}{$Re_c$}  & 
\multirow{2}{*}{$\lp\dfrac{Ra_{\mathcal{Q}}E^{2/3}}{Pr^2}\rp^{1/2}$}    
& Equation~(\ref{eq:revac})  & 
\multirow{2}{*}{$\lp\dfrac{Ra_{\mathcal{Q}}E^{1/2}}{Pr^2}\rp^{1/2}$ to 
$\lp\dfrac{Ra_{\mathcal{Q}}E^{1/2}}{Pr^2}\rp^{2/5}$} &
Equation~(\ref{eq:reybounds})
&  \multirow{2}{*}{$Ra^{0.46}$ to $Ra^{1/2}$} & 
\multirow{2}{*}{Figure~\ref{fig:reytransition}(\textit{b})} \\
  & & Figure~\ref{fig:wNL}(\textit{b}) & & Figure~\ref{fig:rey_scaling} && 
  \\ \\
   \multirow{2}{*}{$-\tempgrad$} & \multirow{2}{*}{$\dfrac{4\eta}{(1+\eta)^2}$} 
& 
\multirow{2}{*}{Equation~(\ref{eq:dtcdr})} & 
\multirow{2}{*}{$\dfrac{Nu}{Re_c^{3/2}E^{1/2}}$?} & Equation~(\ref{eq:betacia}) 
& \multirow{2}{*}{0} & \multirow{2}{*}{Figures~\ref{fig:beta} and 
\ref{fig:lbetatransition}(\textit{a})} \\ 
  & & & & Figure~\ref{fig:beta} & & \\ \\
  \multirow{2}{*}{$\ell/L$} & \multirow{2}{*}{$E^{1/3}$} & 
Equation~(\ref{eq:lperpvac}) & \multirow{2}{*}{$Re_c^{1/2}\,E^{1/2}$} & 
Equation~(\ref{eq:lperpcia}) & \multirow{2}{*}{$\dfrac{Ra^{1/2}}{Nu^{5/2}}$} &
Equation~(\ref{eq:lengthscalesRBC}) \\ 
   & & Figure~\ref{fig:lengthscalesOnset} & & Figure~\ref{fig:lengthscalesNL} & 
& Figure~\ref{fig:lbetatransition}(\textit{b}) \\ \\
   \multirow{2}{*}{$\lambda_T/L$} & \multirow{2}{*}{Undefined} & 
\multirow{2}{*}{} &  \multirow{2}{*}{$0.2\,Nu^{-1+o(E)}$?} & 
Equation~(\ref{eq:tblrot})& \multirow{2}{*}{$0.5\,Nu^{-1}$} & 
Equation~(\ref{eq:tblrot}) \\
  & & & & Figure~\ref{fig:tempLayers} & &  Figures~\ref{fig:tempLayers} and 
\ref{fig:bLayersTransition}(\textit{a})\\ \\
   \multirow{2}{*}{$\lambda_U/L$} & \multirow{2}{*}{$ E^{1/2}$} & 
Equation~(\ref{eq:ekmanLayers}) & \multirow{2}{*}{$E^{1/2}$} & 
Equation~(\ref{eq:ekmanLayers})  & \multirow{2}{*}{$Re^{-1/2}$} & 
Equation~(\ref{eq:RBCbl})\\
  &&  Figure~\ref{fig:ekLayers} && Figure~\ref{fig:ekLayers} &  & 
Figures~\ref{fig:ekLayers} and \ref{fig:bLayersTransition}(\textit{b}) \\
  \hline \\
 \end{tabular}
 \caption{Summary table of the scaling properties of the quantities of interest 
in the different regimes of rotating convection derived in this work. The 
acronyms employed on the third line stand for the leading-order force balance: 
\emph{Visco-Archimedean-Coriolis} (VAC), \emph{Coriolis-Inertia-Archimedean} 
(CIA) and \emph{Inertia-Archimedean} (IA). The notation $o(E)$ 
designates a weak-dependence on the Ekman number, while the question marks 
highlight possible uncertainties on the scaling laws. See text for the 
successive derivation of scalings.}
 \label{tab:results}
\end{sidewaystable}

For $Ra$ just above critical (i.e. $Ra \gtrsim Ra_c$), our numerical 
simulations have confirmed the scaling relation of the form $Nu-1\sim Ra/Ra_c 
-1$, predicted by the perturbation analysis by \cite{Busse86} and 
\cite{Gillet06}. In this weakly non-linear regime of rotating convection, the 
convective flow is laminar and takes the form of a drifting thermal Rossby wave 
with a typical size of $\ell /L \sim E^{1/3}$. The triple force balance 
between viscosity, Coriolis force and buoyancy (the so-called VAC balance), 
suggests the scaling for \modif{the flow velocity $Re_c \sim 
Ra_{\mathcal{Q}}^{1/2}E^{1/3}$}, which is in good agreement with the numerical 
data. 

In the limit of small Ekman numbers, an increase of the supercriticality is 
accompanied by a gradual transition to a turbulent quasi-geostrophic regime 
(when  $Ra \geq 6\,Ra_c$ and $Nu >2$). The heat transport scaling is then 
expected to become independent of the thermal and viscous diffusivities and 
to depend only on the supercriticality $Ra/Ra_c$, \modif{yielding $Nu \sim 
Ra^{3/2}E^2$} \citep[see][]{Gillet06,Julien12,Stellmach14}. A small 
subset of our numerical data has been found to approach this asymptotic scaling 
in a narrow fraction of the parameter space delimited by $6\,Ra_c \leq Ra \leq 
0.4\,E^{-8/5}$. In good agreement with the theory by \cite{Julien12}, we have 
observed a breakdown of the $Ra^{3/2}$ scaling law when the thermal boundary 
layer is not dominated by rotational effects any longer, i.e. when 
$RaE^{8/5}=\mathcal{O}(1)$. 

Thanks to a decomposition of the dimensionless viscous dissipation rate 
$\tilde{\epsilon}_U$ into bulk and boundary layer contributions, we derived a 
theoretical scaling of 
the form $\tilde{\epsilon}_U\,E^{1/2} \sim (a\,Re^{5/2}+b\,Re^2)$, which
accurately describes the numerical data when adjusting the two fit parameters 
$a$ and $b$. A sizeable fraction of the 
dissipation occurs in the fluid bulk, which is dominated by a triple force 
balance between Coriolis, Inertia and buoyancy \citep[the so-called inertial 
theory of rotating convection or CIA balance, e.g.][]{Aubert01}. The remaining 
fraction of the 
dissipation can be attributed to the viscous friction in the Ekman boundary 
layers. In contrast to the existing scalings that neglect the boundary layer 
dissipation \citep[][]{Aubert01,Gillet06,KingBuffett13,Barker14}, this 
scaling law accurately captures the scaling behaviour of the Reynolds number.
Our scaling further predicts that the bulk dissipation will dominate when 
$Re_c > 5000$. Beyond this value, the inertial scaling for rotating convection  
$Re_c\sim Ra_{\mathcal{Q}}^{2/5}E^{1/5}$ and $\ell\sim \sqrt{Re_c 
E}\,L$ (Rhines scaling) should be gradually approached.

Beyond $Ra = 0.4\,E^{-8/5}$, the rotational constraint on the convective flow 
gradually decreases until the dynamics resembles non-rotating convection.
Within this parameter range, that we designate as the 
\emph{transitional regime}, we have observed continuous changes of the flow 
properties. The heat transfer scaling exponents show continuous variations that 
depend on $Ra$ and $E$ rather than simple polynomial laws. This makes the 
derivation of asymptotic scalings inherent to this physical regime extremely 
difficult. From the intersection between the steep $Nu\sim Ra^{3/2}E^{2}$ 
scaling for rapidly-rotating convection and the shallow exponent for 
non-rotating convection $Nu\sim Ra^{1/3}$, we have defined a transition 
Rayleigh number $Ra_T\sim E^{-12/7}$, which indeed allows to separate the 
rotation-influenced solutions from those resembling non-rotating convection.
Beyond $RaE^{12/7} \sim \mathcal{O}(10^2)$, all the diagnostic quantities 
studied here follow the scalings for non-rotating convection
\citep{Gastine15}. This defines the upper bound of the transitional regime 
displayed in the regime diagram (figure~\ref{fig:regime}).

Our systematic study of rotating convection in spherical shells 
revealed interesting differences to the local simulations carried out in 
cartesian coordinates \citep{King12,Stellmach14,Cheng15}. In the limit of small 
Ekman numbers ($E=\mathcal{O}(10^{-7}$), these studies have obtained 
much steeper heat transfer scaling laws (from $Nu \sim Ra^3E^4$ to $Nu \sim
Ra^{3.6}E^{4.8}$) than our findings. This has been attributed to an active 
role of the Ekman boundary layers, which supposedly promotes very efficient 
heat transfer, much steeper than the diffusivity-free asymptotic scaling 
\citep{Julien16}. In spherical geometry, the Ekman pumping might play a 
significant role and may indeed
affect the heat transport in the polar regions where gravity is aligned with 
the rotation axis. As shown by \cite{Yadav16}, the heat transport in spherical 
shells with rigid mechanical boundaries is however dominated by the 
equatorial regions, where the influence of Ekman pumping on the heat transfer 
might be negligible. A regional analysis of the heat transport in 
spherical shell models as well as the computation of new cartesian models 
in which gravity is orthogonal to the rotation axis could possibly help to 
ascertain this scenario. 

Dynamo processes and convection in planetary and stellar interiors 
frequently operate at Prandtl numbers much smaller than unity. The parameter 
study presented here has been focused on the peculiar case of $Pr=1$. 
It would be interesting to complement our study with simulations with $Pr = 
\mathcal{O}(10^{-2}-10^{-1})$ to verify the theoretical $Pr$-scalings derived 
here (see table~\ref{tab:results}). Recent studies by \cite{King13PNAS} and 
\cite{Guervilly16} indeed reveal interesting new physical phenomena inherent 
to small Prandtl number fluids that could possibly impact the scaling 
properties.

\begin{acknowledgments}
The authors wish to thank Jonathan Aurnou and Keith Julien for fruitful 
scientific discussions and constructive criticisms at different stages of this 
research work. All the computations have been carried out on the GWDG computer 
facilities in G\"ottingen, on the IBM iDataPlex HPC System Hydra at the MPG 
Rechenzentrum Garching and on the S-CAPAD platform at IPGP. TG has been partly 
supported by the Special Priority Program 1488 (PlanetMag, 
\url{www.planetmag.de}) of the German Science Foundation. This is IPGP 
contribution \modif{3780}.
\end{acknowledgments}

\appendix

\section{Table of results}

\setcounter{table}{0}
\renewcommand{\thetable}{A\arabic{table}}
\LTcapwidth=\textwidth

\begin{center}
\begin{longtable}{ccccccccccc}
\hline
\multicolumn{10}{c}{} \\
 $\#$ & $Ra$ & $Nu$ & $Re_c$ & $\beta_T$ & $\ell$ & $\lambda_T^{i}/L$ & 
$\lambda_U^{i}/L$ & $\chi_{T}$ & $\chi_{U}$ & $N_r\times \ell_{max}$ \\
\hline
\multicolumn{11}{c}{} \\
\multicolumn{11}{c}{$E=10^{-1}$} \\
1 & $3\times10^{3}$ & 1.71 & 0.0 & -0.18 & 0.704 & -- & -- & 1.000 & 1.000 & $61 \times 64$ \\
2 & $5\times10^{3}$ & 2.04 & 12.7 & -0.07 & 0.665 & -- & -- & 1.002 & 1.003 & $61 \times 133$ \\
3 & $1\times10^{4}$ & 2.47 & 21.6 & 0.03 & 0.638 & -- & -- & 1.000 & 1.000 & $61 \times 133$ \\
4 & $3\times10^{4}$ & 3.39 & 42.6 & 0.02 & 0.481 & $1.2\times10^{-1}$ & $5.5\times10^{-2}$ & 1.001 & 1.002 & $61 \times 133$ \\
5 & $5\times10^{4}$ & 3.88 & 55.1 & 0.01 & 0.420 & $1.0\times10^{-1}$ & $4.9\times10^{-2}$ & 1.000 & 1.000 & $61 \times 133$ \\
6 & $1\times10^{5}$ & 4.71 & 79.1 & 0.00 & 0.361 & $8.5\times10^{-2}$ & $4.3\times10^{-2}$ & 1.002 & 1.003 & $61 \times 133$ \\
7 & $3\times10^{5}$ & 6.38 & 137.8 & 0.00 & 0.298 & $6.3\times10^{-2}$ & $3.3\times10^{-2}$ & 1.001 & 1.001 & $61 \times 133$ \\
8 & $5\times10^{5}$ & 7.37 & 176.8 & 0.00 & 0.270 & $5.4\times10^{-2}$ & $3.0\times10^{-2}$ & 1.000 & 1.000 & $61 \times 133$ \\
9 & $1\times10^{6}$ & 8.90 & 250.5 & -0.01 & 0.239 & $4.5\times10^{-2}$ & $2.6\times10^{-2}$ & 1.000 & 1.000 & $81 \times 170$ \\
10 & $3\times10^{6}$ & 12.08 & 426.2 & -0.01 & 0.206 & $3.3\times10^{-2}$ & $2.0\times10^{-2}$ & 1.000 & 1.000 & $97 \times 213$ \\
11 & $5\times10^{6}$ & 13.97 & 539.5 & -0.01 & 0.192 & $2.8\times10^{-2}$ & $1.8\times10^{-2}$ & 1.000 & 1.001 & $97 \times 256$ \\
12 & $7\times10^{6}$ & 15.45 & 646.7 & -0.01 & 0.184 & $2.5\times10^{-2}$ & $1.7\times10^{-2}$ & 1.000 & 1.003 & $97 \times 213$ \\
13 & $1\times10^{7}$ & 17.12 & 763.0 & -0.01 & 0.174 & $2.3\times10^{-2}$ & $1.6\times10^{-2}$ & 1.001 & 1.006 & $97 \times 266$ \\
14 & $2\times10^{7}$ & 20.85 & 1045.5 & -0.01 & 0.161 & $1.9\times10^{-2}$ & $1.3\times10^{-2}$ & 1.000 & 1.005 & $129 \times 426$ \\
15 & $3\times10^{7}$ & 23.44 & 1288.5 & -0.01 & 0.155 & $1.7\times10^{-2}$ & $1.2\times10^{-2}$ & 1.000 & 1.007 & $129 \times 426$ \\
16 & $5\times10^{7}$ & 27.41 & 1611.5 & -0.03 & 0.144 & $1.4\times10^{-2}$ & $1.1\times10^{-2}$ & 1.001 & 1.025 & $129 \times 426$ \\
\multicolumn{11}{c}{} \\
\multicolumn{11}{c}{$E=10^{-2}$} \\
17 & $2.5\times10^{3}$ & 1.062 & 2.3 & -0.87 & 0.404 & -- & -- & 1.000 & 1.000 & $61 \times 85$ \\
18 & $3\times10^{3}$ & 1.19 & 4.5 & -0.74 & 0.400 & -- & -- & 1.000 & 1.000 & $61 \times 85$ \\
19 & $5\times10^{3}$ & 1.47 & 8.0 & -0.54 & 0.298 & -- & -- & 1.000 & 1.000 & $61 \times 85$ \\
20 & $7\times10^{3}$ & 1.69 & 12.1 & -0.40 & 0.376 & -- & -- & 1.000 & 1.000 & $61 \times 85$ \\
21 & $1\times10^{4}$ & 2.03 & 17.1 & -0.25 & 0.366 & -- & -- & 0.999 & 0.998 & $61 \times 85$ \\
22 & $1.5\times10^{4}$ & 2.38 & 23.4 & -0.16 & 0.362 & -- & -- & 1.000 & 1.001 & $61 \times 85$ \\
23 & $2\times10^{4}$ & 2.64 & 28.6 & -0.13 & 0.362 & -- & -- & 0.999 & 0.999 & $61 \times 85$ \\
24 & $3\times10^{4}$ & 3.05 & 37.1 & -0.09 & 0.353 & $1.2\times10^{-1}$ & $5.1\times10^{-2}$ & 1.001 & 1.001 & $61 \times 85$ \\
25 & $5\times10^{4}$ & 3.66 & 51.1 & -0.06 & 0.348 & $1.1\times10^{-1}$ & $4.8\times10^{-2}$ & 1.000 & 1.000 & $61 \times 85$ \\
26 & $7\times10^{4}$ & 4.09 & 62.3 & -0.04 & 0.342 & $9.5\times10^{-2}$ & $4.5\times10^{-2}$ & 1.000 & 1.001 & $61 \times 85$ \\
27 & $1\times10^{5}$ & 4.59 & 76.5 & -0.03 & 0.331 & $8.6\times10^{-2}$ & $4.3\times10^{-2}$ & 0.998 & 0.997 & $61 \times 133$ \\
28 & $1.5\times10^{5}$ & 5.18 & 95.7 & -0.02 & 0.319 & $7.7\times10^{-2}$ & $3.8\times10^{-2}$ & 1.000 & 1.000 & $61 \times 133$ \\
29 & $2\times10^{5}$ & 5.61 & 111.0 & -0.01 & 0.310 & $7.1\times10^{-2}$ & $3.6\times10^{-2}$ & 0.999 & 0.998 & $61 \times 133$ \\
30 & $3\times10^{5}$ & 6.36 & 137.9 & -0.00 & 0.291 & $6.3\times10^{-2}$ & $3.3\times10^{-2}$ & 1.001 & 1.001 & $49 \times 85$ \\
31 & $5\times10^{5}$ & 7.33 & 178.5 & -0.00 & 0.270 & $5.4\times10^{-2}$ & $3.0\times10^{-2}$ & 1.000 & 1.000 & $61 \times 133$ \\
32 & $7\times10^{5}$ & 8.05 & 210.4 & -0.00 & 0.257 & $4.9\times10^{-2}$ & $2.8\times10^{-2}$ & 1.000 & 1.000 & $61 \times 133$ \\
33 & $1\times10^{6}$ & 8.90 & 246.8 & -0.01 & 0.242 & $4.5\times10^{-2}$ & $2.6\times10^{-2}$ & 1.000 & 1.000 & $61 \times 133$ \\
34 & $1.7\times10^{6}$ & 10.30 & 319.4 & -0.01 & 0.226 & $3.8\times10^{-2}$ & $2.3\times10^{-2}$ & 1.000 & 1.000 & $97 \times 256$ \\
35 & $3\times10^{6}$ & 12.08 & 423.8 & -0.01 & 0.202 & $3.3\times10^{-2}$ & $2.0\times10^{-2}$ & 1.000 & 1.000 & $97 \times 256$ \\
36 & $5\times10^{6}$ & 13.98 & 537.5 & -0.01 & 0.191 & $2.8\times10^{-2}$ & $1.8\times10^{-2}$ & 1.000 & 1.001 & $97 \times 256$ \\
37 & $1.5\times10^{7}$ & 19.25 & 922.2 & -0.01 & 0.168 & $2.1\times10^{-2}$ & $1.5\times10^{-2}$ & 0.998 & 1.017 & $97 \times 341$ \\
38 & $2\times10^{7}$ & 20.82 & 1074.3 & -0.00 & 0.166 & $1.9\times10^{-2}$ & $1.3\times10^{-2}$ & 0.999 & 1.006 & $121 \times 426$ \\
\multicolumn{11}{c}{} \\
\multicolumn{11}{c}{$E=3\times 10^{-3}$} \\
39 & $6\times10^{3}$ & 1.022 & 1.9 & -0.92 & 0.380 & -- & -- & 1.000 & 1.000 & $61 \times 85$ \\
40 & $7\times10^{3}$ & 1.097 & 4.3 & -0.87 & 0.376 & -- & -- & 1.000 & 1.000 & $61 \times 85$ \\
41 & $1\times10^{4}$ & 1.26 & 8.2 & -0.77 & 0.365 & -- & -- & 1.000 & 1.000 & $61 \times 85$ \\
42 & $1.5\times10^{4}$ & 1.52 & 11.9 & -0.59 & 0.238 & -- & -- & 1.000 & 1.000 & $61 \times 85$ \\
43 & $2\times10^{4}$ & 1.70 & 15.9 & -0.54 & 0.254 & -- & -- & 1.003 & 1.007 & $61 \times 85$ \\
44 & $3\times10^{4}$ & 2.12 & 23.9 & -0.40 & 0.260 & -- & -- & 1.000 & 1.000 & $61 \times 85$ \\
45 & $4\times10^{4}$ & 2.54 & 31.5 & -0.29 & 0.261 & -- & -- & 1.000 & 1.000 & $61 \times 85$ \\
46 & $5\times10^{4}$ & 2.86 & 38.2 & -0.24 & 0.263 & -- & -- & 1.000 & 1.000 & $61 \times 85$ \\
47 & $6\times10^{4}$ & 3.13 & 44.2 & -0.20 & 0.265 & $1.0\times10^{-1}$ & $3.8\times10^{-2}$ & 1.000 & 1.000 & $61 \times 85$ \\
48 & $7\times10^{4}$ & 3.35 & 49.7 & -0.18 & 0.267 & $9.9\times10^{-2}$ & $3.8\times10^{-2}$ & 1.000 & 1.000 & $61 \times 85$ \\
49 & $8\times10^{4}$ & 3.55 & 54.9 & -0.17 & 0.269 & $9.5\times10^{-2}$ & $3.7\times10^{-2}$ & 0.999 & 0.999 & $61 \times 106$ \\
50 & $9\times10^{4}$ & 3.72 & 59.6 & -0.16 & 0.269 & $9.1\times10^{-2}$ & $3.7\times10^{-2}$ & 0.999 & 0.998 & $61 \times 106$ \\
51 & $1\times10^{5}$ & 3.87 & 64.0 & -0.15 & 0.270 & $8.8\times10^{-2}$ & $3.6\times10^{-2}$ & 1.000 & 1.000 & $73 \times 106$ \\
52 & $1.3\times10^{5}$ & 4.28 & 76.1 & -0.13 & 0.267 & $8.0\times10^{-2}$ & $3.5\times10^{-2}$ & 1.000 & 1.000 & $73 \times 106$ \\
53 & $1.5\times10^{5}$ & 4.50 & 83.2 & -0.12 & 0.265 & $7.7\times10^{-2}$ & $3.4\times10^{-2}$ & 1.000 & 1.000 & $73 \times 106$ \\
54 & $2\times10^{5}$ & 4.99 & 98.7 & -0.11 & 0.259 & $7.0\times10^{-2}$ & $3.3\times10^{-2}$ & 1.000 & 1.000 & $73 \times 106$ \\
55 & $3\times10^{5}$ & 5.76 & 123.8 & -0.10 & 0.250 & $6.2\times10^{-2}$ & $3.0\times10^{-2}$ & 1.000 & 1.000 & $81 \times 128$ \\
56 & $5\times10^{5}$ & 6.88 & 159.8 & -0.08 & 0.239 & $5.3\times10^{-2}$ & $2.8\times10^{-2}$ & 1.000 & 1.000 & $81 \times 133$ \\
57 & $7\times10^{5}$ & 7.70 & 190.0 & -0.06 & 0.234 & $4.8\times10^{-2}$ & $2.7\times10^{-2}$ & 0.999 & 0.999 & $81 \times 133$ \\
58 & $9\times10^{5}$ & 8.36 & 217.0 & -0.05 & 0.230 & $4.5\times10^{-2}$ & $2.5\times10^{-2}$ & 1.000 & 0.999 & $97 \times 170$ \\
59 & $1.3\times10^{6}$ & 9.37 & 266.6 & -0.04 & 0.222 & $4.1\times10^{-2}$ & $2.3\times10^{-2}$ & 1.000 & 1.000 & $97 \times 170$ \\
60 & $2.5\times10^{6}$ & 11.39 & 380.8 & -0.02 & 0.207 & $3.4\times10^{-2}$ & $2.0\times10^{-2}$ & 1.000 & 1.000 & $97 \times 213$ \\
\multicolumn{11}{c}{} \\
\multicolumn{11}{c}{$E=10^{-3}$} \\
61 & $2\times10^{4}$ & 1.076 & 4.8 & -0.90 & 0.224 & -- & -- & 1.000 & 1.000 & $65 \times 85$ \\
62 & $2.5\times10^{4}$ & 1.16 & 7.5 & -0.86 & 0.223 & -- & -- & 1.000 & 1.000 & $65 \times 85$ \\
63 & $3\times10^{4}$ & 1.20 & 8.5 & -0.85 & 0.194 & -- & -- & 1.000 & 1.000 & $65 \times 85$ \\
64 & $4\times10^{4}$ & 1.28 & 11.0 & -0.83 & 0.177 & -- & -- & 1.000 & 1.000 & $65 \times 85$ \\
65 & $5\times10^{4}$ & 1.44 & 15.4 & -0.74 & 0.186 & -- & -- & 1.000 & 1.000 & $65 \times 85$ \\
66 & $6\times10^{4}$ & 1.60 & 19.1 & -0.64 & 0.177 & -- & -- & 1.000 & 1.000 & $65 \times 85$ \\
67 & $8\times10^{4}$ & 1.87 & 26.4 & -0.59 & 0.190 & -- & -- & 1.001 & 1.003 & $65 \times 85$ \\
68 & $1\times10^{5}$ & 2.24 & 34.5 & -0.52 & 0.191 & -- & -- & 1.000 & 0.999 & $65 \times 85$ \\
69 & $1.3\times10^{5}$ & 2.77 & 45.7 & -0.43 & 0.190 & -- & -- & 0.999 & 0.999 & $65 \times 85$ \\
70 & $1.5\times10^{5}$ & 3.08 & 52.3 & -0.38 & 0.190 & $8.5\times10^{-2}$ & $2.7\times10^{-2}$ & 1.000 & 1.001 & $65 \times 85$ \\
71 & $2\times10^{5}$ & 3.73 & 67.3 & -0.32 & 0.190 & $7.5\times10^{-2}$ & $2.6\times10^{-2}$ & 1.000 & 0.999 & $65 \times 85$ \\
72 & $2.5\times10^{5}$ & 4.22 & 80.0 & -0.28 & 0.191 & $6.9\times10^{-2}$ & $2.6\times10^{-2}$ & 1.000 & 1.000 & $65 \times 85$ \\
73 & $3\times10^{5}$ & 4.64 & 91.6 & -0.26 & 0.191 & $6.4\times10^{-2}$ & $2.5\times10^{-2}$ & 1.000 & 1.000 & $65 \times 85$ \\
74 & $4\times10^{5}$ & 5.44 & 115.4 & -0.20 & 0.193 & $5.8\times10^{-2}$ & $2.5\times10^{-2}$ & 1.000 & 0.999 & $65 \times 85$ \\
75 & $5\times10^{5}$ & 5.97 & 133.7 & -0.19 & 0.192 & $5.4\times10^{-2}$ & $2.4\times10^{-2}$ & 1.000 & 0.999 & $65 \times 85$ \\
76 & $6\times10^{5}$ & 6.42 & 150.6 & -0.18 & 0.191 & $5.0\times10^{-2}$ & $2.3\times10^{-2}$ & 1.000 & 0.999 & $65 \times 85$ \\
77 & $7\times10^{5}$ & 6.77 & 165.9 & -0.17 & 0.190 & $4.8\times10^{-2}$ & $2.3\times10^{-2}$ & 1.000 & 1.000 & $81 \times 106$ \\
78 & $1\times10^{6}$ & 7.68 & 206.7 & -0.15 & 0.190 & $4.3\times10^{-2}$ & $2.1\times10^{-2}$ & 1.000 & 1.000 & $81 \times 133$ \\
79 & $1.5\times10^{6}$ & 8.91 & 264.2 & -0.13 & 0.187 & $3.8\times10^{-2}$ & $2.0\times10^{-2}$ & 1.001 & 1.001 & $81 \times 133$ \\
80 & $2\times10^{6}$ & 9.86 & 310.7 & -0.12 & 0.184 & $3.4\times10^{-2}$ & $1.9\times10^{-2}$ & 1.000 & 1.000 & $81 \times 133$ \\
81 & $3.3\times10^{6}$ & 11.84 & 405.1 & -0.10 & 0.180 & $2.9\times10^{-2}$ & $1.8\times10^{-2}$ & 1.000 & 1.001 & $81 \times 133$ \\
82 & $5\times10^{6}$ & 13.38 & 476.1 & -0.09 & 0.170 & $2.7\times10^{-2}$ & $1.7\times10^{-2}$ & 1.000 & 1.001 & $97 \times 213$ \\
83 & $7\times10^{6}$ & 14.95 & 556.7 & -0.07 & 0.167 & $2.4\times10^{-2}$ & $1.6\times10^{-2}$ & 1.000 & 1.003 & $97 \times 213$ \\
84 & $1\times10^{7}$ & 16.91 & 673.2 & -0.05 & 0.163 & $2.2\times10^{-2}$ & $1.5\times10^{-2}$ & 1.000 & 1.007 & $97 \times 213$ \\
85 & $1.4\times10^{7}$ & 18.60 & 807.9 & -0.05 & 0.160 & $2.0\times10^{-2}$ & $1.3\times10^{-2}$ & 0.999 & 1.007 & $97 \times 266$ \\
86 & $1.9\times10^{7}$ & 20.49 & 966.2 & -0.03 & 0.153 & $1.9\times10^{-2}$ & $1.2\times10^{-2}$ & 1.000 & 1.006 & $121 \times 341$ \\
87 & $3.5\times10^{7}$ & 24.52 & 1231.8 & -0.02 & 0.146 & $1.6\times10^{-2}$ & $1.1\times10^{-2}$ & 0.999 & 1.011 & $129 \times 426$ \\
\multicolumn{11}{c}{} \\
\multicolumn{11}{c}{$E=3\times 10^{-4}$} \\
88 & $6\times10^{4}$ & 1.022 & 3.4 & -0.94 & 0.173 & -- & -- & 1.000 & 1.000 & $65 \times 106$ \\
89 & $7\times10^{4}$ & 1.060 & 6.0 & -0.94 & 0.173 & -- & -- & 1.000 & 1.000 & $65 \times 106$ \\
90 & $8\times10^{4}$ & 1.090 & 7.8 & -0.93 & 0.174 & -- & -- & 1.000 & 1.000 & $65 \times 106$ \\
91 & $1\times10^{5}$ & 1.14 & 10.3 & -0.92 & 0.173 & -- & -- & 1.000 & 1.000 & $65 \times 106$ \\
92 & $1.3\times10^{5}$ & 1.18 & 13.2 & -0.93 & 0.171 & -- & -- & 1.000 & 1.000 & $65 \times 106$ \\
93 & $1.5\times10^{5}$ & 1.22 & 15.3 & -0.92 & 0.171 & -- & -- & 1.000 & 1.000 & $65 \times 106$ \\
94 & $2\times10^{5}$ & 1.39 & 21.2 & -0.82 & 0.140 & -- & -- & 1.000 & 1.000 & $65 \times 106$ \\
95 & $2.5\times10^{5}$ & 1.53 & 27.1 & -0.76 & 0.136 & -- & -- & 1.000 & 1.000 & $65 \times 106$ \\
96 & $3\times10^{5}$ & 1.75 & 34.9 & -0.70 & 0.135 & -- & -- & 1.000 & 1.000 & $65 \times 106$ \\
97 & $3.5\times10^{5}$ & 2.04 & 43.7 & -0.64 & 0.136 & -- & -- & 1.000 & 1.000 & $65 \times 106$ \\
98 & $4\times10^{5}$ & 2.33 & 52.2 & -0.59 & 0.136 & -- & -- & 1.000 & 1.000 & $65 \times 106$ \\
99 & $5\times10^{5}$ & 2.91 & 68.4 & -0.51 & 0.138 & -- & -- & 1.000 & 1.001 & $73 \times 128$ \\
100 & $6\times10^{5}$ & 3.49 & 83.6 & -0.45 & 0.139 & $6.5\times10^{-2}$ & $1.8\times10^{-2}$ & 1.000 & 1.000 & $73 \times 128$ \\
101 & $7\times10^{5}$ & 4.03 & 97.9 & -0.40 & 0.140 & $5.9\times10^{-2}$ & $1.7\times10^{-2}$ & 1.000 & 1.000 & $73 \times 128$ \\
102 & $8\times10^{5}$ & 4.51 & 111.2 & -0.36 & 0.140 & $5.5\times10^{-2}$ & $1.7\times10^{-2}$ & 1.000 & 1.000 & $73 \times 128$ \\
103 & $1\times10^{6}$ & 5.36 & 135.4 & -0.31 & 0.142 & $4.9\times10^{-2}$ & $1.7\times10^{-2}$ & 1.001 & 1.001 & $73 \times 128$ \\
104 & $1.3\times10^{6}$ & 6.38 & 167.0 & -0.27 & 0.142 & $4.4\times10^{-2}$ & $1.6\times10^{-2}$ & 0.999 & 0.999 & $73 \times 128$ \\
105 & $1.5\times10^{6}$ & 6.92 & 185.5 & -0.26 & 0.144 & $4.1\times10^{-2}$ & $1.6\times10^{-2}$ & 1.000 & 1.000 & $73 \times 128$ \\
106 & $2\times10^{6}$ & 8.04 & 227.9 & -0.23 & 0.144 & $3.7\times10^{-2}$ & $1.5\times10^{-2}$ & 1.000 & 0.999 & $73 \times 128$ \\
107 & $2.5\times10^{6}$ & 8.89 & 264.3 & -0.22 & 0.144 & $3.4\times10^{-2}$ & $1.5\times10^{-2}$ & 0.999 & 0.998 & $73 \times 128$ \\
108 & $3\times10^{6}$ & 9.63 & 298.4 & -0.21 & 0.144 & $3.1\times10^{-2}$ & $1.5\times10^{-2}$ & 1.000 & 0.999 & $81 \times 133$ \\
109 & $4\times10^{6}$ & 10.87 & 358.2 & -0.20 & 0.143 & $2.8\times10^{-2}$ & $1.4\times10^{-2}$ & 1.000 & 0.999 & $81 \times 133$ \\
110 & $5\times10^{6}$ & 11.97 & 412.1 & -0.20 & 0.140 & $2.6\times10^{-2}$ & $1.4\times10^{-2}$ & 1.001 & 0.999 & $81 \times 133$ \\
111 & $7\times10^{6}$ & 13.42 & 501.0 & -0.18 & 0.136 & $2.4\times10^{-2}$ & $1.3\times10^{-2}$ & 1.000 & 1.000 & $97 \times 170$ \\
112 & $1\times10^{7}$ & 15.16 & 614.0 & -0.16 & 0.132 & $2.1\times10^{-2}$ & $1.2\times10^{-2}$ & 1.000 & 1.003 & $97 \times 256$ \\
113 & $1.5\times10^{7}$ & 17.49 & 779.4 & -0.13 & 0.133 & $1.9\times10^{-2}$ & $1.1\times10^{-2}$ & 1.001 & 1.004 & $97 \times 341$ \\
114 & $2\times10^{7}$ & 19.40 & 908.0 & -0.13 & 0.133 & $1.7\times10^{-2}$ & $1.1\times10^{-2}$ & 1.001 & 1.010 & $97 \times 341$ \\
115 & $3\times10^{7}$ & 22.33 & 1076.3 & -0.10 & 0.121 & $1.6\times10^{-2}$ & $1.0\times10^{-2}$ & 1.000 & 1.009 & $121 \times 426$ \\
116 & $5\times10^{7}$ & 26.15 & 1377.0 & -0.09 & 0.123 & $1.4\times10^{-2}$ & $9.4\times10^{-3}$ & 1.001 & 1.010 & $161 \times 512$ \\
\multicolumn{11}{c}{} \\
\multicolumn{11}{c}{$E=10^{-4}$} \\
117 & $2\times10^{5}$ & 1.014 & 3.9 & -0.94 & 0.133 & -- & -- & 1.000 & 1.000 & $73 \times 85$ \\
118 & $2.5\times10^{5}$ & 1.052 & 8.1 & -0.96 & 0.134 & -- & -- & 1.000 & 1.000 & $73 \times 85$ \\
119 & $3\times10^{5}$ & 1.081 & 10.7 & -0.96 & 0.134 & -- & -- & 1.000 & 1.000 & $73 \times 85$ \\
120 & $3.5\times10^{5}$ & 1.10 & 12.7 & -0.96 & 0.134 & -- & -- & 1.000 & 1.000 & $73 \times 85$ \\
121 & $4\times10^{5}$ & 1.12 & 14.4 & -0.96 & 0.133 & -- & -- & 1.000 & 1.000 & $73 \times 85$ \\
122 & $5\times10^{5}$ & 1.18 & 19.5 & -0.92 & 0.131 & -- & -- & 1.000 & 1.000 & $73 \times 85$ \\
123 & $6\times10^{5}$ & 1.25 & 24.4 & -0.87 & 0.124 & -- & -- & 1.000 & 1.000 & $73 \times 85$ \\
124 & $8\times10^{5}$ & 1.32 & 30.7 & -0.86 & 0.124 & -- & -- & 1.000 & 1.000 & $73 \times 85$ \\
125 & $1\times10^{6}$ & 1.50 & 40.3 & -0.79 & 0.113 & -- & -- & 1.001 & 1.002 & $73 \times 85$ \\
126 & $1.15\times10^{6}$ & 1.69 & 49.5 & -0.74 & 0.109 & -- & -- & 1.002 & 1.005 & $73 \times 85$ \\
127 & $1.3\times10^{6}$ & 1.95 & 60.5 & -0.69 & 0.107 & -- & -- & 1.000 & 1.002 & $73 \times 85$ \\
128 & $1.5\times10^{6}$ & 2.29 & 74.7 & -0.65 & 0.108 & -- & -- & 1.000 & 1.000 & $81 \times 128$ \\
129 & $1.8\times10^{6}$ & 2.87 & 95.8 & -0.60 & 0.107 & -- & -- & 1.000 & 0.999 & $73 \times 85$ \\
130 & $2\times10^{6}$ & 3.19 & 109.1 & -0.55 & 0.108 & $6.1\times10^{-2}$ & $1.2\times10^{-2}$ & 1.000 & 1.000 & $81 \times 128$ \\
131 & $2.5\times10^{6}$ & 4.10 & 141.4 & -0.48 & 0.110 & $5.0\times10^{-2}$ & $1.1\times10^{-2}$ & 1.000 & 1.000 & $81 \times 128$ \\
132 & $3\times10^{6}$ & 4.99 & 172.0 & -0.43 & 0.114 & $4.3\times10^{-2}$ & $1.1\times10^{-2}$ & 1.000 & 1.000 & $81 \times 128$ \\
133 & $3.5\times10^{6}$ & 5.82 & 200.4 & -0.39 & 0.115 & $3.9\times10^{-2}$ & $1.1\times10^{-2}$ & 1.000 & 0.999 & $81 \times 128$ \\
134 & $4\times10^{6}$ & 6.58 & 226.5 & -0.36 & 0.116 & $3.5\times10^{-2}$ & $1.0\times10^{-2}$ & 1.000 & 0.999 & $81 \times 128$ \\
135 & $5\times10^{6}$ & 7.92 & 274.7 & -0.33 & 0.117 & $3.1\times10^{-2}$ & $1.0\times10^{-2}$ & 1.000 & 0.999 & $81 \times 128$ \\
136 & $6\times10^{6}$ & 9.01 & 317.7 & -0.31 & 0.118 & $2.8\times10^{-2}$ & $10.0\times10^{-3}$ & 1.000 & 0.998 & $81 \times 128$ \\
137 & $7\times10^{6}$ & 9.90 & 355.9 & -0.30 & 0.118 & $2.6\times10^{-2}$ & $9.8\times10^{-3}$ & 1.000 & 0.998 & $81 \times 128$ \\
138 & $8\times10^{6}$ & 10.68 & 391.8 & -0.29 & 0.119 & $2.5\times10^{-2}$ & $9.8\times10^{-3}$ & 1.000 & 0.998 & $81 \times 133$ \\
139 & $1\times10^{7}$ & 12.02 & 457.9 & -0.26 & 0.119 & $2.3\times10^{-2}$ & $9.6\times10^{-3}$ & 1.000 & 1.000 & $81 \times 170$ \\
140 & $1.5\times10^{7}$ & 14.41 & 590.3 & -0.24 & 0.119 & $2.0\times10^{-2}$ & $9.6\times10^{-3}$ & 1.000 & 1.001 & $97 \times 213$ \\
141 & $2\times10^{7}$ & 16.38 & 707.5 & -0.23 & 0.119 & $1.8\times10^{-2}$ & $9.3\times10^{-3}$ & 1.000 & 1.003 & $97 \times 213$ \\
142 & $2.5\times10^{7}$ & 18.16 & 816.2 & -0.22 & 0.118 & $1.6\times10^{-2}$ & $8.9\times10^{-3}$ & 1.001 & 1.008 & $97 \times 213$ \\
143 & $4\times10^{7}$ & 21.54 & 1094.5 & -0.19 & 0.111 & $1.4\times10^{-2}$ & $8.0\times10^{-3}$ & 1.000 & 1.006 & $121 \times 341$ \\
144 & $6\times10^{7}$ & 25.18 & 1371.3 & -0.17 & 0.104 & $1.3\times10^{-2}$ & $7.4\times10^{-3}$ & 1.000 & 1.010 & $161 \times 426$ \\
145 & $8\times10^{7}$ & 28.02 & 1583.3 & -0.16 & 0.100 & $1.1\times10^{-2}$ & $7.0\times10^{-3}$ & 0.999 & 1.003 & $201 \times 512$ \\
146 & $1\times10^{8}$ & 30.75 & 1785.6 & -0.14 & 0.097 & $1.1\times10^{-2}$ & $6.8\times10^{-3}$ & 1.001 & 1.011 & $201 \times 682$ \\
\multicolumn{11}{c}{} \\
\multicolumn{11}{c}{$E=3\times 10^{-5}$} \\
147 & $8\times10^{5}$ & 1.010 & 4.7 & -0.94 & 0.089 & -- & -- & 1.000 & 1.000 & $65 \times 128$ \\
148 & $1\times10^{6}$ & 1.036 & 9.6 & -0.96 & 0.093 & -- & -- & 1.000 & 1.000 & $65 \times 128$ \\
149 & $1.5\times10^{6}$ & 1.086 & 17.0 & -0.97 & 0.091 & -- & -- & 1.000 & 1.000 & $65 \times 128$ \\
150 & $2\times10^{6}$ & 1.15 & 25.9 & -0.95 & 0.091 & -- & -- & 1.000 & 1.000 & $65 \times 128$ \\
151 & $3\times10^{6}$ & 1.25 & 40.2 & -0.91 & 0.096 & -- & -- & 1.000 & 1.000 & $65 \times 128$ \\
152 & $4\times10^{6}$ & 1.38 & 54.3 & -0.85 & 0.086 & -- & -- & 1.000 & 1.000 & $65 \times 128$ \\
153 & $5\times10^{6}$ & 1.60 & 71.1 & -0.78 & 0.077 & -- & -- & 1.000 & 1.000 & $65 \times 133$ \\
154 & $6\times10^{6}$ & 1.96 & 94.8 & -0.71 & 0.075 & -- & -- & 1.000 & 1.000 & $65 \times 133$ \\
155 & $7\times10^{6}$ & 2.39 & 121.0 & -0.67 & 0.076 & -- & -- & 1.000 & 0.999 & $65 \times 133$ \\
156 & $8\times10^{6}$ & 2.80 & 146.2 & -0.63 & 0.078 & -- & -- & 1.000 & 1.000 & $65 \times 170$ \\
157 & $1\times10^{7}$ & 3.72 & 197.0 & -0.56 & 0.082 & $5.0\times10^{-2}$ & $7.0\times10^{-3}$ & 1.000 & 1.000 & $97 \times 170$ \\
158 & $1.3\times10^{7}$ & 5.12 & 270.5 & -0.48 & 0.085 & $3.8\times10^{-2}$ & $6.5\times10^{-3}$ & 1.001 & 1.001 & $65 \times 170$ \\
159 & $1.5\times10^{7}$ & 6.04 & 315.7 & -0.45 & 0.087 & $3.2\times10^{-2}$ & $6.3\times10^{-3}$ & 1.001 & 1.001 & $65 \times 170$ \\
160 & $2\times10^{7}$ & 8.30 & 423.0 & -0.38 & 0.090 & $2.6\times10^{-2}$ & $6.2\times10^{-3}$ & 1.000 & 1.000 & $97 \times 213$ \\
161 & $2.5\times10^{7}$ & 10.31 & 514.9 & -0.34 & 0.091 & $2.2\times10^{-2}$ & $6.1\times10^{-3}$ & 1.000 & 1.001 & $97 \times 256$ \\
162 & $3\times10^{7}$ & 11.95 & 596.9 & -0.32 & 0.092 & $2.0\times10^{-2}$ & $6.0\times10^{-3}$ & 1.000 & 1.001 & $97 \times 256$ \\
163 & $4\times10^{7}$ & 14.65 & 745.2 & -0.29 & 0.093 & $1.7\times10^{-2}$ & $5.9\times10^{-3}$ & 1.001 & 1.002 & $97 \times 341$ \\
164 & $5\times10^{7}$ & 16.58 & 864.1 & -0.28 & 0.094 & $1.5\times10^{-2}$ & $5.8\times10^{-3}$ & 1.003 & 1.005 & $97 \times 341$ \\
165 & $7\times10^{7}$ & 19.71 & 1086.3 & -0.26 & 0.094 & $1.4\times10^{-2}$ & $5.7\times10^{-3}$ & 1.002 & 1.004 & $129 \times 426$ \\
166 & $1\times10^{8}$ & 23.34 & 1350.8 & -0.25 & 0.095 & $1.2\times10^{-2}$ & $5.6\times10^{-3}$ & 1.003 & 1.007 & $129 \times 512$ \\
167 & $1.5\times10^{8}$ & 27.96 & 1722.8 & -0.23 & 0.096 & $1.0\times10^{-2}$ & $5.2\times10^{-3}$ & 1.004 & 1.012 & $161 \times 682$ \\
\multicolumn{11}{c}{} \\
\multicolumn{11}{c}{$E=10^{-5}$} \\
168 & $2.75\times10^{6}$ & 1.001 & 1.9 & -0.94 & 0.061 & -- & -- & 1.000 & 0.999 & $97 \times 133$ \\
169 & $2.85\times10^{6}$ & 1.004 & 4.2 & -0.94 & 0.061 & -- & -- & 1.000 & 0.999 & $97 \times 133$ \\
170 & $2.9\times10^{6}$ & 1.006 & 4.9 & -0.94 & 0.061 & -- & -- & 1.000 & 0.999 & $97 \times 133$ \\
171 & $2.95\times10^{6}$ & 1.008 & 5.6 & -0.94 & 0.061 & -- & -- & 1.000 & 1.000 & $97 \times 133$ \\
172 & $3\times10^{6}$ & 1.009 & 6.2 & -0.94 & 0.061 & -- & -- & 1.000 & 1.000 & $97 \times 133$ \\
173 & $3.5\times10^{6}$ & 1.023 & 10.3 & -0.95 & 0.061 & -- & -- & 1.000 & 1.000 & $97 \times 133$ \\
174 & $4\times10^{6}$ & 1.035 & 13.0 & -0.96 & 0.061 & -- & -- & 1.000 & 0.999 & $97 \times 170$ \\
175 & $6\times10^{6}$ & 1.089 & 25.4 & -0.99 & 0.063 & -- & -- & 1.000 & 1.000 & $97 \times 170$ \\
176 & $8\times10^{6}$ & 1.15 & 37.4 & -0.99 & 0.066 & -- & -- & 1.000 & 1.000 & $97 \times 170$ \\
177 & $1\times10^{7}$ & 1.19 & 47.9 & -0.95 & 0.066 & -- & -- & 1.000 & 1.001 & $97 \times 170$ \\
178 & $1.3\times10^{7}$ & 1.28 & 65.6 & -0.90 & 0.066 & -- & -- & 1.000 & 1.002 & $97 \times 170$ \\
179 & $1.5\times10^{7}$ & 1.33 & 75.6 & -0.87 & 0.064 & -- & -- & 1.000 & 1.000 & $97 \times 170$ \\
180 & $2\times10^{7}$ & 1.61 & 106.5 & -0.80 & 0.056 & -- & -- & 1.000 & 0.998 & $97 \times 170$ \\
181 & $2.5\times10^{7}$ & 1.94 & 141.8 & -0.74 & 0.055 & -- & -- & 1.000 & 0.999 & $97 \times 213$ \\
182 & $3\times10^{7}$ & 2.47 & 190.5 & -0.69 & 0.057 & -- & -- & 1.000 & 1.000 & $97 \times 213$ \\
183 & $4\times10^{7}$ & 3.64 & 291.0 & -0.61 & 0.061 & $4.8\times10^{-2}$ & $4.6\times10^{-3}$ & 1.000 & 1.000 & $97 \times 213$ \\
184 & $5\times10^{7}$ & 4.90 & 390.4 & -0.54 & 0.065 & $3.6\times10^{-2}$ & $4.4\times10^{-3}$ & 1.000 & 1.001 & $97 \times 213$ \\
185 & $6\times10^{7}$ & 6.17 & 486.3 & -0.49 & 0.067 & $2.9\times10^{-2}$ & $4.2\times10^{-3}$ & 1.002 & 1.002 & $97 \times 213$ \\
186 & $7\times10^{7}$ & 7.45 & 576.1 & -0.45 & 0.069 & $2.6\times10^{-2}$ & $4.1\times10^{-3}$ & 1.002 & 1.005 & $97 \times 213$ \\
187 & $8\times10^{7}$ & 8.88 & 675.7 & -0.41 & 0.073 & $2.2\times10^{-2}$ & $4.1\times10^{-3}$ & 1.003 & 1.004 & $97 \times 341$ \\
188 & $1\times10^{8}$ & 11.40 & 838.6 & -0.37 & 0.074 & $1.8\times10^{-2}$ & $4.0\times10^{-3}$ & 1.006 & 1.006 & $97 \times 341$ \\
189 & $1.5\times10^{8}$ & 16.21 & 1171.5 & -0.33 & 0.076 & $1.4\times10^{-2}$ & $3.6\times10^{-3}$ & 1.002 & 1.007 & $129 \times 341$ \\
190 & $2\times10^{8}$ & 20.07 & 1457.3 & -0.31 & 0.078 & $1.2\times10^{-2}$ & $3.6\times10^{-3}$ & 1.006 & 1.011 & $129 \times 426$ \\
191 & $3\times10^{8}$ & 25.66 & 1903.0 & -0.29 & 0.081 & $9.6\times10^{-3}$ & $3.5\times10^{-3}$ & 1.006 & 1.015 & $161 \times 682$ \\
\multicolumn{11}{c}{} \\
\multicolumn{11}{c}{$E=3\times 10^{-6}$} \\
192 & $1.3\times10^{7}$ & 1.007 & 7.6 & -0.94 & 0.040 & -- & -- & 1.000 & 0.999 & $97 \times 192$ \\
193 & $1.5\times10^{7}$ & 1.015 & 12.2 & -0.95 & 0.043 & -- & -- & 1.000 & 0.997 & $97 \times 192$ \\
194 & $2\times10^{7}$ & 1.043 & 22.6 & -0.97 & 0.041 & -- & -- & 1.000 & 0.999 & $97 \times 192$ \\
195 & $3\times10^{7}$ & 1.10 & 41.5 & -0.99 & 0.041 & -- & -- & 1.000 & 1.000 & $97 \times 266$ \\
196 & $5\times10^{7}$ & 1.21 & 81.0 & -0.95 & 0.044 & -- & -- & 1.000 & 1.001 & $129 \times 341$ \\
197 & $7\times10^{7}$ & 1.32 & 115.2 & -0.88 & 0.044 & -- & -- & 1.000 & 1.000 & $129 \times 341$ \\
198 & $1.1\times10^{8}$ & 1.71 & 188.7 & -0.78 & 0.040 & -- & -- & 1.000 & 1.000 & $129 \times 426$ \\
199 & $1.5\times10^{8}$ & 2.52 & 311.5 & -0.70 & 0.042 & -- & -- & 0.999 & 1.004 & $129 \times 341$ \\
200 & $1.8\times10^{8}$ & 3.26 & 415.9 & -0.65 & 0.045 & $5.3\times10^{-2}$ & $2.7\times10^{-3}$ & 1.002 & 1.003 & $129 \times 341$ \\
201 & $2\times10^{8}$ & 3.80 & 493.1 & -0.62 & 0.047 & $4.6\times10^{-2}$ & $2.7\times10^{-3}$ & 1.003 & 1.003 & $129 \times 426$ \\
202 & $2.5\times10^{8}$ & 5.30 & 678.2 & -0.56 & 0.051 & $3.4\times10^{-2}$ & $2.5\times10^{-3}$ & 1.001 & 1.003 & $129 \times 512$ \\
203 & $3\times10^{8}$ & 6.78 & 854.4 & -0.50 & 0.054 & $2.6\times10^{-2}$ & $2.4\times10^{-3}$ & 1.005 & 1.003 & $129 \times 426$ \\
204 & $4\times10^{8}$ & 9.87 & 1190.1 & -0.43 & 0.057 & $1.9\times10^{-2}$ & $2.3\times10^{-3}$ & 1.011 & 1.009 & $129 \times 512$ \\
205 & $5\times10^{8}$ & 12.96 & 1504.7 & -0.39 & 0.059 & $1.4\times10^{-2}$ & $2.2\times10^{-3}$ & 1.004 & 1.009 & $161 \times 512$ \\
206 & $6\times10^{8}$ & 16.10 & 1794.0 & -0.36 & 0.060 & $1.2\times10^{-2}$ & $2.2\times10^{-3}$ & 1.015 & 1.009 & $201 \times 682^\star$ \\
207 & $8\times10^{8}$ & 20.99 & 2209.9 & -0.34 & 0.063 & $9.5\times10^{-3}$ & $2.1\times10^{-3}$ & 1.004 & 1.013 & $201 \times 682^{\star\star\star}$ \\
208 & $1\times10^{9}$ & 25.17 & 2622.5 & -0.32 & 0.064 & $8.5\times10^{-3}$ & $2.1\times10^{-3}$ & 1.008 & 1.018 & $201 \times 768^{\star\star\star}$ \\
209 & $1.5\times10^{9}$ & 34.27 & 3520.6 & -0.29 & 0.061 & $6.9\times10^{-3}$ & $2.1\times10^{-3}$ & 0.994 & 0.982 & $401 \times 1024^{\star\star\star}$ \\
\multicolumn{11}{c}{} \\
\multicolumn{11}{c}{$E=10^{-6}$} \\
210 & $4\times10^{8}$ & 1.50 & 232.8 & -0.84 & 0.030 & -- & -- & 1.003 & 1.001 & $181 \times 512$ \\
211 & $5\times10^{8}$ & 1.79 & 308.8 & -0.78 & 0.029 & -- & -- & 1.000 & 0.998 & $181 \times 576$ \\
212 & $6.5\times10^{8}$ & 2.49 & 471.3 & -0.71 & 0.032 & -- & -- & 1.000 & 1.000 & $161 \times 576^{\star\star}$ \\
213 & $8\times10^{8}$ & 3.41 & 678.0 & -0.65 & 0.035 & $5.0\times10^{-2}$ & $1.6\times10^{-3}$ & 1.001 & 0.998 & $181 \times 640$ \\
214 & $1\times10^{9}$ & 4.77 & 952.5 & -0.59 & 0.039 & $3.6\times10^{-2}$ & $1.5\times10^{-3}$ & 0.985 & 0.986 & $193 \times 682^{\star\star}$ \\
215 & $1.2\times10^{9}$ & 6.19 & 1208.0 & -0.53 & 0.043 & $2.9\times10^{-2}$ & $1.4\times10^{-3}$ & 1.012 & 1.022 & $193 \times 853^{\star\star}$ \\
216 & $1.5\times10^{9}$ & 8.58 & 1598.0 & -0.48 & 0.046 & $2.1\times10^{-2}$ & $1.4\times10^{-3}$ & 1.010 & 1.010 & $201 \times 853$ \\
217 & $2\times10^{9}$ & 12.30 & 2156.9 & -0.43 & 0.050 & $1.4\times10^{-2}$ & $1.3\times10^{-3}$ & 1.003 & 1.009 & $241 \times 853^{\star\star}$ \\
218 & $2.5\times10^{9}$ & 16.17 & 2715.3 & -0.39 & 0.053 & $1.1\times10^{-2}$ & $1.2\times10^{-3}$ & 1.002 & 1.009 & $321 \times 1024^{\star\star}$ \\
219 & $3.2\times10^{9}$ & 20.76 & 3341.7 & -0.36 & 0.054 & $8.8\times10^{-3}$ & $1.2\times10^{-3}$ & 1.005 & 1.016 & $321 \times 1024^\star$ \\
220 & $4\times10^{9}$ & 25.65 & 4075.4 & -0.36 & 0.056 & $7.4\times10^{-3}$ & $1.2\times10^{-3}$ & 1.003 & 1.008 & $401 \times 1024^{\star\star}$ \\
\multicolumn{11}{c}{} \\
\multicolumn{11}{c}{$E=3\times 10^{-7}$} \\
221 & $3.23\times10^{9}$ & 2.42 & 744.0 & -0.72 & 0.023 & -- & -- & 1.006 & 1.008 & $241 \times 896$ \\
222 & $4\times10^{9}$ & 3.36 & 1075.2 & -0.67 & 0.026 & $5.3\times10^{-2}$ & $9.1\times10^{-4}$ & 0.996 & 0.995 & $241 \times 896^{\star\star\star}$ \\
223 & $5.5\times10^{9}$ & 5.47 & 1605.4 & -0.59 & 0.034 & $3.3\times10^{-2}$ & $8.3\times10^{-4}$ & 1.018 & 1.010 & $241 \times 896^{\star\star\star}$ \\
224 & $7\times10^{9}$ & 7.85 & 2254.7 & -0.52 & 0.038 & $2.3\times10^{-2}$ & $7.5\times10^{-4}$ & 1.012 & 1.019 & $301 \times 1024^\star$ \\
225 & $9\times10^{9}$ & 10.96 & 3047.8 & -0.47 & 0.043 & $1.6\times10^{-2}$ & $7.2\times10^{-4}$ & 1.009 & 1.024 & $401 \times 1280^{\star\star}$ \\
226 & $1.3\times10^{10}$ & 17.02 & 4317.7 & -0.42 & 0.047 & $1.0\times10^{-2}$ & $7.0\times10^{-4}$ & 1.000 & 1.014 & $513 \times 1280^{\star\star}$ \\
227 & $1.8\times10^{10}$ & 24.78 & 5933.6 & -0.40 & 0.050 & $7.2\times10^{-3}$ & 
$6.9\times10^{-4}$ & 1.006 & 1.028 & $641 \times 1365^{\star\star\star}$ \\ \\
\caption{{\footnotesize Summary table of the quantities of interest for the 
numerical models computed in this study. These simulations all have $Pr=1$ and 
$r_i/r_o=0.6$. The thermal and viscous boundary layer thicknesses are only 
given for the cases where boundary layers can be clearly identified. The 
stars in some selected cases of the last column indicate that those simulations 
have been computed with an azimuthal symmetry: one star corresponds to a 
two-fold symmetry, two stars to a four-fold symmetry and three stars to a 
eight-fold symmetry.}} \\
 \hline \\
\label{tab:runs}
\end{longtable}
\end{center}

\newpage

\section{Critical Rayleigh numbers}

\setcounter{table}{0}
\renewcommand{\thetable}{B\arabic{table}}

\begin{table}
\centering
 \begin{tabular}{ccc}
  $E$ & $Ra_c$ & $m_c$\\ \\
        $10^{-1}$ & $1.130\times 10^3$ &   5 \\
$3\times 10^{-2}$ & $1.324\times 10^3$ &   6 \\
        $10^{-2}$ & $2.298\times 10^3$ &   7 \\
$3\times 10^{-3}$ & $5.711\times 10^3$ &   8 \\
        $10^{-3}$ & $1.581\times 10^4$ &  10 \\
$3\times 10^{-4}$ & $5.482\times 10^4$ &  14 \\
        $10^{-4}$ & $1.838\times 10^5$ &  20 \\
$3\times 10^{-5}$ & $7.348\times 10^5$ &  30 \\
        $10^{-5}$ & $2.714\times 10^6$ &  43 \\
$3\times 10^{-6}$ & $1.178\times 10^7$ &  64 \\ 
        $10^{-6}$ & $4.616\times 10^7$ &  92 \\
$3\times 10^{-7}$ & $2.108\times 10^8$ & 137 \\
        $10^{-7}$ & $8.560\times 10^8$ & 198 \\
 \end{tabular}
  \caption{Critical Rayleigh numbers $Ra_c$ and critical azimuthal wavenumbers 
$m_c$ for the different Ekman numbers employed here. The exact values have been 
obtained using the open-source eigenmode solver \textit{Singe} \citep{Vidal15}, 
available at \url{https://bitbucket.org/nschaeff/singe}. }
\label{tab:rac}
\end{table}


\begin{thebibliography}{85}
\expandafter\ifx\csname natexlab\endcsname\relax\def\natexlab#1{#1}\fi

\bibitem[{Ahlers} {\em et~al.\/}(2009){Ahlers}, {Grossmann} \&
  {Lohse}]{Ahlers09}
{\sc {Ahlers}, G., {Grossmann}, S. \& {Lohse}, D.} 2009 {Heat transfer and
  large scale dynamics in turbulent Rayleigh-B{\'e}nard convection}. {\em
  Reviews of Modern Physics\/} {\bf 81}, 503--537.

\bibitem[{Amati} {\em et~al.\/}(2005){Amati}, {Koal}, {Massaioli},
  {Sreenivasan} \& {Verzicco}]{Amati05}
{\sc {Amati}, G., {Koal}, K., {Massaioli}, F., {Sreenivasan}, K.~R. \&
  {Verzicco}, R.} 2005 {Turbulent thermal convection at high Rayleigh numbers
  for a Boussinesq fluid of constant Prandtl number}. {\em Physics of Fluids\/}
  {\bf 17}~(12), 121701.

\bibitem[{Aubert} {\em et~al.\/}(2001){Aubert}, {Brito}, {Nataf}, {Cardin} \&
  {Masson}]{Aubert01}
{\sc {Aubert}, J., {Brito}, D., {Nataf}, H.-C., {Cardin}, P. \& {Masson},
  J.-P.} 2001 {A systematic experimental study of rapidly rotating spherical
  convection in water and liquid gallium}. {\em Physics of the Earth and
  Planetary Interiors\/} {\bf 128}, 51--74.

\bibitem[{Aubert} {\em et~al.\/}(2003){Aubert}, {Gillet} \& {Cardin}]{Aubert03}
{\sc {Aubert}, J., {Gillet}, N. \& {Cardin}, P.} 2003 {Quasigeostrophic models
  of convection in rotating spherical shells}. {\em Geochemistry, Geophysics,
  Geosystems\/} {\bf 4}, 1.

\bibitem[{Aurnou}(2007)]{Aurnou07}
{\sc {Aurnou}, J.~M.} 2007 {Planetary core dynamics and convective heat
  transfer scaling}. {\em Geophysical and Astrophysical Fluid Dynamics\/} {\bf
  101}, 327--345.

\bibitem[{Aurnou} {\em et~al.\/}(2015){Aurnou}, {Calkins}, {Cheng}, {Julien},
  {King}, {Nieves}, {Soderlund} \& {Stellmach}]{Aurnou15}
{\sc {Aurnou}, J.~M., {Calkins}, M.~A., {Cheng}, J.~S., {Julien}, K., {King},
  E.~M., {Nieves}, D., {Soderlund}, K.~M. \& {Stellmach}, S.} 2015 {Rotating
  convective turbulence in Earth and planetary cores}. {\em Physics of the
  Earth and Planetary Interiors\/} {\bf 246}, 52--71.

\bibitem[{Barker} {\em et~al.\/}(2014){Barker}, {Dempsey} \&
  {Lithwick}]{Barker14}
{\sc {Barker}, A.~J., {Dempsey}, A.~M. \& {Lithwick}, Y.} 2014 {Theory and
  Simulations of Rotating Convection}. {\em \apj\/} {\bf 791}, 13.

\bibitem[{Bercovici} {\em et~al.\/}(1989){Bercovici}, {Schubert}, {Glatzmaier}
  \& {Zebib}]{Bercovici89}
{\sc {Bercovici}, D., {Schubert}, G., {Glatzmaier}, G.~A. \& {Zebib}, A.} 1989
  {Three-dimensional thermal convection in a spherical shell}. {\em Journal of
  Fluid Mechanics\/} {\bf 206}, 75--104.

\bibitem[{Blasius}(1908)]{Blasius08}
{\sc {Blasius}, H.} 1908 {Grenzschichten in Fl\"ussigkeiten mit kleiner
  Reibung}. {\em Z. Math. Phys.\/} {\bf 56}, 1--37.

\bibitem[{Boubnov} \& {Golitsyn}(1990)]{Boubnov90}
{\sc {Boubnov}, B.~M. \& {Golitsyn}, G.~S.} 1990 {Temperature and velocity
  field regimes of convective motions in a rotating plane fluid layer}. {\em
  Journal of Fluid Mechanics\/} {\bf 219}, 215--239.

\bibitem[{Breuer} {\em et~al.\/}(2004){Breuer}, {Wessling}, {Schmalzl} \&
  {Hansen}]{Breuer04}
{\sc {Breuer}, M., {Wessling}, S., {Schmalzl}, J. \& {Hansen}, U.} 2004 {Effect
  of inertia in Rayleigh-B{\'e}nard convection}. {\em \pre\/} {\bf 69}~(2),
  026302.

\bibitem[{Busse}(1970)]{Busse70}
{\sc {Busse}, F.~H.} 1970 {Thermal instabilities in rapidly rotating systems.}
  {\em Journal of Fluid Mechanics\/} {\bf 44}, 441--460.

\bibitem[{Busse} \& {Carrigan}(1974)]{Busse74}
{\sc {Busse}, F.~H. \& {Carrigan}, C.~R.} 1974 {Convection induced by
  centrifugal buoyancy}. {\em Journal of Fluid Mechanics\/} {\bf 62}, 579--592.

\bibitem[{Busse} \& {Or}(1986)]{Busse86}
{\sc {Busse}, F.~H. \& {Or}, A.~C.} 1986 {Convection in a rotating cylindrical
  annulus - Thermal Rossby waves}. {\em Journal of Fluid Mechanics\/} {\bf
  166}, 173--187.

\bibitem[{Cardin} \& {Olson}(1994)]{Cardin94}
{\sc {Cardin}, P. \& {Olson}, P.} 1994 {Chaotic thermal convection in a rapidly
  rotating spherical shell: consequences for flow in the outer core}. {\em
  Physics of the Earth and Planetary Interiors\/} {\bf 82}, 235--259.

\bibitem[{Cardin} \& {Olson}(2015)]{Cardin15}
{\sc {Cardin}, P. \& {Olson}, P.} 2015 {8.13 Experiments on Core Dynamics}. In
  {\em Treatise on Geophysics (Second Edition)\/}, Second edition edn. (ed.
  Gerald Schubert), pp. 317--339. Oxford: Elsevier.

\bibitem[{Chandrasekhar}(1961)]{Chandra61}
{\sc {Chandrasekhar}, S.} 1961 {\em {Hydrodynamic and Hydrodynamic
  stability}\/}. Oxford University Press.

\bibitem[{Cheng} {\em et~al.\/}(2015){Cheng}, {Stellmach}, {Ribeiro},
  {Grannan}, {King} \& {Aurnou}]{Cheng15}
{\sc {Cheng}, J.~S., {Stellmach}, S., {Ribeiro}, A., {Grannan}, A., {King},
  E.~M. \& {Aurnou}, J.~M.} 2015 {Laboratory-numerical models of rapidly
  rotating convection in planetary cores}. {\em Geophysical Journal
  International\/} {\bf 201}, 1--17.

\bibitem[{Chill\`a} \& {Schumacher}(2012)]{Chilla12}
{\sc {Chill\`a}, F. \& {Schumacher}, J.} 2012 {New perspectives in turbulent
  Rayleigh-B\'enard convection}. {\em The European Physical Journal E\/} {\bf
  35}~(7).

\bibitem[{Christensen}(2002)]{Christensen02}
{\sc {Christensen}, U.~R.} 2002 {Zonal flow driven by strongly supercritical
  convection in rotating spherical shells}. {\em Journal of Fluid Mechanics\/}
  {\bf 470}, 115--133.

\bibitem[{Christensen} \& {Aubert}(2006)]{Christensen06}
{\sc {Christensen}, U.~R. \& {Aubert}, J.} 2006 {Scaling properties of
  convection-driven dynamos in rotating spherical shells and application to
  planetary magnetic fields}. {\em Geophysical Journal International\/} {\bf
  166}, 97--114.

\bibitem[{Christensen} {\em et~al.\/}(2001){Christensen}, {Aubert}, {Cardin},
  {Dormy}, {Gibbons}, {Glatzmaier}, {Grote}, {Honkura}, {Jones}, {Kono},
  {Matsushima}, {Sakuraba}, {Takahashi}, {Tilgner}, {Wicht} \&
  {Zhang}]{Christensen01}
{\sc {Christensen}, U.~R., {Aubert}, J., {Cardin}, P., {Dormy}, E., {Gibbons},
  S., {Glatzmaier}, G.~A., {Grote}, E., {Honkura}, Y., {Jones}, C., {Kono}, M.,
  {Matsushima}, M., {Sakuraba}, A., {Takahashi}, F., {Tilgner}, A., {Wicht}, J.
  \& {Zhang}, K.} 2001 {A numerical dynamo benchmark}. {\em Physics of the
  Earth and Planetary Interiors\/} {\bf 128}, 25--34.

\bibitem[{Christensen} \& {Wicht}(2015)]{Christensen15}
{\sc {Christensen}, U.~R. \& {Wicht}, J.} 2015 {8.10 - Numerical Dynamo
  Simulations}. In {\em Treatise on Geophysics (Second Edition)\/}, Second
  edition edn. (ed. Gerald Schubert), pp. 245 -- 277. Oxford: Elsevier.

\bibitem[{Cordero} \& {Busse}(1992)]{Cordero92}
{\sc {Cordero}, S. \& {Busse}, F.~H.} 1992 {Experiments on convection in
  rotating hemispherical shells - Transition to a quasi-periodic state}. {\em
  \grl\/} {\bf 19}, 733--736.

\bibitem[{Davidson}(2013)]{Davidson13}
{\sc {Davidson}, P.~A.} 2013 {Scaling laws for planetary dynamos}. {\em
  Geophysical Journal International\/} {\bf 195}, 67--74.

\bibitem[{Davidson}(2015)]{DavidsonBook}
{\sc {Davidson}, P.~A.} 2015 {\em {Turbulence: an introduction for scientists
  and engineers}\/}. Oxford University Press.

\bibitem[{Dormy} {\em et~al.\/}(2004){Dormy}, {Soward}, {Jones}, {Jault} \&
  {Cardin}]{Dormy04}
{\sc {Dormy}, E., {Soward}, A.~M., {Jones}, C.~A., {Jault}, D. \& {Cardin}, P.}
  2004 {The onset of thermal convection in rotating spherical shells}. {\em
  Journal of Fluid Mechanics\/} {\bf 501}, 43--70.

\bibitem[{Ecke} \& {Niemela}(2014)]{Ecke14}
{\sc {Ecke}, R.~E. \& {Niemela}, J.~J.} 2014 {Heat Transport in the Geostrophic
  Regime of Rotating Rayleigh-B{\'e}nard Convection}. {\em Physical Review
  Letters\/} {\bf 113}~(11), 114301.

\bibitem[{Egbers} {\em et~al.\/}(2003){Egbers}, {Beyer}, {Bonhage},
  {Hollerbach} \& {Beltrame}]{Egbers03}
{\sc {Egbers}, C., {Beyer}, W., {Bonhage}, A., {Hollerbach}, R. \& {Beltrame},
  P.} 2003 {The geoflow-experiment on ISS (part I): Experimental preparation
  and design of laboratory testing hardware}. {\em Advances in Space
  Research\/} {\bf 32}, 171--180.

\bibitem[{Funfschilling} {\em et~al.\/}(2005){Funfschilling}, {Brown},
  {Nikolaenko} \& {Ahlers}]{Funfschilling05}
{\sc {Funfschilling}, D., {Brown}, E., {Nikolaenko}, A. \& {Ahlers}, G.} 2005
  {Heat transport by turbulent Rayleigh B{\'e}nard convection in cylindrical
  samples with aspect ratio one and larger}. {\em Journal of Fluid Mechanics\/}
  {\bf 536}, 145--154.

\bibitem[{Garcia} {\em et~al.\/}(2014){Garcia}, {S{\'a}nchez} \&
  {Net}]{Garcia14}
{\sc {Garcia}, F., {S{\'a}nchez}, J. \& {Net}, M.} 2014 {Numerical simulations
  of thermal convection in rotating spherical shells under laboratory
  conditions}. {\em Physics of the Earth and Planetary Interiors\/} {\bf 230},
  28--44.

\bibitem[{Gastine} \& {Wicht}(2012)]{Gastine12}
{\sc {Gastine}, T. \& {Wicht}, J.} 2012 {Effects of compressibility on driving
  zonal flow in gas giants}. {\em \icarus\/} {\bf 219}, 428--442.

\bibitem[{Gastine} {\em et~al.\/}(2015){Gastine}, {Wicht} \&
  {Aurnou}]{Gastine15}
{\sc {Gastine}, T., {Wicht}, J. \& {Aurnou}, J.~M.} 2015 {Turbulent
  Rayleigh-B{\'e}nard convection in spherical shells}. {\em Journal of Fluid
  Mechanics\/} {\bf 778}, 721--764.

\bibitem[{Gastine} {\em et~al.\/}(2016){Gastine}, {Wicht}, {Barik}, {Putigny}
  \& {Duarte}]{MagIC}
{\sc {Gastine}, T., {Wicht}, J., {Barik}, A., {Putigny}, B. \& {Duarte},
  L.~D.~V.} 2016 {MagIC v5.4, doi:10.5281/zenodo.51723}.

\bibitem[{Gillet} \& {Jones}(2006)]{Gillet06}
{\sc {Gillet}, N. \& {Jones}, C.~A.} 2006 {The quasi-geostrophic model for
  rapidly rotating spherical convection outside the tangent cylinder}. {\em
  Journal of Fluid Mechanics\/} {\bf 554}, 343--369.

\bibitem[{Gilman}(1977)]{Gilman77}
{\sc {Gilman}, P.~A.} 1977 {Nonlinear Dynamics of Boussinesq Convection in a
  Deep Rotating Spherical Shell. I.} {\em GAFD\/} {\bf 8}, 93--135.

\bibitem[{Gilman} \& {Glatzmaier}(1981)]{Glatz1}
{\sc {Gilman}, P.~A. \& {Glatzmaier}, G.~A.} 1981 {Compressible convection in a
  rotating spherical shell - I - Anelastic equations}. {\em \apjs\/} {\bf 45},
  335--349.

\bibitem[{Greenspan}(1968)]{Greenspan68}
{\sc {Greenspan}, H.~P.} 1968 {\em {The theory of rotating fluids}\/}.
  Cambridge University Press.

\bibitem[{Grossmann} \& {Lohse}(2000)]{Grossmann00}
{\sc {Grossmann}, S. \& {Lohse}, D.} 2000 {Scaling in thermal convection: a
  unifying theory}. {\em Journal of Fluid Mechanics\/} {\bf 407}, 27--56.

\bibitem[{Guervilly}(2010)]{Guervilly10}
{\sc {Guervilly}, C.} 2010 {Dynamos num\'eriques plan\'etaires g\'en\'er\'ees
  par cisaillement en surface ou chauffage interne}. PhD thesis, {Sciences de
  la Terre, univers et environnement, Universit\'e de Grenoble, France}.

\bibitem[{Guervilly} \& {Cardin}(2016)]{Guervilly16}
{\sc {Guervilly}, C. \& {Cardin}, P.} 2016 {Subcritical convection in a rapidly
  rotating sphere at low Prandtl number}. {\em ArXiv e-prints\/} .

\bibitem[{Hart} {\em et~al.\/}(1986){Hart}, {Glatzmaier} \& {Toomre}]{Hart86}
{\sc {Hart}, J.~E., {Glatzmaier}, G.~A. \& {Toomre}, J.} 1986 {Space-laboratory
  and numerical simulations of thermal convection in a rotating hemispherical
  shell with radial gravity}. {\em Journal of Fluid Mechanics\/} {\bf 173},
  519--544.

\bibitem[{Horn} \& {Shishkina}(2015)]{Horn15}
{\sc {Horn}, S. \& {Shishkina}, O.} 2015 {Toroidal and poloidal energy in
  rotating Rayleigh-B{\'e}nard convection}. {\em Journal of Fluid Mechanics\/}
  {\bf 762}, 232--255.

\bibitem[{Ingersoll} \& {Pollard}(1982)]{Ingersoll82}
{\sc {Ingersoll}, A.~P. \& {Pollard}, D.} 1982 {Motion in the interiors and
  atmospheres of Jupiter and Saturn - Scale analysis, anelastic equations,
  barotropic stability criterion}. {\em \icarus\/} {\bf 52}, 62--80.

\bibitem[{Jarvis}(1993)]{Jarvis93}
{\sc {Jarvis}, G.~T.} 1993 {Effects of curvature on two-dimensional models of
  mantle convection - Cylindrical polar coordinates}. {\em \jgr\/} {\bf 98},
  4477--4485.

\bibitem[{Jones}(2015)]{Jones15}
{\sc {Jones}, C.~A.} 2015 {8.05 Thermal and Compositional Convection in the
  Outer Core}. In {\em Treatise on Geophysics (Second Edition)\/}, Second
  edition edn. (ed. Gerald Schubert), pp. 115 --159. Oxford: Elsevier.

\bibitem[{Jones} {\em et~al.\/}(2011){Jones}, {Boronski}, {Brun}, {Glatzmaier},
  {Gastine}, {Miesch} \& {Wicht}]{Jones11}
{\sc {Jones}, C.~A., {Boronski}, P, {Brun}, A.~S., {Glatzmaier}, G.~A.,
  {Gastine}, T., {Miesch}, M.~S. \& {Wicht}, J.} 2011 {Anelastic
  convection-driven dynamo benchmarks}. {\em \icarus\/} {\bf 216}, 120--135.

\bibitem[{Julien} {\em et~al.\/}(2016){Julien}, {Aurnou}, {Calkins},
  {Knobloch}, {Marti}, {Stellmach} \& {Vasil}]{Julien16}
{\sc {Julien}, K., {Aurnou}, J.~M., {Calkins}, M.~A., {Knobloch}, E., {Marti},
  P., {Stellmach}, S. \& {Vasil}, G.~M.} 2016 {A nonlinear model for
  rotationally constrained convection with Ekman pumping}. {\em Journal of
  Fluid Mechanics\/} {\bf 798}, 50--87.

\bibitem[{Julien} {\em et~al.\/}(2012{\natexlab{{\em a\/}}}){Julien},
  {Knobloch}, {Rubio} \& {Vasil}]{Julien12a}
{\sc {Julien}, K., {Knobloch}, E., {Rubio}, A.~M. \& {Vasil}, G.~M.}
  2012{\natexlab{{\em a\/}}} {Heat Transport in Low-Rossby-Number
  Rayleigh-B{\'e}nard Convection}. {\em Physical Review Letters\/} {\bf
  109}~(25), 254503.

\bibitem[{Julien} {\em et~al.\/}(1996){Julien}, {Legg}, {McWilliams} \&
  {Werne}]{Julien96}
{\sc {Julien}, K., {Legg}, S., {McWilliams}, J. \& {Werne}, J.} 1996 {Rapidly
  rotating turbulent Rayleigh-Benard convection}. {\em Journal of Fluid
  Mechanics\/} {\bf 322}, 243--273.

\bibitem[{Julien} {\em et~al.\/}(2012{\natexlab{{\em b\/}}}){Julien}, {Rubio},
  {Grooms} \& {Knobloch}]{Julien12}
{\sc {Julien}, K., {Rubio}, A.~M., {Grooms}, I. \& {Knobloch}, E.}
  2012{\natexlab{{\em b\/}}} {Statistical and physical balances in low Rossby
  number Rayleigh-B{\'e}nard convection}. {\em Geophysical and Astrophysical
  Fluid Dynamics\/} {\bf 106}, 392--428.

\bibitem[{King} \& {Aurnou}(2013)]{King13PNAS}
{\sc {King}, E.~M. \& {Aurnou}, J.~M.} 2013 {Turbulent convection in liquid
  metal with and without rotation}. {\em Proceedings of the National Academy of
  Sciences\/} {\bf 110}~(17), 6688--6693.

\bibitem[{King} \& {Buffett}(2013)]{KingBuffett13}
{\sc {King}, E.~M. \& {Buffett}, B.~A.} 2013 {Flow speeds and length scales in
  geodynamo models: The role of viscosity}. {\em Earth and Planetary Science
  Letters\/} {\bf 371}, 156--162.

\bibitem[{King} {\em et~al.\/}(2010){King}, {Soderlund}, {Christensen}, {Wicht}
  \& {Aurnou}]{King10}
{\sc {King}, E.~M., {Soderlund}, K.~M., {Christensen}, U.~R., {Wicht}, J. \&
  {Aurnou}, J.~M.} 2010 {Convective heat transfer in planetary dynamo models}.
  {\em Geochemistry, Geophysics, Geosystems\/} {\bf 11}, 6016.

\bibitem[{King} {\em et~al.\/}(2012){King}, {Stellmach} \& {Aurnou}]{King12}
{\sc {King}, E.~M., {Stellmach}, S. \& {Aurnou}, J.~M.} 2012 {Heat transfer by
  rapidly rotating Rayleigh-B{\'e}nard convection}. {\em Journal of Fluid
  Mechanics\/} {\bf 691}, 568--582.

\bibitem[{King} {\em et~al.\/}(2013){King}, {Stellmach} \& {Buffett}]{King13}
{\sc {King}, E.~M., {Stellmach}, S. \& {Buffett}, B.} 2013 {Scaling behaviour
  in Rayleigh-B{\'e}nard convection with and without rotation}. {\em Journal of
  Fluid Mechanics\/} {\bf 717}, 449--471.

\bibitem[{King} {\em et~al.\/}(2009){King}, {Stellmach}, {Noir}, {Hansen} \&
  {Aurnou}]{KingNature09}
{\sc {King}, E.~M., {Stellmach}, S., {Noir}, J., {Hansen}, U. \& {Aurnou},
  J.~M.} 2009 {Boundary layer control of rotating convection systems}. {\em
  \nat\/} {\bf 457}, 301--304.

\bibitem[{Kraichnan}(1962)]{Kraichnan62}
{\sc {Kraichnan}, R.~H.} 1962 {Turbulent Thermal Convection at Arbitrary
  Prandtl Number}. {\em Physics of Fluids\/} {\bf 5}, 1374--1389.

\bibitem[{Kunnen} {\em et~al.\/}(2010){Kunnen}, {Geurts} \& {Clercx}]{Kunnen10}
{\sc {Kunnen}, R.~P.~J., {Geurts}, B.~J. \& {Clercx}, H.~J.~H.} 2010
  {Experimental and numerical investigation of turbulent convection in a
  rotating cylinder}. {\em Journal of Fluid Mechanics\/} {\bf 642}, 445.

\bibitem[{Kunnen} {\em et~al.\/}(2016){Kunnen}, {Ostilla-M\'onico}, {van der
  Poel}, {Verzicco} \& {Lohse}]{Kunnen16}
{\sc {Kunnen}, R.~P.~J., {Ostilla-M\'onico}, R., {van der Poel}, E.~P.,
  {Verzicco}, R. \& {Lohse}, D.} 2016 {Transition to geostrophic convection:
  the role of the boundary conditions}. {\em Journal of Fluid Mechanics\/} {\bf
  799}, 413--432.

\bibitem[{Lakkaraju} {\em et~al.\/}(2012){Lakkaraju}, {Stevens}, {Verzicco},
  {Grossmann}, {Prosperetti}, {Sun} \& {Lohse}]{Lakkaraju12}
{\sc {Lakkaraju}, R., {Stevens}, R.~J.~A.~M., {Verzicco}, R., {Grossmann}, S.,
  {Prosperetti}, A., {Sun}, C. \& {Lohse}, D.} 2012 {Spatial distribution of
  heat flux and fluctuations in turbulent Rayleigh-B{\'e}nard convection}. {\em
  \pre\/} {\bf 86}~(5), 056315.

\bibitem[{Liu} \& {Ecke}(1997)]{Liu97}
{\sc {Liu}, Y. \& {Ecke}, R.~E.} 1997 {Heat Transport Scaling in Turbulent
  Rayleigh-B{\'e}nard Convection: Effects of Rotation and Prandtl Number}. {\em
  Physical Review Letters\/} {\bf 79}, 2257--2260.

\bibitem[{Liu} \& {Ecke}(2011)]{Liu11}
{\sc {Liu}, Y. \& {Ecke}, R.~E.} 2011 {Local temperature measurements in
  turbulent rotating Rayleigh-B{\'e}nard convection}. {\em \pre\/} {\bf
  84}~(1), 016311.

\bibitem[{Oruba} \& {Dormy}(2014)]{Oruba14}
{\sc {Oruba}, L. \& {Dormy}, E.} 2014 {Predictive scaling laws for spherical
  rotating dynamos}. {\em Geophysical Journal International\/} {\bf 198},
  828--847.

\bibitem[{Plumley} {\em et~al.\/}(2016){Plumley}, {Julien}, {Marti} \&
  {Stellmach}]{Plumley16}
{\sc {Plumley}, M., {Julien}, K., {Marti}, P. \& {Stellmach}, S.} 2016 {The
  effects of Ekman pumping on quasi-geostrophic Rayleigh-Benard convection}.
  {\em Journal of Fluid Mechanics\/} {\bf 803}, 51--71.

\bibitem[{Prandtl}(1905)]{Prandtl04}
{\sc {Prandtl}, L.} 1905 {\em Verhandlungen des III. Int. Math. Kongr.,
  Heidelberg, 1904\/}. Leipzig: Teubner, p. 484--491.

\bibitem[{Rhines}(1975)]{Rhines75}
{\sc {Rhines}, P.~B.} 1975 {Waves and turbulence on a beta-plane}. {\em Journal
  of Fluid Mechanics\/} {\bf 69}, 417--443.

\bibitem[{Rossby}(1969)]{Rossby69}
{\sc {Rossby}, H.~T.} 1969 {A study of Benard convection with and without
  rotation}. {\em Journal of Fluid Mechanics\/} {\bf 36}, 309--335.

\bibitem[{Schaeffer}(2013)]{Schaeffer13}
{\sc {Schaeffer}, N.} 2013 {Efficient spherical harmonic transforms aimed at
  pseudospectral numerical simulations}. {\em Geochemistry, Geophysics,
  Geosystems\/} {\bf 14}, 751--758.

\bibitem[{Schmitz} \& {Tilgner}(2009)]{Schmitz09}
{\sc {Schmitz}, S. \& {Tilgner}, A.} 2009 {Heat transport in rotating
  convection without Ekman layers}. {\em \pre\/} {\bf 80}~(1), 015305.

\bibitem[{Shew} \& {Lathrop}(2005)]{Shew05}
{\sc {Shew}, W.~L. \& {Lathrop}, D.~P.} 2005 {Liquid sodium model of
  geophysical core convection}. {\em Physics of the Earth and Planetary
  Interiors\/} {\bf 153}, 136--149.

\bibitem[{Shishkina} {\em et~al.\/}(2010){Shishkina}, {Stevens}, {Grossmann} \&
  {Lohse}]{Shishkina10}
{\sc {Shishkina}, O., {Stevens}, R.~J.~A.~M., {Grossmann}, S. \& {Lohse}, D.}
  2010 {Boundary layer structure in turbulent thermal convection and its
  consequences for the required numerical resolution}. {\em New Journal of
  Physics\/} {\bf 12}~(7), 075022.

\bibitem[{Soderlund} {\em et~al.\/}(2012){Soderlund}, {King} \&
  {Aurnou}]{Soderlund12}
{\sc {Soderlund}, K.~M., {King}, E.~M. \& {Aurnou}, J.~M.} 2012 {The influence
  of magnetic fields in planetary dynamo models}. {\em Earth and Planetary
  Science Letters\/} {\bf 333}, 9--20.

\bibitem[{Stellmach} {\em et~al.\/}(2014){Stellmach}, {Lischper}, {Julien},
  {Vasil}, {Cheng}, {Ribeiro}, {King} \& {Aurnou}]{Stellmach14}
{\sc {Stellmach}, S., {Lischper}, M., {Julien}, K., {Vasil}, G., {Cheng},
  J.~S., {Ribeiro}, A., {King}, E.~M. \& {Aurnou}, J.~M.} 2014 {Approaching the
  Asymptotic Regime of Rapidly Rotating Convection: Boundary Layers versus
  Interior Dynamics}. {\em Physical Review Letters\/} {\bf 113}~(25), 254501.

\bibitem[{Stevens} {\em et~al.\/}(2013){Stevens}, {Clercx} \&
  {Lohse}]{Stevens13a}
{\sc {Stevens}, R.~J.~A.~M., {Clercx}, H.~J.~H. \& {Lohse}, D.} 2013 {Heat
  transport and flow structure in rotating Rayleigh-B{\'e}nard convection}.
  {\em European Journal of Mechanics B Fluids\/} {\bf 40}, 41--49.

\bibitem[{Stevens} {\em et~al.\/}(2010){Stevens}, {Verzicco} \&
  {Lohse}]{Stevens10}
{\sc {Stevens}, R.~J.~A.~M., {Verzicco}, R. \& {Lohse}, D.} 2010 {Radial
  boundary layer structure and Nusselt number in Rayleigh-B{\'e}nard
  convection}. {\em Journal of Fluid Mechanics\/} {\bf 643}, 495--507.

\bibitem[{Stevenson}(1979)]{Stevenson79}
{\sc {Stevenson}, D.~J.} 1979 {Turbulent thermal convection in the presence of
  rotation and a magnetic field - A heuristic theory}. {\em Geophysical and
  Astrophysical Fluid Dynamics\/} {\bf 12}, 139--169.

\bibitem[{Sumita} \& {Olson}(2003)]{SumitaOlson03}
{\sc {Sumita}, I. \& {Olson}, P.} 2003 {Experiments on highly supercritical
  thermal convection in a rapidly rotating hemispherical shell}. {\em Journal
  of Fluid Mechanics\/} {\bf 492}, 271--287.

\bibitem[{Tilgner} \& {Busse}(1997)]{Tilgner97}
{\sc {Tilgner}, A. \& {Busse}, F.~H.} 1997 {Finite-amplitude convection in
  rotating spherical fluid shells}. {\em Journal of Fluid Mechanics\/} {\bf
  332}, 359--376.

\bibitem[{Verzicco} \& {Camussi}(1999)]{Verzicco99}
{\sc {Verzicco}, R. \& {Camussi}, R.} 1999 {Prandtl number effects in
  convective turbulence}. {\em Journal of Fluid Mechanics\/} {\bf 383}, 55--73.

\bibitem[{Vidal} \& {Schaeffer}(2015)]{Vidal15}
{\sc {Vidal}, J. \& {Schaeffer}, N.} 2015 {Quasi-geostrophic modes in the
  Earth's fluid core with an outer stably stratified layer}. {\em Geophysical
  Journal International\/} {\bf 202}~(3), 2182--2193.

\bibitem[{Wicht}(2002)]{Wicht02}
{\sc {Wicht}, J.} 2002 {Inner-core conductivity in numerical dynamo
  simulations}. {\em Physics of the Earth and Planetary Interiors\/} {\bf 132},
  281--302.

\bibitem[{Yadav} {\em et~al.\/}(2016){Yadav}, {Gastine}, {Christensen},
  {Duarte} \& {Reiners}]{Yadav16}
{\sc {Yadav}, R.~K., {Gastine}, T., {Christensen}, U.~R., {Duarte}, L.~D.~V. \&
  {Reiners}, A.} 2016 {Effect of shear and magnetic field on the heat-transfer
  efficiency of convection in rotating spherical shells}. {\em Geophysical
  Journal International\/} {\bf 204}, 1120--1133.

\bibitem[{Zhong} \& {Ahlers}(2010)]{Zhong10}
{\sc {Zhong}, J.-Q. \& {Ahlers}, G.} 2010 {Heat transport and the large-scale
  circulation in rotating turbulent Rayleigh-B{\'e}nard convection}. {\em
  Journal of Fluid Mechanics\/} {\bf 665}, 300--333.

\bibitem[{Zhong} {\em et~al.\/}(2009){Zhong}, {Stevens}, {Clercx}, {Verzicco},
  {Lohse} \& {Ahlers}]{Zhong09}
{\sc {Zhong}, J.-Q., {Stevens}, R.~J.~A.~M., {Clercx}, H.~J.~H., {Verzicco},
  R., {Lohse}, D. \& {Ahlers}, G.} 2009 {Prandtl-, Rayleigh-, and Rossby-Number
  Dependence of Heat Transport in Turbulent Rotating Rayleigh-B{\'e}nard
  Convection}. {\em Physical Review Letters\/} {\bf 102}~(4), 044502.

\end{thebibliography}

\end{document}